\documentclass[12pt]{article}
\usepackage{psfig,epsfig,supertabular,cite,amsmath,amssymb}
\usepackage{graphics}

\input epsf

\newcommand{\be}{\begin{equation}}
\newcommand{\ee}{\end{equation}}
\newcommand{\bea}{\begin{eqnarray}}
\newcommand{\eea}{\end{eqnarray}}
\def\barr{\begin{array}}
\def\earr{\end{array}}

\newcommand{\nn}{\nonumber}
\newcommand{\eq}[1]{Eq.~(\ref{#1})}
\newcommand{\eqs}[1]{Eqs.~(\ref{#1})}

\def\vep{\varepsilon}

\def\rhobar{\bar \rho}
\def\etabar{\bar \eta}

\def\diag{\mathrm{Diag}}

\begin{document}


\newcommand{\sheptitle}
{Study of theory and phenomenology of some classes of
family symmetry and unification models}

\newcommand{\shepauthor}
{G. L. Kane$^{1 *}$, S. F. King$^{2 \dagger}$, I. N. R. Peddie$^{2 **}$ and L.
Velasco-Sevilla$^{1 \dagger\dagger}$}

\newcommand{\shepaddress}
{$^1$Michigan Center for Theoretical Physics, Randall Laboratory,\\
University of Michigan, 500 E University Avenue,\\
Ann Arbor, MI 48109, U.S.A. \\
\vspace{0.25in}
$^2$School of Physics and Astronomy, University of Southampton, \\
        Southampton, SO17 1BJ, U.K}

\vspace{0.25in}

\newcommand{\shepabstract} {We review and compare theoretically and
phenomenologically a number of possible family symmetries, which
when combined with unification, could
be important in explaining quark, lepton and neutrino masses and mixings,
providing new results in several cases. Theoretical possibilities
include Abelian or non-Abelian, symmetric or non symmetric Yukawa matrices,
Grand Unification or not. Our main focus is on anomaly-free
$U(1)$ family symmetry
combined with $SU(5)$ unification, although we also discuss
other possibilities. We provide a detailed phenomenological fit
of the fermion masses and mixings for several examples,
and discuss the supersymmetric flavour issues in such theories,
including a detailed analysis of lepton flavour violation.
We show that it is not possible to quantitatively and decisively
discriminate between these different theoretical possibilities 
at the present time.}

\begin{titlepage}
\begin{flushright}
hep-ph/0504038 \\
SHEP-05-12\\
MCTP-05-70
\end{flushright}
\begin{center}
{\large{\bf \sheptitle}}
\\ \shepauthor \\ \mbox{} \\ {\it \shepaddress} \\
{\bf Abstract} \bigskip \end{center} \setcounter{page}{0}
\shepabstract
\begin{flushleft}
\today
\end{flushleft}

\vskip 0.5in
\noindent
$*$ ~e-mail: gkane@umich.edu\\
$\dagger$ ~e-mail: sfk@hep.phys.soton.ac.uk\\
$**$ e-mail: inrp@hep.phys.soton.ac.uk\\
$\dagger\dagger$ e-mail: lvelsev@umich.edu
\end{titlepage}

\newpage

\section{Introduction}
The hierarchy of quark and charged lepton masses and the small quark mixing angles
has been one of the most puzzling aspects left unresolved by the Standard Model.
The recent discovery of neutrino masses and mixings has provided further clues
in the search for the new physics Beyond the Standard Model which must be
responsible for the pattern of fermion masses and mixing angles.
One promising approach to understanding the fermion spectrum is
the idea of family symmetry, and in particular the idea of a
$U(1)$ family symmetry as originally proposed by Froggatt and Nielsen \cite{Froggatt:1978nt}.
Such an approach was given considerable impetus by the observation
that in many string constructions additional $U(1)$ symmetries
are ubiquitous, and furthermore such a gauged broken $U(1)$
could provide a phenomenologically viable candidate
family symmetry by virtue of the Green-Schwartz anomaly cancellation
mechanism \cite{Green:1984sg} which provides a string solution to the no-go theorem
that anomaly freedom requires such symmetries to be family independent \cite{Weinberg:anomalies}.
As a result of this a considerable literature has developed in recent
years based on string-inspired $U(1)$ family symmetries
\cite{Chankowski:2005qp,Babuetal}.

Many non-abelian family symmetries have also been considered, 
for example based on $SU(3)$ family symmetry \cite{King:2001uz},
and also
textures and analyses of fermion masses have been done not using any family
symmetry.  At the present time some very successful approaches exist, and
others that may with modification also be effective.  Family symmetries can
be abelian or non-abelian, they can require symmetric Yukawa matrices or
not, they can be imposed with or without an associated grand unified theory,
and so on.  Criteria that could be used to choose among possible approaches
include not only describing the quark masses and mixings, and the charged
lepton masses, but also neutrino masses and mixings, supersymmetry soft
breaking effects (since particularly the trilinear couplings are affected by
the Yukawa couplings), how many parameters are used to describe the data,
whether some results such as the Cabibbo angle are generic or fitted, and
more.  One of our main goals here is to look at the various possibilities
systematically and see if some seem to be favoured by how well they do on a
set of criteria such as the above listed ones.  Presumably family
symmetries originate in string theories, and are different for different
string constructions that lead to a description of nature, so identifying a
unique family symmetry (or a subset of possible ones) could point strongly
toward a class of string theories and away from other classes. At the
present time this approach is not very powerful, though it gives some
interesting insights, but better analyses and additional data may improve it.

In this paper we shall consider $U(1)$ family symmetries and
unification as a viable framework for quark and lepton masses and
mixing angles in the light of neutrino mass and mixing data
\cite{King:2003jb}, using
sequential right-hand neutrino dominance \cite{King:1998jw} as a guide
to constructing hierarchical neutrino mass models with bi-large
mixing.  As has been pointed earlier \cite{Ross:2000fn}, models which
satisfy the Gatto-Sartori-Tonin relations (GST
\cite{Gatto:1968ss}){\footnote{$V_{us}=|\sqrt{\frac{m_d}{m_s}}-e^{i\Phi_1}\sqrt{\frac{m_u}{m_c}}|$}}
require the presence of both positive and negative Abelian charges.
As we will discuss, the sequential dominance conditions require also
the presence of both positive and negative Abelian charges, and hence
at least two flavon fields of equal and opposite charges. These models
however result in complicated $U(1)$ charges, on the other hand
Non-GST models have a simpler charge structure and may be possible to
realize in a more general context. In this work we also consider non
GST cases.

We shall consider $U(1)$ family symmetry combined with unified gauge groups
based on $SU(5)$ and $SO(10)$, assuming a Georgi-Jarlskog relation,
and also consider non-unified models without such a relation.
We will present new classes of solutions
to the anomaly cancellation conditions and perform phenomenological fits,
and we will compare the different classes of $U(1)$
to each other and to non-Abelian family symmetry models based on
$SU(3)$ \cite{King:2001uz}, 
by performing specific phenomenological fits to the
undetermined coefficients of the operators.
Finally we will consider the implications of such
an approach on flavour-changing processes in the framework
of supersymmetry, leaving a detailed analysis for a future reference.

The layout of the paper is as follows. In Section \ref{sec:anomconst} we consider the
general conditions for Green-Schwartz anomaly cancellation, and move on to describe
the classes of solutions, by whether they are consistent with $SU(5)$, $SO(10)$, Pati-Salam
unification of representations, generalized non-unified relations, or not at all consistent
with unification. Having found these solutions, we move on in section \ref{sec:new-paramaterisation}
to re-parametrize in terms of differences in $U(1)_F$ charges. In section \ref{sec:su5q} we consider
the constraints on the Yukawa textures from requiring acceptable quark mixings and quark and lepton
masses. Then in section \ref{sec:neuts}, the constraints from getting acceptable neutrino masses
and mixings from single right-handed neutrino dominance (SRHND) models, which are a class of see-saw
models. In section \ref{sec:su5-solut-satisfy-GST} we construct solutions which are consistent with
$SU(5)$ unification, the Gatto-Satori-Tonin (GST) relation \cite{Gatto:1968ss}, and correct fermion masses and mixings. 
In section \ref{sec:su5-solutions-not-GST} we construct solutions which are consistent with $SU(5)$ unification,
correct fermion masses and mixing angles but which are not consistent with the GST relation. In section \ref{sec:non-su5-cases}
we construct solutions which are not consistent with $SU(5)$ unification. In section \ref{sec:fitsmasses}, we take
some of the solutions constructed in section \ref{sec:su5-solut-satisfy-GST} and section \ref{sec:su5-solutions-not-GST}
and fit the arbitrary $O(1)$ parameters to try to closely predict the observed fermion masses and mixing angles.
Then in section \ref{sec:susyconst} we briefly consider whether flavour changing processes will be dangerously high
in these models, presenting two specific scenarios: a non minimal sugra possibility and a string-inspired mSUGRA-like scenario which is expected to be (or be close to) the best-case scenario for flavour-changing and for which we check explicitly $\mu\rightarrow e\gamma$  Finally, we conclude in
section \ref{sec:conclusions}.

\section{Anomaly Constraints on $U(1)$ Family symmetries\label{sec:anomconst}}

\subsection{Green-Schwartz anomaly cancellation}
\label{sec:green-schw-anom}

Consider an arbitrary $U(1)$ symmetry which extends the Standard 
Model gauge group. If
we were to insist that it does not contribute to mixed anomalies with 
the Standard Model,
we would find that the generators of $U(1)$ would be a linear 
combination of Weak hypercharge
and $B-L$ \cite{Weinberg:anomalies}. This clearly is not useful for 
family symmetries, so we need to use a more sophisticated
way of removing the anomalies, Green-Schwartz anomaly cancellation \cite{Green:1984sg}. In 
this case, we can cancel the mixed
$U(1) - SU(3) - SU(3)$, $U(1) - SU(2) - SU(2)$ and 
$U(1) - U(1)_Y - U(1)_Y$ anomalies, $A_3$, $A_2$, and $A_1$
if they appear in the ratio:
\begin{equation}
  \label{eq:aratio}
  A_3 : A_2 : A_1: A_{U(1)}:A_G = k_3 : k_2 : k_1: 3 k_{U(1)}:24,
\end{equation}
where we have included the relations to the anomalies of the anomalous flavour groups $A_{U(1)}$ and the gravitational anomaly;  $k_i$ are the Kac-Moody levels of the gauge groups, defined by the GUT-scale relation:
\begin{equation}
  \label{eq:g2ratio}
  g_3^2 k_3 = g_2^2 k_2 = g_1^2 k_1
\end{equation}
If we work with a GUT that has the canonical GUT normalization, we 
find:
\begin{equation}
  \label{eq:arelation}
  A_3 = A_2 = \frac{3}{5} A_1
\end{equation}
But we still require that the $U(1) - U(1) - U(1)_Y$ 
anomaly, $A_1^\prime$ vanishes.
Now, the anomalies are given by:
\begin{equation}
  \label{eq:4}
  A_i = \frac{1}{2}\mathrm{Tr}\left[ \left\{ T^{(i)}_a , 
T^{(i)}_c\right\} T^\prime_c \right].
\end{equation}
We then use the fact that $\left\{T_a, T_b\right\} = \delta_{ab} 
\mathbf{1}$ for $SU(N)$ and
$\left\{ Y, Y\right\} = 2Y^2$ for $U(1)_Y$ to obtain:
\begin{eqnarray}
  \label{eq:A3}
  A_3 &=& \frac{1}{2} \left[ \sum_{i = 1}^3 ( 2 q_i + u_i + d_i ) 
\right] \\
  \label{eq:A2}
  A_2 &=& \frac{1}{2} \left[ \sum_{i = 1}^3 ( 3 q_i + l_i) + h_u + h_d 
\right] \\
  \frac{3}{5} A_1 &=& \frac{1}{2}
  \left[
    \sum_{i = 1}^3 ( \frac{q_i}{5} + \frac{8 u_i}{5} + \frac{2}{5} d_i 
+ \frac{3 l_i}{5}
    + \frac{6 e_i}{5} ) + \frac{3}{5}( h_u + h_d )
  \right]\\
  \label{eq:A1p}
  A_1^\prime &=& \sum_{i=1}^3 ( -q_i^2 + 2 u_i^2 - d_i^2 + l_i^2 - 
e_i^2 ) + ( h_d^2 - h_u^2 ) = 0
\end{eqnarray}
\begin{table}[ht]
  \centering
  \begin{tabular}{|c|cccccccc|}
    \hline
    Field & $Q_i$ & $\overline{U}_i$ & $\overline{D}_i$ & $L_i$ & 
$\overline{E}_i$ & $\overline{N}_i$ & $H_u$ & $H_d$ \\
    \hline
    Charge & $q_i$ & $u_i$ & $d_i$ & $l_i$ & $e_i$ & $n_i$ & $h_u$ & 
$h_d$ \\
    \hline
  \end{tabular}
  \caption{Fields and family charges}
  \label{tab:charges}
\end{table}
Since in the mixed anomalies of the $U(1)$ group with the SM gauge group that cancel via 
the Green-Schwartz mechanism wherever a charge
appears, it appears in a sum, we parameterize the sums as follows \cite{Jain:1994hd}:
\begin{eqnarray}
  \label{eq:sumqi}
  \sum_{i=1}^3 q_i \!&=&\! x + u,\quad \sum_{i=1}^3 u_i \ =\ x + 2u, \\
  \label{eq:sumdi}
  \sum_{i=1}^3 d_i \!&=&\! y + v,\quad \sum_{i=1}^3 l_i \ =\ y, \\
  \label{eq:sumei}
  \sum_{i=1}^3 e_i \!&=&\! x, \\
  \label{eq:hd}
  h_u \!&=&\! -z,\quad  h_d \ =\ z + ( u + v ).
\end{eqnarray}
Substituting  \eq{eq:sumqi}-\eq{eq:hd} into \eq{eq:A3}-\eq{eq:A1p}
we find that they satisfy
Eq.~(\ref{eq:arelation}):
\begin{equation}
  \label{eq:anomalies}
  A_3 = A_2 = \frac{3}{5} A_1 = \frac{1}{2} \left[ 3x + 4u + y + 
v\right],
\end{equation}
which shows that the parameterization is consistent. However we need to find those solutions which also satisfy $A'_{1}=0$.
We will see how we can achieve this for different cases. Since the proposal of the GS anomaly mechanism it has been known that the easiest solution, 
$u=v=0$, leads to a $SU(5)$ or Pati-Salam group realization of mass matrices. Another possible solution is to have $u = -v \ne 0$. Both these forms
 admit a  SUSY $\mu$ term in the tree level
 superpotential at the gravitational scale. However given the form of  \eq{eq:sumqi}-\eq{eq:hd} one can try to use the flavour symmetry in order 
to forbid this term, allowing it just in the K\"ahler potential and thus invoking the Giudice-Masiero \cite{Giudice:1988yz} mechanism in order to generate the
 $\mu$ of the desired phenomenological order. Therefore apart from the cases $u+v=0$ we examine plausible cases for  $u \ne -v \ne 0$. Of course 
in the cases $u=v=0, u=-v\ne 0$ one can use another symmetry to forbid the $\mu$ term in the superpotential, however it is appealing if the flavour
 symmetry forbids the $\mu$ term at high scales.

\subsection{Anomaly free $A_1^\prime$ with $u = v = 0$ solutions\label{sec:yukawa-textures-uv-zero}}
In this case the parameterization simplifies and in fact we can
decompose the $U(1)$ charges in flavour independent and flavour dependent parts
\begin{equation}
\label{eq:FIaFDch}
f_i = \frac{1}{3}f + f_i^\prime.
\end{equation}
The first term is flavour independent because it just depends on the total sum of the individual charges and the  $f_i^\prime$ are flavour dependent charges. We can always find $x$ and $y$ which satisfy
\begin{equation}
  \label{eq:sumfip}
\sum_{i=1}^3 f_i^\prime = 0.
\end{equation}
In this way  $A'_{1}$ can be expressed in flavour independent plus flavour dependent terms
\begin{eqnarray}
A'_{1}=A'_{1FI}+ A'_{1FD}.
\end{eqnarray}
Following this, with the unfortunate notation that we have a new $u$, 
completely unrelated to the $u$ that we have already set to zero, we then have:
\begin{eqnarray}
  \label{eq:A1pFIFD}
  A_1^\prime &=& A'_{1FI}+ A'_{1FD}\nn\\
 &=& \frac{1}{3} \left[ - q^2 + 2 u^2 - d^2 + l^2 - e^2 
\right] 
  + \sum_{i = 1}^3 ( -q_i^{\prime\; 2} + 2 u_i^{\prime\; 2} - 
d_i^{\prime\;2} + l_i^{\prime\;2} - e_i^{\prime\;2} )
\end{eqnarray}
Now it is clear that the terms in the
square bracket in \eq{eq:A1pFIFD} are family 
independent. It turns out that the square bracket term is
automatically
zero in this case, since from Eqs.\ref{eq:sumqi}-\ref{eq:sumei},
we have: $q = u = e = x$ and $l = d = y$. Then 
we have to make the family dependent part (the second term in \eq{eq:A1pFIFD}) vanish.

\subsubsection{$SU(5)$ and $SO(10)$ type cases}
One way to make the family dependent part  vanish,  $A'_{1FD}=0$ ,   
is to set $l_i = d_i$ and $q_i = u_i = e_i$ 
\footnote{The reason that the charges are unprimed
here is that if it is true for the primed charges, it is also true for 
the unprimed charges}. This condition would be automatic in 
$SU(5)$, but in general such a condition on the charges
does not necessarily imply a field theory $SU(5)$ GUT to actually be
present, although it may be.

Since the generic Yukawa structure is of the form:
\begin{equation}
  \label{eq:upsymcaspar}
  Y^f \approx 
  \left[
    \begin{array}{ccc}
      \epsilon^{|f_1 + q_1+h_f| } & 
      \epsilon^{|f_2 + q_1+h_f|} &
      \epsilon^{|f_3 + q_1+h_f| } \\
      \epsilon^{|f_3 + q_2+h_f| } &
      \epsilon^{|f_2 + q_2+h_f| } &
      \epsilon^{|f_3 + q_2+h_f| } \\
      \epsilon^{|f_1 + q_3+h_f|} &
      \epsilon^{|f_2 + q_3+h_f| } &
      \epsilon^{|f_3 + q_3+h_f| }
    \end{array}
  \right].
\end{equation}
it is clear that the $SU(5)$ relations $d_i = l_i$, $q_i = u_i 
= e_i$ lead to Yukawa textures of the form:
\begin{eqnarray}
  \label{eq:Yusu5}
  Y^u &\approx&
  \left[
    \begin{array}{ccc}
      \epsilon^{|2 e_1 -2e_3|} &
      \epsilon^{|e_1 + e_2-2e_3|} &
      \epsilon^{|e_1 - e_3|} \\
      \epsilon^{|e_1 + e_2 - 2e_3|} &
      \epsilon^{|2 e_2- 2 e_3|} &
      \epsilon^{|e_2 - e_3|} \\
      \epsilon^{|e_1 - e_3 |} &
      \epsilon^{|e_2 - e_3    |} &
      \epsilon^{| 0       |}
    \end{array}
  \right], \\
\label{eq:Ydsu5}
  Y^d &\approx&
  \left[
    \begin{array}{ccc}
      \epsilon^{|l_1 + e_1+h_d| } &
      \epsilon^{|l_2 + e_1+h_d| } &
      \epsilon^{|l_3 + e_1+h_d| } \\
      \epsilon^{|l_1 + e_2+h_d| } &
      \epsilon^{|l_2 + e_2+h_d| } &
      \epsilon^{|l_3 + e_2+h_d| } \\
      \epsilon^{|l_1 + e_3+h_d| } &
      \epsilon^{|l_2 + e_3+h_d| } &
      \epsilon^{|l_3 + e_3+h_d| }
    \end{array}
  \right], \\
 Y^e &\approx& Y^{d\ T}.
\end{eqnarray}
Note that the up matrix is approximately symmetric,
due to the assumed $SU(5)$ relation of charges.
The reason why the textures above are approximate is that 
each entry in each matrix contains an undetermined order unity
flavour dependent coefficient, generically denoted as $a^f_{ij}=O(1)$.
We shall continue to suppress such coefficients
in order to make the discussion less cumbersome, 
but will return to this question when 
we discuss the numerical fits later in the paper.
We have also assumed that the up and down Yukawa matrices are
described by a single expansion parameter $\epsilon$.
The possibility of having two different expansion parameters,
one for the up sector and one for the down sector, 
will also be discussed later in the paper.
In order to
have an acceptable top quark mass, we have required that $h_u+2e_3 = 0$, in which 
case the smallness of the bottom quark mass can be due
to $h_d+e_3+l_3 \ne 0$, and we are free to have a small $\tan\beta$, because we 
don't need large $\tan\beta$ to explain the ratio $\frac{m_t}{m_b}$ on its own.

Also note that, as expected from the $SU(5)$ relation of charges, the
down and electron textures are the approximate transposes of each
other, $Y^d \approx (Y^e)^T$. Such a relation implies bad mass
relations for between the down type quarks and charged leptons, but
may be remedied by using Clebsch factors such as a Georgi-Jarlskog
factor of 3 in the (2,2) position of the charged lepton Yukawa
matrix.

If we were to look at the case $x = y$, then we would have a solution
suggestive of unified $SO(10)$ GUT symmetry, for which $l_i = q_i =
u_i = d_i = e_i$. The same comments above also apply here, namely that 
such a condition on the charges, though consistent with 
an $SO(10)$ GUT does not necessarily imply a field theory realization of it. 
The matrices \eq{eq:Yusu5}-\eq{eq:Ydsu5} would all become equal to 
the same symmetric texture in Eq.\ref{eq:Yusu5}, in the $SO(10)$ 
case that $x=y$.

\subsubsection{Pati-Salam type cases}
In this case, applying the Pati-Salam constraints on the charges,
\begin{eqnarray}
\label{eq:pscharg}
q_i=l_i\equiv  q^L_i,\quad u_i=d_i=e_i=n_i\equiv q^R_i,
\end{eqnarray}
so we can immediately see that also for this choice of charges 
both the the
flavour independent and 
dependent parts in \eq{eq:A1pFIFD} vanishes.  We have also
included the right-handed neutrino charges, which do not enter into
the anomaly cancellation conditions,\ \eq{eq:A3}-\eq{eq:A1p}, but with
a Pati-Salam group should obey the relation of \eq{eq:pscharg}. Thus
in this case all the mass matrices have the form
\begin{eqnarray}
 Y^{f} &=&
  \left(
    \begin{array}{ccc}
      \epsilon^{|l_1 + e_1+h_{f}| } &
      \epsilon^{|l_1 + e_2+h_{f}| } &
      \epsilon^{|l_1 + e_3+h_{f}|} \\
      \epsilon^{|l_2 + e_1+h_{f}|} &
      \epsilon^{|l_2 + e_2+h_{f}|} &
      \epsilon^{|l_2 + e_3+h_{f}| } \\
      \epsilon^{|l_3 + e_1+h_{f}| } &
      \epsilon^{|l_3 + e_2+h_{f}| } &
      \epsilon^{|l_3 + e_3+h_{f}| }
    \end{array}
\right)
\label{PStexture}
\end{eqnarray}
for $h_{f}=h_u,\ h_d$. In this case we always need to satisfy $x=y$, in contrast with the generic case of $SU(5)$ where it is not necessary $x=y$. So we can put one of the charges in terms of the other two and the parameters $x=y$
\begin{eqnarray}
e_1=x-(e_2+e_3),\quad l_1=x-(l_2+l_3),\quad \Rightarrow
e_1+e_2+e_3=l_1+l_2+l_3.
\label{PScondn}
\end{eqnarray}
We have already noted that the Pati-Salam constraints 
on the charges imply that the anomaly $A_1'$
automatically vanishes. It is also a remarkable fact that 
the constraints in Eq.\ref{PScondn} do not in practice lead to 
any physical constraints on the form of the Yukawa texture
in Eq.\ref{PStexture}. In practice, assuming only that $u+v=0$,
one can start with any set of charges $l_i$, $e_i$ which 
lead to any desired Yukawa texture, where the charges do not
satisfy the anomaly free constraint in Eq.\ref{PScondn}.
Then from any set of non-anomaly-free charges one can construct
a set of anomaly-free charges which do satisfy Eq.\ref{PScondn},
but do not change the form of the Yukawa matrix in Eq.\ref{PStexture},
by simply making an equal and opposite flavour-independent shift 
on the charges as follows \cite{King:2000ge}:
$e_i\rightarrow e_i +\Delta$, $l_i\rightarrow l_i -\Delta$.
In this paper we shall not consider the Pati-Salam approach in detail.

\subsection{Solutions with anomaly free $A_1^\prime$ with $u + v = 0 \
  \ (u,v \ne 0)$ \label{sec:u-=-v}}
In this case, we can repeat the analysis of the previous subsection, 
but with the general constraints. Note however,
that since $u+v = 0$, $h_u = -z$ and $h_d = +z$.

Then we are left with the result that 
\begin{eqnarray}
  \label{eq:12}
  A_1^\prime = \frac{1}{3} \left[
    6 u^2 + 6 x u + 2 y u
    \right] -
    \sum_{i=1}^3 \left( q_i^{\prime\;2} - 2 u_i^{\prime\;2} + 
d_i^{\prime\;2} - l_i^{\prime\;2} + e_i^{\prime\;2} \right).
\end{eqnarray}
Note that the family independent part will vanish if 
\begin{equation}
\label{eq:13}
u = -v = -\left( x + \frac{y}{3} \right).
\end{equation}

Having done this, we may substitute Eq.~(\ref{eq:13}) into 
Eqs.~(\ref{eq:sumqi}-~\ref{eq:hd}) Then we find that:
\begin{eqnarray}
  \label{eq:14}
  \sum_{i=1}^3 q_i &=& -\frac{y}{3},\quad \quad \
  \sum_{i=1}^3 u_i \ =\ - ( x + \frac{2y}{3} ), \nn\\
  \sum_{i=1}^3 d_i &=& x + \frac{4y}{3},\quad
  \sum_{i=1}^3 l_i \ \ =\ y,\\
  \sum_{i=1}^3 e_i &=& x.
\end{eqnarray}

\subsubsection{Yukawa textures for a sample solution \label{sec:yukawa-textures-uv-nonzero}}
At this point, we note that there will be a large number of solutions. 
However, one class of solutions that will easily be satisfied
will be:
\begin{equation}
  \label{eq:19}
  q_i = -\frac{l_i}{3} \;,\; u_i = - ( \frac{2 l_i}{3} + e_i ) \; , \; 
d_i = \frac{4 l_i}{3} + e_i.
\end{equation}
The same equation will hold for the primed charges:
\begin{equation}
  \label{eq:20}
  q_i^\prime = -\frac{l_i^\prime}{3} \;,\; u_i^\prime = - ( \frac{2 
l_i^\prime}{3} + e_i^\prime ) \; , \; 
  d_i^\prime = \frac{4 l_i^\prime}{3} + e_i^\prime.
\end{equation}

We can now put Eq.~(\ref{eq:20}) into the anomaly, Eq.~(\ref{eq:12}). 
In this case we find that:
\begin{eqnarray}
  \nonumber
  A_1^\prime &=& \frac{1}{3} \left[ x^2 ( 6 - 6 ) + \frac{2}{3} y^2( 1 
- 1 ) + xy ( 4- 2 - 2) \right] \\
  && - \sum_{i=1}^3 \left( l_i^{\prime\;2} \frac{1}{9} ( -1 + 8 - 16 + 
9 ) + e_i^{\prime\;2}( 2 - 1 -1 ) \right)
    = 0.
  \label{eq:21}\end{eqnarray}
So we see that for this particular relation of leptonic and quark 
charges, we are automatically anomaly-free.

Again, we see that, just as for the $u = v = 0$ case, we can specify 
everything by the leptonic charges $l_i$ and $e_i$.
However, in this case we will get three different textures. 
Specifically, we will get:
\begin{eqnarray}
  \label{eq:yu-umvn0}
  Y^u &\approx& 
  \left[
    \begin{array}{ccc}
      \epsilon^{|l_1 + e_1 + h_u|} & 
      \epsilon^{|\frac{1}{3}(l_2 + 2l_1)+e_1+ h_u|} &
      \epsilon^{|\frac{1}{3}(l_3 + 2l_1)+e_1+ h_u|} \\
      \epsilon^{|\frac{1}{3}(l_1+ 2l_2)+e_2 + h_u|} &
      \epsilon^{|l_2+e_2 + h_u|} &
      \epsilon^{|\frac{1}{3}(l_3+2l_2)+e_2 + h_u|} \\
      \epsilon^{|\frac{1}{3}(l_1+2l_2)+e_3 + h_u|} &
      \epsilon^{|\frac{1}{3}(l_2+2l_3)+e_3 + h_u|} &
      \epsilon^{|l_3 + e_3 + h_u|}
  \end{array}    
  \right] \\
  \label{eq:yd-umvn0}
  Y^d &\approx& 
  \left[
    \begin{array}{ccc}
      \epsilon^{|l_1+e_1 - h_u|} &
      \epsilon^{|\frac{1}{3}(-l_1+4l_2)+e_2 - h_u|} &
      \epsilon^{|\frac{1}{3}(-l_1+4l_3)+e_3 - h_u|} \\
      \epsilon^{|\frac{1}{3}(-l_2+4l_1)+e_1 - h_u|} &
      \epsilon^{|l_2+e_2 - h_u|} &
      \epsilon^{|\frac{1}{3}(-l_2+4l_3)+e_2 - h_u|} \\
      \epsilon^{|\frac{1}{3}(-l_1+4l_3)+e_3 - h_u|} &
      \epsilon^{|\frac{1}{3}(-l_2+4l_3)+e_3 - h_u|} &
      \epsilon^{|l_3+e_3 - h_u|}
  \end{array}    
  \right] \\
\label{eq:ye-umvn0}
  Y^e &\approx& 
  \left[
    \begin{array}{ccc}
      \epsilon^{|l_1+e_1 - h_u|} & 
      \epsilon^{|l_1+e_2 - h_u|} &
      \epsilon^{|l_1+e_3 - h_u|} \\
      \epsilon^{|l_2+e_1 - h_u|} &
      \epsilon^{|l_2+e_2 - h_u|} &
      \epsilon^{|l_2+e_3 - h_u|} \\
      \epsilon^{|l_3+e_1 - h_u|} &
      \epsilon^{|l_3+e_2 - h_u|} &
      \epsilon^{|l_3+e_3 - h_u|}
  \end{array}    
  \right]
\end{eqnarray}
We note that this is a rather predictive scheme; 
we require that the diagonal elements are of
the same order in the between the down and electron Yukawa matrices constrained by the 
anomalies. Also, we require (at the very least)
$l_3 + e_3 +h_u= 0$ to get a correct top quark mass.

\subsection{Anomaly free $A_1^\prime$ with $u + v \neq 0$ solutions\label{sec:yukawa-textures-upv-notzero}}

In this case we can not decompose the expression of   $A_1^\prime$ into flavour independent and flavour dependent parts, but we can use for example the relation $\left(\sum f_i\right)^2=\sum f_i^2+2(f_1(f_2+f_3)+f_2f_3)$ such that we have
\begin{eqnarray}
A_1^\prime=-2(4u^2+u(v+3x+z)+v(z-y))-\!2\!\!\!\!\!\!\!\sum_{f=u,d,l,e,q}\!\!\!\!\!\! g_f (f_1(f_2+f_3)+f_2f_3),
\end{eqnarray}
where $g_f=1,-2,1,-1,1$ respectively for $f=q,u,d,l,e$. However it is difficult to depart from here in order to find some ansatz which cancels the $A_1^\prime$ anomaly. Instead we can generalize the kind of relations which in the limit of $u=v=0$ would give the $SU(5)$ cases or the Pati-Salam cases.
\subsubsection{An extended $SU(5)$ case} 
\label{sec:genrzsu5like}
Here a non-GUT case is considered, taken by generalizing the $SU(5)$ relation between the charges. In the $SU(5)$ case,
we had $q_i = u_i = e_i$ and $d_i = l_i$. If instead we have the linear relations:
\begin{eqnarray}
\label{eq:chargrelgensu5like}
q_i=u_i+\alpha=e_i+\gamma,\quad d_i=l_i+\beta,\quad
\end{eqnarray}
From the parameterization of Eqs.~(\ref{eq:A3}-\ref{eq:A1p}), we see that 
in the limit of the $u=v=0$ we recover the $SU(5)$ case. In agreement with the cancellation of anomalies then one should have
\begin{eqnarray}
q_i=u_i-\frac{u}{3}=e_i+\frac{u}{3},\quad d_i=l_i+\frac{v}{3}.
\label{eq:1}
\end{eqnarray}
In the expression of the  $A_1^\prime$ anomaly, as given in \eq{eq:A1p}, the sums of squared charges cancel and we can write  it just in terms of sum of charges, which we have parameterized in terms of $u,v,x,y$,
\begin{eqnarray}
A_1^\prime=-10 \frac{u^2}{3}-\frac{2}{3}v^2+2u(x+v)+2y\frac{v}{3}-2z(u+v)=0.
\end{eqnarray}
Thus we need to satisfy this equation in order to have anomaly free solutions. Requiring the condition of $O(1)$ top coupling we have
\begin{eqnarray}
\label{eq:charggensu5like}
h_u&=&-z=-2e_3-u,\nn\\
h_d&=&2u+v+2e_3,\nn\\
{\mathcal{C}}(Y^u_{ij})&=&|e_i+e_j-2e_3|,\nn\\
{\mathcal{C}}(Y^d_{ij})&=&|e_i+l_j+2e_3+\frac{7u}{3}+\frac{4v}{3}|,\nn\\
{\mathcal{C}}(Y^e_{ij})&=&|l_i+e_j+2e_3+2u+v|,
\end{eqnarray}
where ${\mathcal{C}}(Y^u_{ij})$ denotes the power of $\epsilon$ for the $(i,j)$ element of the correspondent Yukawa matrix. Note that although we did not begin with an {\it a priori} condition of having $Y^u$ symmetric, the requirement of the $O(1)$ top coupling cancels the parameter $u$ in all the entries of $Y^u$ and so we end up with a symmetric matrix. 
\subsubsection{An extended Pati-Salam case}
\label{sec:pati-salam-like-case}
Following the extended $SU(5)$ case, we look for solutions which in the
$u=v=0$ limit reproduce the Pati-Salam case, so we should have the
relations
\begin{eqnarray}
\label{eq:PSgenrel}
q_i=l_i+\alpha,\quad u_i=d_i+\beta.
\end{eqnarray}
Also $e_i$ and $n_i$ need to be related to $u_i$ by a constant, as in \eq{eq:PSgenrel}. In these case in order to satisfy  the G-S anomaly conditions we need
\begin{eqnarray}
q_i=l_i+\frac{u+(x-y)}{3},\quad u_i=e_i+\frac{2u}{3},\quad d_i=e_i+\frac{v+(y-x)}{3}.
\label{eq:8}
\end{eqnarray}
Thus the expression for the $A_1^\prime$ anomaly is
\begin{eqnarray}
A_1^\prime&=&-\frac{2}{9}\left[8u^2+4v^2+u(9v+11x-2y)+2(x-y)^2 -v(2x+y)\right]\nn\\
&&-2z(u+v),
\end{eqnarray}
and finally requiring the condition of $O(1)$ top Yukawa coupling we have
\begin{eqnarray}
h_u&=&-z=-(l_3+e_3+u+\frac{x-y}{3}),\nn\\
h_d&=&l_3+e_3+2u+v+\frac{x-y}{3},\nn\\
{\mathcal{C}}(Y^u_{ij})&=&|l_i-l_3+e_j-e_3|,\nn\\
{\mathcal{C}}(Y^d_{ij})&=&|l_i+e_j+l_3+e_3+\frac{4v+7u+(x-y)}{3}+\frac{4v}{3}|,\nn\\
{\mathcal{C}}(Y^e_{ij})&=&|l_i+e_j+2e_3+2u+v|.
\end{eqnarray}
%

\section{A useful phenomenological parameterization}
\label{sec:new-paramaterisation}
So far we have discussed the anomaly cancellation conditions in $U(1)$ family
symmetry models, and some of the possible solutions to these
conditions, including some new solutions not previously
discussed in the literature. It turns out however that the 
anomaly free charges themselves do not provide the most convenient
parameters for discussing the phenomenological constraints on the
Yukawa matrices arising from the quark and lepton spectrum.
It is more convenient to introduce a 
new parameterization for the Yukawa matrices as follows:
\begin{equation}
  \label{eq:6}
  Y^f \approx \left(
    \begin{array}{ccc}
      \epsilon^{|s'_f + r'_f + k_f|} & \epsilon^{|s'_f + r_f + k_f|} & \epsilon^{|s'_f + k_f|} \\
      \epsilon^{|s_f  + r'_f + k_f|} & \epsilon^{|s_f +  r_f + k_f|} & \epsilon^{|s_f  + k_f|} \\
      \epsilon^{|      r'_f + k_f|} & \epsilon^{|        r_f + k_f|} & \epsilon^{|       k_f|}
    \end{array}
    \right)
\end{equation}
where $f=u,d,e,\nu$, and we have introduced the
parameters $r_f, r'_f, s_f, s'_f, k_f$ which are defined 
in terms of the charges in Table 1 as:
\begin{eqnarray}
  \nonumber
  r_f = f_2 - f_3 & r'_f = f_1 - f_3 & k_u = q_3 + u_3 + h_u \\
  \nonumber
  s_{u,d} = q_2 - q_3 & s'_{u,d} = q_1 - q_3 & k_d = q_3 + d_3 + h_d \\
  \nonumber
  s_{e,\nu} = l_2 - l_3 & s'_{e,\nu} = l_1 - l_3 & k_e = l_3 + e_3 + h_d \\
  \label{eq:gyukpar}
 & & k_\nu = l_3 + n_3 + h_u 
\end{eqnarray}
In order to get an acceptable top quark mass, we require that $k_u = 0$. 
Note that the parametrization above is 
completely general, there is no information loss from the form of
Eq.~(\ref{eq:upsymcaspar}), and thus far we have not imposed any
constraints on the charges arising from either anomaly cancellation
or from GUTs. We now consider the simplifications 
which arise in the new parametrization 
when the charges are constrained by considerations of anomaly cancellation and 
GUTs, as discussed in the previous section.

\subsubsection*{Simplification in $SU(5)$ type case}

Consider the case where the family charges are consistent with the representations in an $SU(5)$ GUT,  $d_i = l_i$, and $q_i = u_i = e_i$:
\begin{eqnarray}
  \nonumber
  k_e = k_d\ & s_{u,d} = r_{u,e} & s'_{u,d} = r'_{u,e} \\
  \label{eq:11}
  s_{e,\nu} = r_d & s'_{e,\nu} = r'_d 
\end{eqnarray}
In this case, all of the parameters can be expressed purely
in terms of the lepton charges:
\begin{eqnarray}
  \nonumber
  s_{u,d}=r_{u,e} = e_2 - e_3 & s'_{u,d} = r'_{u,e} = e_1 - e_3 \\
  s_{e, \nu} = r_d = l_2 - l_3 & s'_{e,\nu} = r'_{d} = l_1 - l_3
\label{eq:24}  
\end{eqnarray}
Note that this leads directly to the fact that $Y^e \approx (Y^d)^T$. The equality is broken by the arbitrary $O(1)$ coefficients.
As discussed, the $SU(5)$ charge conditions are sufficient to
guarantee anomaly cancellation for the case $u=v=0$.

\subsubsection*{Simplification in the extended $SU(5)$ case}

In the case $u+v\neq 0$, anomalies can again be cancelled by assuming
the charge conditions in Eq.~(\ref{eq:chargrelgensu5like}).
If we take Eq.~(\ref{eq:chargrelgensu5like}), we can again simplify Eq.~(\ref{eq:gyukpar}). In this case we find:
\begin{eqnarray}
  \nonumber
  s_{u,d} = r_{u,e} & s'_{u,d} = r'_{u,e} \\
  \label{eq:16}
  s_{e,\nu} = r_d & s'_{e,\nu} = r'_d
\end{eqnarray}

In this case we have that the texture of $Y^e$ can be attained from $Y^d$ by replacing $k_d$ with $k_e$ and then
transposing.

\subsubsection*{Simplification in the Pati-Salam case}

In the case of having charge relations consistent with a Pati-Salam theory,
$q_i = l_i$ and $u_i = d_i = e_i = n_i$, we can simplify:
\begin{eqnarray}
  \nonumber
  k_e = k_d & s_{u,d} = s_{e,\nu} & s'_{u,d} = s'_{e,\nu} \\
  \label{eq:15}
  k_u = k_\nu & r_u = r_d = r_e = r_\nu & r'_u = r'_d = r'_e = r'_\nu 
\end{eqnarray}

\section{Quark masses and mixings in $SU(5)$ \label{sec:su5q}}
In this section we shall provide some constraints on the
phenomenological parameters introduced in the last section,
arising from the quark masses and mixings,
assuming the simplification in the $SU(5)$ type case mentioned above.
In $SU(5)$  Eqs.~(\ref{eq:6}),(\ref{eq:11}) imply the quark Yukawa
matrices are explicitly of the form:
\begin{eqnarray}
\label{eq:su5matparam}
Y^u\approx 
\left(
\begin{array}{ccc}
\vep^{|2s'|}&\vep^{|s'+s|}&\vep^{|s'|}\\
\vep^{|s'+s|}&\vep^{|2s|}&\vep^{|s|}\\
\vep^{|s'|}&\vep^{|s|}&1
\end{array}
\right),\ \ \ \ 
Y^d\approx 
\left(
\begin{array}{ccc}
\vep^{|s'+r'_{d}+k_d|}&\vep^{|s'+r_{d}+k_d|}&\vep^{|s'+k_d|}\\
\vep^{|s+r'_{d}+k_d|}&\vep^{|s+r_{d}+k_d|}&\vep^{|s+k_d|}\\
\vep^{|r'_{d}+k_d|}&\vep^{|r_{d}+k_d|}&\vep^{|k_d|}
\end{array}
\right).
\end{eqnarray}
where we have written $s=s_{u,d}=r_{u,e}$, $s' = s'_{u,d}=r'_{u,e}$.
\footnote{Note that the extended $SU(5)$ anomaly free solutions examined
in section \ref{sec:genrzsu5like} leave the parameters 
$s,s',r_d,r'_d,k_d$ invariant, as is clear by comparing
Eqs.\ref{eq:11} and \ref{eq:16}.
Hence the results in this section for the quark
sector apply not only to the $SU(5)$ type case
but also the extended $SU(5)$ anomaly free cases.}
Note that we are assuming a single expansion parameter $\vep$, 
and are suppressing $O(1)$ coefficients. Clebsch factors are also not
considered, and only leading order operators are discussed.

In order to determine the possible solutions for $s,\ s',\ r_d, \
r'_d$ and $k_d$ which successfully reproduce quark 
masses and mixings one can numerically diagonalize Yukawa matrices and
obtain the CKM matrix. However, in order to understand the behaivour
of this structure it is quite useful to use the technique of
diagonalization by blocks in the $(2,3)$, $(1,3)$ and $(1,2)$ sectors
\footnote{This only works if there is an appropriate hierarchy among the elements}. The results are presented in the next subsections.

\subsection{Quark Masses}

\noindent Barring accidental cancellations the down quark Yukawa
matrix $Y^d$ may be diagonalized, leading to the following 
eigenvalues:
\begin{eqnarray}
\label{eq:yukeigen}
y_1\!\!\!\!&\approx&\!\!\! a_{11}\vep^{|s'+r'+k|}-
\frac{(a_{31}\vep^{|r'+k|}+a_{23}a_{21}\vep^{|s+k|+|s+r'+k|-|k|}e^{2i(\beta^L_2-\beta^L_1)})}{c^{R}_{23}(\vep^{|k|}+a^{2}_{32}\vep^{2|r+k|-|k|}e^{-2i(\beta^R_2-\beta^R_1)} )}\times\nn\\ 
&&\times (a_{13}\vep^{|s'+k|}\!+a_{23}a_{12}\vep^{|r+k|+|s'+r+k|-|k|}e^{-2i(\beta^R_2-\beta^R_1)})+\nn\\
&&\!\!\!\!\!\!\!\! \frac{-(a_{12}\vep^{|s'+r+k|}\!-\!a_{32}a_{13}\vep^{|r+k|+|s'+k|-|k|})(a_{21}\vep^{|s+r'+k|}\!-\!a_{23}a_{31}\vep^{|s+k|+|r'+k|-|k|} )}{(a_{22}\vep^{|s+r+k|}-a_{23}a_{32}\vep^{|s+k|+|r+k|-|k|})e^{-i(\beta^L_3-\beta^R_3)}},\nn
\end{eqnarray}
\begin{eqnarray}
y_2\!\!\!\!&\approx&\!\!\!c^R_{23}\left(a_{22} \vep^{|s+r+k|} -a_{23}a_{32}\vep^{|r+k|+|s+k|-|k|}\right)e^{2i(\beta^L_2-\beta^R_2)},\nn\\
y_3\!\!\!\!&\approx&\!\!\!c^{R}_{23}\left(\vep^{|k|} +a^{2}_{32}\vep^{2|r+k|-|k|}e^{2i(\beta^R_1-\beta^R_2)} \right)e^{i(\beta^L_1-\beta^R_1)},
\end{eqnarray}
where we have suppressed the index $d$ in order to make clearer the
notation and re-scaled all the (complex) coefficients by $1/a_{33}$,
so that instead of having $a_{33}$ we have 1. 
Note that the down quark masses are given by:
$m^d_i=y^d_i v_d/\sqrt{2}$.
Analogous results also apply to the up quark sector,
with the replacements
$r\rightarrow s$, $r'\rightarrow s'$, $k\rightarrow 0$. 
The phases $\beta^L_i$
correspond to the diagonalization matrices of the Yukawa matrices,
whose notation is given in Appendix (\ref{ap:diagmat}).
 
It is important to remark that in the case of positive charges all the
elements of the first row of the Yukawa matrix contribute at the same
order, $s'+r'+k$, to their correspondent lightest eigenvalue, so in
these cases it is not possible to have the Gatto-Sartori-Tonin (GST)
relation. However in the cases of having $s$ and $s'$ (analogous for
$r$ and $r'$) with different sign, as in the example of
\eq{eq:textibross}, we can have a cancellation in powers of $\vep$ to
the contribution to $y_1$ coming from the diagonalization in the
$(1,2)$ sector, which is the third term in the expression for $y_1$ in
\eq{eq:yukeigen}. On the other hand we can have an enhancement in the
power of $\vep$ of the contributions from the $(1,1)$ entry and the
rotation in the $(1,3)$ sectors, which correspond to the first and
second term of $y_1$, respectively, in \eq{eq:yukeigen}. This together
with the condition ${\mathcal{C}}(Y_{21})={\mathcal{C}}(Y_{12})$ are
the requirements to achieve the GST relation. We will present examples
satisfying and not satisfying the GST relation.

We remark here the constraints from the bottom mass are
\begin{eqnarray}
\label{eq:tanb}
m_b \tan\beta=\vep^{|k_d|} m_t,\qquad  k_d=q_3+d_3+h_d
\end{eqnarray}
since $m_t=O(\langle H_u\rangle)$ and $\tan\beta=\langle H_u\rangle/\langle H_d\rangle$. Thus in terms of charges we have $h_u=-(q_3+u_3)$ and $h_d=q_3+u_3$, for $u=v=0$, $k=2q_3+d_3+u_3$.

\subsection{Quark Mixings}

We can also obtain the mixing angles in this approximation and compare
to the required experimental values (see Appendix \ref{ap:compinf}).
The mixing angles in the down sector, again dropping flavour indices,
are as follows:
\begin{eqnarray}
t^L_{23}&=&e^{i(\beta^L_2-\beta^L_1)}a_{23}\vep^{|s+k|-|k|}+a_{23}a_{22}\vep^{|s+r+k|+|s+k|-2|k|}e^{i\xi_L}\nn\\
t^R_{23}&=&e^{i(\beta^R_2-\beta^R_1)}a_{32}\vep^{|r+k|-|k|}+a_{23}a_{22}\vep^{|s+r+k|+|s+k|-2|k|}e^{i\xi_R}\nn
\end{eqnarray}
\begin{eqnarray}
t^L_{13}&=&\frac{a_{13}\vep^{|s'+k|}+a_{32}a_{12}\vep^{|r+k|+|s'+r+k|-|k|}e^{-i2(\beta^R_2-\beta^R_1)}}{\left(\vep^{|k|}+a^2_{32}\vep^{2|r+k|-|k|}e^{2i(\beta^R_1-\beta^R_2)}
\right) e^{i\beta^L_1}}\nn\\
t^R_{13}&=&\frac{a_{31}\vep^{|r'+k|}+a_{23}a_{21}\vep^{|s+k|+|s+r'+k|-|k|}e^{2i(\beta^L_2-\beta^L_1)}}{\left(\vep^{|k|}+a^2_{32}\vep^{2|r+k|-|k|}e^{2i(\beta^R_1-\beta^R_2)}
\right) e^{-i\beta^R_1}}\sqrt{1+|a^2_{32}|\vep^{2|r+k|-2|k|}}\nn\\
t^L_{12}&=&\frac{\left(a_{12}\vep^{|s'+r+k|}-a_{32}a_{13}\vep^{|r+k|+|s'+k|-|k|}\right)e^{-i(\beta^R_3+\beta^L_2)}}{\left(
a_{22}\vep^{|s+r+k|}-a_{23}a_{32}\vep^{|s+k|+|r+k|-|k|} \right)}\nn\\
t^R_{12}&=&\frac{\left(a_{21}\vep^{|s+r'+k|}-a_{23}a_{31}\vep^{|s+k|+|r'+k|-|k|}\right)e^{i(\beta^L_3+\beta^R_2)}}{\left(
a_{22}\vep^{|s+r+k|}-a_{23}a_{32}\vep^{|s+k|+|r+k|-|k|} \right)}\nn\\
\xi_L\!\!\!&\!=\!&\!\!\!-(\beta^L_2-\beta^L_1)-2(\beta^R_2-\beta^R_1),\
\xi_R=-(\beta^R_2-\beta^R_1)-2(\beta^L_2-\beta^L_1)\label{eq:mixsgeral}.
\end{eqnarray}
Analogous results also apply to the up quark sector, with the
replacements $r_d\rightarrow s$, $r_d'\rightarrow s'$, $k_d\rightarrow 0$.
Note that in the case of positive $s,s',r,r'$ and $k$, the angles
$t^L_{12}$ and $t^L_{23}$, of the left sector do not depend on
$r_d,r'_d$, 
so they are equal, at first approximation, for the up and down
sectors. Having the tangent of the angles expressed in terms of the
Yukawa elements we can see directly their contributions to the CKM
elements ($V_{\mathrm{CKM}}=L^uL^{d\dagger}$ in the notation of
Appendix (\ref{ap:diagmat}))
\begin{eqnarray}
\frac{|V_{ub}|}{|V_{cb}|}&=&\frac{|s^{u}_{12}s^Q_{23}-s^Q_{13}e^{i(\Phi_1-\Phi_2)}|}{|s^Q_{23}|}\approx 0.09\sim (\lambda^2,\lambda) \nn\\
\frac{|V_{td}|}{|V_{ts}|}&=&\frac{|s^{d}_{12}s^Q_{23}-s^Q_{13}e^{i(\Phi_2)}|}{|s^Q_{23}|}\sim \lambda\nn\\
|V_{us}|&=&|s^d_{12}-s^u_{12}e^{i\Phi_1}|=\lambda \approx 0.224\nn\\
{\rm{Im}}\{J\}&=&s^Q_{23}(s^Q_{23}s^d_{12}s^u_{12}\sin(\Phi_1)-s^Q_{13}(s^d_{12}\sin(\Phi_2))- s^u_{12}\sin(\Phi_2-\Phi_1)),
\label{eq:Vsasyu1}
\end{eqnarray}
with $s^Q_{ij}=|s^d_{ij}-e^{i\Phi_{X_{ij}}}s^u_{ij}|$. The phases $\Phi_1$, $\Phi_2$ and $\Phi_{X_{ij}}$ depending on the contributions that the mixing angles receive from the different elements of the Yukawa matrix and have a different expression in terms of the phases of the Yukaw matrix for different cases.  For example when the elements $(1,2)$ and $(1,3)$ are of the same order and the right handed mixing angle in the $(2,3)$ sector is large, the
$\Phi_2$ phase will be
\begin{eqnarray}
\label{eq:phi2}
\Phi_2={\rm{Arg}}\left[\frac{Y^d_{12}+Y^d_{13}t^R_{23}}{Y^d_{33}+Y^d_{23}t^R_{23}} \right]
\end{eqnarray}
As we can see from the expressions in \eq{eq:Vsasyu1} involving $\Phi_1$, this can be associated to the $U$ sector. When all the signalization angles in this sector are small, then this phase takes the form
\begin{eqnarray}
\label{eq:phi1}
\Phi_1=\phi^u_{12}-\phi^u_{22}
\end{eqnarray}
where $\phi_{12}$ and $\phi_{22}$ are the phases of the $Y^u_{12}$ and $Y^u_{22}$ elements.
Finally the phases $\Phi_{ij}$, which appear in $s^Q_{ij}$,  can be associated either with the $U$ or with the $D$ sector.
\begin{table}[ht] \centering%
\begin{center}
\begin{tabular}{|l l |c||l l| c|}
\hline
\!$U(1)$ relations\!\!& \!Constraint &\!Reason\!\!& $U(1)$ relations\!\!& \!Constraint &\!Reason\!\\
\hline
$\vep^{|s+k_d|-|k_d|}$ & $\!\sim \lambda^2$ & $s^Q_{23}$ & $\vep^{|3q_3+d_3|}$&$\!\sim (1,\lambda^3)$&$m_b$\\
$\vep^{|s'+k_d|-|k_d|}$ & $\!\gtrsim \lambda^3$ & $s^Q_{13}$ & $\vep^{|s+r_d+k_d|-|k_d|}$ & $\!\sim (\lambda^2,\  \lambda^3)  $ & $\frac{m_s}{m_b}$\\
$\vep^{|s'+r_d+k_d|-|s+r_d+k_d|}\!$ & $\!\sim\lambda$ & $s^{Q}_{12}$ & $\vep^{|s'+r'_d+k_d|-|k_d|}$ & $\!\sim (\lambda^4,\  \lambda^5)$ & $\frac{m_d}{m_s}$\\
$$ & $$ & $$ & $\vep^{|2s+k_d|-|k_d|}$ & $\!\sim \lambda^4 $ & $\frac{m_c}{m_t}$\\
$$ & $$ & $$ & $\vep^{|2s'+k_d|-|k_d|}$ & $\! \geq \lambda^6$ & $\frac{m_u}{m_c}$\\
\hline
\end{tabular}
\end{center}
\caption{{\small Constraints on the parameters $s,\ s',\ r_d, \ r'_d$
and $k_d$ from quark mixing angles and mass ratios. For the mixing
angles we need to satisfy the conditions for up or down sector,
where the analogous conditions for the up sector are obtained by
making the replacements 
$r_d\rightarrow s$, $r_d'\rightarrow s'$, $k_d\rightarrow 0$. They
do not need to be satisfied for both as long as for the sector in
which they are not satisfied they do not give a bigger contribution
than the indicated power.} }
\label{table:phen1}
\end{table}

With the requirements of Table (\ref{table:phen1}) and the values of quark masses in Appendix (\ref{ap:compinf}), we can identify the viable solutions in the quark sector. 
One solution which has been widely explored is the up-down symmetric case for which we have $x=y$ thus, $f_i=q_i=u_i=e_i=d_i=l_i$. In this case $h_u=-2e_3=-h_d$ so $k_u=0$, $k_d=k_l=4e_3$, but in this case we need two expansion parameters $\vep_u$ and $\vep_d$ to reproduce appropriate mass ratios and mixings, thus we have
\begin{eqnarray}
Y^f=
 \left[
\begin{array}{ccc}
\vep^{|2s'+k_f|}_f&\vep^{|s+s'+k_f|}_f&\vep^{|s'+k_f|}_f\\
\vep^{|s+s'+k_f|}_f&\vep^{|2s+k_f|}_f&\vep^{|s+k_f|}_f\\
\vep^{|s'+k_f|}_f&\vep^{|s+k_f|}_f&\vep^{|k_f|}
\end{array}
\right].
\end{eqnarray} 
We can think of fixing $s+s'$, and then check for which choice of $s$ we have appropriate phenomenological solutions. For example if we take $s+s'=\pm 3$ 
and $e_3=0$ ($k_f=0$, $\forall f$) we have
\begin{eqnarray}
\label{eq:textibross}
Y^f=
 \left[
\begin{array}{ccc}
\vep^{|6-2f_2|}_f&\vep^{|3|}_f&\vep^{|3-f_2|}_f\\
\vep^{|3|}_f&\vep^{|2f_2|}_f&\vep^{|f_2|}_f\\
\vep^{|3-f_2|}_f&\vep^{|f_2|}_f&1
\end{array}
\right]
\end{eqnarray} 
The viable phenomenological fit for the case of quarks is for $f_2=-1$ and $f_1=4$ or  $f_2=1$ and $f_1=-4$
\cite{Ibanez:1994ig}. In this case we have then  $x=y=\pm 3$
respectively.

%

\section{Neutrino masses and mixings in SRHND\label{sec:neuts}}
%
%
In this section we apply the requirements of getting acceptable
neutrino masses and mixings by using a class seesaw model where 
$l_2 =l_3$. These are a subset of a class of seesaw models called single
right-handed neutrino dominance (SRHND) or sequential dominance
\cite{King:1998jw}. 
This additional constraint $l_2 =l_3$ will henceforth be 
applied in obtaining phenomenological solutions in the
lepton sector. 

Apart from the obvious benefit of considering the neutrino sector, it
will turn out that the neutrino sector will constrain the absolute 
values of the charges
under the $U(1)$ family symmetry, (not the charge differences,) due to
the Majorana nature of neutrinos.
This is due to the relations between the
charges imposed by the relevant GUT constraints, or the extended GUT
constraints, eq.~(\ref{eq:1}) for the extended $SU(5)$ solution of
section \ref{sec:genrzsu5like} and eq.~(\ref{eq:8}) for 
the extended Pati-Salam solution of section \ref{sec:pati-salam-like-case}.
For example the additional constraint $l_2 =l_3$ implies immediately
\begin{equation}
r_d = s_{e, \nu} = l_2 - l_3=0,
\label{l2l3}
\end{equation}
in the $SU(5)$ type cases from Eq.\ref{eq:24}.

Here we would like to study the cases for which large mixing angles in
the atmospheric sector and the neutrino sector can be explained
naturally in terms of the parameters of the $U(1)$ class of symmetries
that we have constructed in the previous sections, under the framework
of the type I see-saw mechanism together with the scenario of the
single right handed neutrino dominance (SRHND). We refer the reader
for a review of this scenario to \cite{King:1998jw}. Here we
make a brief summary of the results and apply them to the present
cases.  In the type I see-saw the mass matrix of the low energy
neutrinos is given by $m_{LL}\approx v^2_u Y^{\nu} M^{-1}_R Y^{\nu
T}$, where $Y^{\nu}$ is the Dirac matrix for neutrinos and $M_R$ is
the Majorana matrix for right-handed neutrinos. If we have three right
handed neutrinos, $M_1$, $M_2$ and $M_3$, then for the right handed
neutrino mass, in terms of $U(1)$ charges we have:
\begin{eqnarray}
\label{eq:Ynu}
Y^\nu &=&  \left[
  \begin{array}{ccc}
    \epsilon^{|l_1+n_1+h_u|} & \epsilon^{|l_1+n_2+h_u|} & \epsilon^{|l_1+n_3+h_u|} \\
    \epsilon^{|l_2+n_1+h_u|} & \epsilon^{|l_2+n_2+h_u|} & \epsilon^{|l_2+n_3+h_u|} \\
    \epsilon^{|l_3+n_1+h_u|} & \epsilon^{|l_3+n_2+h_u|} & \epsilon^{|l_3+n_3+h_u|} 
  \end{array}
\right]\\
\label{eq:MR}
M_{RR}&=& \left[
\begin{array}{ccc}
\vep^{|2n_1+\sigma|}&\vep^{|n_1+n_2+\sigma|}&\vep^{|n_1+n_3+\sigma|}\\
\vep^{|n_1+n_2+\sigma|}&\vep^{|2n_2+\sigma|}&\vep^{|n_2+n_3+\sigma|}\\
\vep^{|n_1+n_3+\sigma|}&\vep^{|n_2+n_3+\sigma|}&\vep^{|2n_3+\sigma|}\\
\end{array}
\right]<\Sigma>
\end{eqnarray}
where the charges $n_i$ are the $U(1)$ charges of the right handed neutrinos, $\nu_{Ri}$ and  $\sigma$ is the $U(1)$ charge  of the field $\Sigma$ giving Majorana masses to the right handed neutrinos. These charges are not constrained by the anomaly cancellation conditions \eq{eq:sumqi}-\eq{eq:hd} of Section  (\ref{sec:anomconst}), at least in the $SU(5)$ case, which gives some freedom in order to find appropriate solutions giving two large mixing angles and one small mixing angle for neutrinos. We expect  $\Sigma$ to be of order the scale at which the $U(1)$ symmetry is broken, for example at $M_P=M_{\rm{Planck}}$, or some other fundamental scale, such as the Grand Unification scale, $M_G$, for the solutions with an underlying GUT theory.

Here we restrict  ourselves to the cases in which \eq{eq:MR} can be considered as diagonal, $M_R\approx \rm{diag}\{M_1,M_2,M_3 \}$, for which we need in the $(2,3)$ block
\begin{eqnarray}
\label{eq:srhnd1}
|n_3+n_2+\sigma|&>& min\{ |2n_3+\sigma|,|2n_2+\sigma|\},\nn\\
2|n_3+n_2+\sigma|&\geq& |2n_3+\sigma| +|2n_2+\sigma|.
\end{eqnarray}
The conditions in the $(1,2)$ block are analogous to the $(2,3)$ and also we need
\begin{eqnarray}
\label{eq:srhnd2}
|n_1+n_3+\sigma|>max\{|2n_2+\sigma|, |2n_3+\sigma|\}.
\end{eqnarray}
Now, there are two cases that we can consider here, which correspond to selecting which of the neutrinos will dominate, $M_1$ or $M_3$. For the later case the SRHND conditions are
\begin{eqnarray}
\frac{|Y^\nu_{i3}Y^\nu_{j3}|}{|M_3|} \gg  \frac{|Y^\nu_{i2}Y^\nu_{j2}|}{|M_2|}\gg
\frac{|Y^{\nu 2}_{31}, Y^{\nu 2}_{21}, Y^{\nu}_{21}, Y^{\nu}_{31} |}{|M_1|}; \quad i,j=1,2,3. 
\end{eqnarray}
For the case in which $M_1$ dominates  we just have to interchange the indices $1$ and $3$ in the neutrino Yukawa terms.

For the case in which $M_3$ dominates, at first order approximation, we have the following expressions for the neutrino mixings \cite{King:1998jw},
\begin{eqnarray}
t^\nu_{23}&=&\frac{Y_{23}^{\nu}}{Y^{\nu}_{33}},\label{eq:tan23}\\
t^\nu_{13} &=&\frac{Y^\nu_{13}} {\sqrt{Y^{\nu 2}_{33}+Y^{\nu 2}_{23}}}+
\frac{M_3}{M_2}\frac{Y^\nu_{12}(s_{23}Y^{\nu}_{22}+c_{23}Y^\nu_{32})}{\sqrt{Y^{\nu 2}_{33}+Y^{\nu 2}_{23}} },\label{eq:tan13}\\
t^\nu_{12}&=&\frac{Y^{\nu}_{12}(Y^{\nu 2}_{33}+Y^{\nu 2}_{23})-Y^{\nu}_{13}(Y^{\nu}_{33}Y^{\nu}_{32}-Y^{\nu}_{22}Y^{\nu}_{23} ) }
{(Y^{\nu}_{33}Y^{\nu}_{33}-Y^{\nu}_{32}Y^{\nu}_{23}) \sqrt{Y^{\nu 2}_{33} +Y^{\nu 2}_{23} +Y^{\nu 2}_{13}  }  }\approx
\frac{Y^\nu_{12}}{c_{23}Y^\nu_{22}-s_{23}Y^\nu_{32}}.\label{eq:tan12}
\end{eqnarray}
In terms of the Abelian charges the Yukawa elements are
\begin{eqnarray}
Y^\nu_{ij}=\vep^{|l_i+n_j+h_u|}\equiv \vep^{| l'_i+n_j|}, 
\quad l'_i\equiv l_i+h_u=l_i-2e_3,
\end{eqnarray}
where we have defined primed lepton doublet charges
which absorb the Higgs charge, as shown.
We can work here in terms of the primed charges, once they are fixed
we can determine the original Abelian charges (unprimed). The approximation in \eq{eq:tan12}
corresponds to the case in which we have enough suppression of the
second term in the expression for $t\nu_{12}$. In \eq{eq:tan13} the
second term can be neglected sometimes, depending on the ratio
$M_3/M_2$. The heaviest low energy neutrino masses are given by
\begin{eqnarray}
m_{\nu_3}=\frac{a^{\nu 2}_3 \vep^{2|l'_2+n_3|}v^2 }{M_3},\quad m_{\nu_2}=\frac{ a^{\nu 2}_2  \vep^{2|l'_2+n_2|}  v^2 }{M_2},
\end{eqnarray}
where we have written $a^{\nu 2}_3  \vep^{2|l'_2+n_3|}= Y^{\nu 2}_{33}+ Y^{\nu 2}_{23}$ and $a^{\nu 2}_2 \vep^{2|l'_2+n_2|}=\left( c_{23} Y^{\nu}_{22} -s_{23}Y^{\nu}_{32} \right)^2$. Thus the ratio of the differences of the solar to atmospheric neutrino can be written as
\begin{eqnarray}
\label{ratmn}
\frac{m_{\nu_2}}{m_{\nu_3}}\approx \frac{M_3}{M_2} \frac{c^2_{23}}{c^2_{12}}\frac{\left( Y^{\nu}_{22} -Y^{\nu}_{32} t\nu_{23}\right)^2}{Y^{\nu  2}_{33} + Y^{\nu 2}_{23}}\sim \vep^{p_2-p_3},
\end{eqnarray}
where
\begin{eqnarray}
\label{eq:p3}
p_{k}=|2l'_2+n_k|-|2n_k+\sigma|,\ {\mathrm{for}}\ k=2,\ 3.
\end{eqnarray}

Note that $p_k$ is then defined such that 
\begin{equation}
  \label{eq:30}
  m_{\nu_k} \approx \frac{v^2}{\left<\Sigma\right>} \epsilon^{p_k}.
\end{equation}

\section{$SU(5)$ solutions satisfying the GST relation}
\label{sec:su5-solut-satisfy-GST}
In this section we shall continue to focus on the 
case of $SU(5)$, where the quark Yukawa matrices take
the form of Eq.\ref{eq:su5matparam},
where, motivated by large atmospheric neutrino mixing,
we shall assume $r_d=0$ from Eq.~(\ref{l2l3})
The purpose of this section is to show how the GST
relation can emerge from $SU(5)$, by imposing additional
constraints on the parameters.
\footnote{Note that results in Section \ref{sec:su5-solut-satisfy-GST} and in 
Section \ref{sec:su5-solutions-not-GST} apply to both $SU(5)$ type and extended $SU(5)$ 
models, as discussed above.}

\subsection{The quark sector}
We have already seen that the GST relation
can be achieved in the u sector, mainly by allowing the
parameters $s$ and $s'$ to have different signs. 
In the down sector to satisfy GST we additionally require:
\begin{eqnarray}
|k_d+r'_d+s|&=&|k_d+s'|\nn\\
|k_d+r'_d+s'|-|k_d|&>&|k_d+r'_d+s|+|k_d+s'|-|k_d+s|\nn\\
|r'_d+k_d|&>&|k_d|.
\end{eqnarray}
The first of these equations ensures the equality of the order of the
elements $(1,2)$ and $(2,1)$ of the $Y^d$ matrix. The second equation
ensures that the element $(1,1)$ is suppressed enough with respect to
the contribution from the signalization of the $(1,2)$ block. This
last condition is usually satisfied whenever
$|k_d+r'_d+s'|>|k_d+r'_d+s|$ is satisfied. Finally the third condition
ensures a small right-handed mixing for d-quarks and a small
left-handed mixing for charged leptons.  Now in order to satisfy the
relations
\begin{eqnarray}
s^u_{12}=\sqrt{\frac{m_u}{m_c}}\approx \lambda^2,\quad s^d_{12}=\sqrt{\frac{m_d}{m_s}}\approx \lambda,
\end{eqnarray}
we need a structure of matrices, in terms of just one expansion parameter  $\vep=O(\lambda)$, such as
\begin{eqnarray}
Y^u= \left[
\begin{array}{ccc}
... &\vep^6&...\\
\vep^6& \vep^4 & \vep^2\\
...&\vep^2 &1
\end{array}
\right],\quad
Y^d= \left[
\begin{array}{ccc}
... &\vep^5& \vep^5\\
\vep^5& \vep^4 & \vep^4\\
...&\vep^2 &\vep^2
\end{array}
\right],
\end{eqnarray}
for which we have
\begin{eqnarray}
s^u_{12}\approx \vep^2 \quad s^d_{12}\approx \vep,\quad s^d_{23}\approx \vep^2,\quad s^d_{13}\approx \vep^3,\nn\\
\frac{m_c}{m_t}\approx \vep^4,\quad \frac{m_s}{m_b}\approx \vep^2, \quad \frac{m_b}{m_t}\approx \vep^2, ~~~~~
\end{eqnarray}
in agreement with observed values for quark masses and mixings for $\vep=\lambda$.

Now we can proceed as in the example of \eq{eq:textibross} where 
$s'+s$ is fixed to be $\pm 3$. In this case we see that we can have plausible solutions in the up sector by allowing half integer solutions
\begin{eqnarray}
\label{eq:solsspgst}
|s'+s|=13/2,\ 6,\ 11/2.
\end{eqnarray}
We will refer to these solutions as Solution 1, 2 and 3 respectively.
Note that only the charge differences are constrained here, the actual charges are not.

\noindent {\bf Solution 1}, $|s+s'|=13/2$,
\begin{eqnarray}
\label{eq:tex1gst}
Y^u= \left[
\begin{array}{ccc}
\vep^{35/2}&\vep^{13/2}&\vep^{35/4}\\
\vep^{13/2}& \vep^{9/2} & \vep^{9/4}\\
\vep^{35/4}&\vep^{9/4} &1
\end{array}
\right],\quad
Y^d= \left[
\begin{array}{ccc}
\vep^{69/4}&\vep^{25/4}& \vep^{25/4}\\
\vep^{25/4}& \vep^{19/4} & \vep^{19/4}\\
\vep^{17/2}&\vep^{5/2} &\vep^{5/2}
\end{array}
\right],
\end{eqnarray}
for
\begin{eqnarray}
&& r'_d=l_1-l_3=11,\ s=-\frac{9}{4},\ s'=\frac{35}{4},\ k_d=-\frac{5}{2},
\quad {\rm or}\nn
\\ && 
r'_d=l_1-l_3=-11, \ s=\frac{9}{4},\ s'=-\frac{35}{4},\ k_d=\frac{5}{2}.
\label{eq:sol1gst}
\end{eqnarray}
{\bf Solution 2}, $|s'+s|=6$,
\begin{eqnarray}
\label{eq:tex2gst}
Y^u= \left[
\begin{array}{ccc}
\vep^{16}&\vep^6&\vep^8\\
\vep^6& \vep^4 & \vep^2\\
\vep^8&\vep^2 &1
\end{array}
\right],\quad
Y^d= \left[
\begin{array}{ccc}
\vep^{31/2}&\vep^{11/2}& \vep^{11/2}\\
\vep^{11/2}& \vep^{9/2} & \vep^{9/2}\\
\vep^{15/2}&\vep^{5/2} &\vep^{5/2}
\end{array}
\right],
\end{eqnarray}
for
\begin{eqnarray}
&&r'_d=l_1-l_3=10, \ s=-2,\ s'=8,\ k_d=-\frac{5}{2},\quad {\rm or}\nn\\
&&r'_d=l_1-l_3=-10, \ s=2,\ s'=-8,\ k_d=\frac{5}{2}.\label{eq:sol2gst}
\end{eqnarray}
{\bf Solution 3}, $|s+s'|=11/2$,
\begin{eqnarray}
\label{tex3gst}
Y^u= \left[
\begin{array}{ccc}
\vep^{29/2}&\vep^{11/2}&\vep^{29/4}\\
\vep^{11/2}& \vep^{7/2} & \vep^{7/4}\\
\vep^{29/4}&\vep^{7/4} &1
\end{array}
\right],\quad
Y^d= \left[
\begin{array}{ccc}
\vep^{31/2}&\vep^{21/4}& \vep^{21/4}\\
\vep^{21/4}& \vep^{15/4} & \vep^{15/4}\\
\vep^{33/4}&\vep^{2} &\vep^{2}
\end{array}
\right],
\end{eqnarray}
for
\begin{eqnarray}
&&r'_d=l_1-l_3=41/4, \ s=-\frac{7}{4},\ s'=\frac{29}{4},\ k_d=-2,
\quad{\rm or}\nn
\\ &&r'_d=l_1-l_3=-41/4, \ s=\frac{7}{4},\ s'=-\frac{29}{4},\ k_d=2. 
\label{eq:sol3gst}
\end{eqnarray}
All the previous solutions \eqs{eq:sol1gst}-\eqs{eq:sol3gst} lead to
small $\tan\beta$ ($O(1)$), due to the choice of $k_d$. To find
solutions such that $\tan\beta$ is $O(10)$ is more difficult, due to
the requirements in the up sector, but we have found the following
solution
\begin{eqnarray}
&&r'_d=l_1-l_3=\frac{19}{2}, \ s=-2,\ s'=\frac{15}{2},\ k_d=-\frac{3}{2},
 \quad {\rm or}\nn
\\ 
&&
r'_d=l_1-l_3=\frac{-19}{2}, \ s=2, \ s'=-\frac{15}{2},\ k_d=\frac{3}{2}
\label{eq:sol4gst}.
\end{eqnarray}
\label{sec:neutgst}

\subsection{The neutrino sector}
\label{sec:neutrino-sector-gst}

Now we construct solutions for the lepton sector constrained by the
requirements from the quark sector in the previous subsection, 
where we assumed $r_d=l_2-l_3=0$, and determined
the charge differences $r'_d = l_1-l_2$ that agree with the GST
relation. Indeed it is convenient to label the solutions
in the previous subsection by the value of $r'_d = l_1-l_2$.
Here we find the charges $n_i,l_i,$ and $\sigma$ which satisfy
the conditions arising from the neutrino sector,
Eqs.~(\ref{eq:tan23}-\ref{eq:tan12}).  {\footnote{The
condition $l_2 =l_3$ is a requirement of the class of see-saw models
that we are looking for, single right-handed neutrino dominance
(SRHND). Note that here we can also have $l'_2=-l'_3$ which then
forces $n_3=0$ for $l'_2 \ne 0$, in which case the solutions will be
even more restricted.}}.  In order to satisfy \eq{eq:tan23}, the most
natural solution to achieve $t^\nu_{12}$ large is to have
\begin{eqnarray}
|l'_1+n_2|=|l'_2+n_2|.
\end{eqnarray}
The simplest solution is to assume that $n_2 = 0$. 
Since $l'_1$ and $l'_2$ are related through $r'_d = l_1 - l_3 = l'_1 - l'_2$ the solutions to this equation are:
\begin{eqnarray}
r'_d&=&0\\
l'_1&=&\frac{r'_d}{2}=-l'_2.\label{eq:soll1l2}
\end{eqnarray}
Since none of the solutions found in the previous subsection had
$r'_d = 0$, we have to work with the second solution in
Eq.~(\ref{eq:soll1l2}).  However, we do not need to solve
Eq.~(\ref{eq:tan12}) exactly, so we are going to perturb away from it,
by keeping $n_2\neq 0$, but we expect it to be small in comparison
with $l'_1 = -l'_2$. Then we write:
\begin{eqnarray}
\label{eq:needp12}
p_{12}=|l'_1+n_2|-|l'_2+n_2|
\end{eqnarray}
So $t^\nu_{12}$ is $O(\vep^{p_{12}})$. The solution \eq{eq:soll1l2} implies that $l'_1$ and $l'_2$ should have opposite sign, so we choose the case $l'_1>0$
(the other case  is similar). Since $r'_d$ is large for all three GST solutions, and $n_2$ should be small in order to satisfy Eq.~(\ref{eq:needp12}),
we can see that $|l'_2 + n_2| = -(l'_2 + n_2)$, and $|l'_1 + n_2| =
l'_1 + n_2$ for all the solutions from the previous subsection.
Putting these relations into Eq.~(\ref{eq:needp12}) we get:
\begin{eqnarray}
\label{eq:condl1l2}
n_2=\frac{p_{12}}{2}.
\end{eqnarray}
So when we choose $p_{12}$, $n_2$ is determined. Now for the $t^\nu_{13}$ mixing, which should be at most $O(\lambda)$, from \eq{eq:tan13} we need
\begin{eqnarray}
\label{eq:n3pos}
|l'_1+n_3|>|l'_2+n_3|\Rightarrow n_3>0,
\end{eqnarray}
hence let us define $p_{13}$ by:
\begin{eqnarray}
\label{eq:needp13}
p_{13}=|l'_1+n_3|-|l'_2+n_3|,
\end{eqnarray}
We assume that the first term in Eq.~(\ref{eq:tan13}) dominates. Then $t^\nu_{13}\approx \vep^{p_{13}}/\sqrt{2}$. 
\footnote{We have checked that this is indeed true for the solutions that we find for $n_2,n_3$ later in this section.}
By applying the same logic that led to Eq.~(\ref{eq:condl1l2}), we achieve:
\begin{eqnarray}
\label{eq:n3zeta}
n_3=\frac{p_{13}}{2}
\end{eqnarray}
So fixing $p_{13} \geq 1$ we fix $n_3$. Now we need to impose the conditions under which we can have an appropriate value of \eq{ratmn}.  
First note that in order to achieve $m_{\nu_3}=O(10^{-2}){\rm eV}$:
\begin{eqnarray}
{\rm{for}} <\Sigma>=M_P,\quad \frac{v^2}{<\Sigma>}\approx 6 \times 10^{-6} \ {\rm eV}\quad {\rm we}\ {\rm need}\ \vep^{p_3}\sim 10^4 \nn\\
{\rm{for}} <\Sigma>=M_G,\quad \frac{v^2}{<\Sigma>}\approx 6 \times 10^{-3} \ {\rm eV}\quad {\rm we}\ {\rm need}\ \vep^{p_3}\sim 10,
\end{eqnarray}
where $p_3$ has been defined in \eq{eq:p3}. In terms of powers of $\lambda$, we have $\lambda^{-4}-\lambda^{-7}=O(10^5)-O(10^4)$ for $<\Sigma>=M_P$ and $\lambda^{-1}, \lambda^{-2}=O(10)$
for $<\Sigma>=M_G$. This corresponds to the following requirements:
\begin{eqnarray}
\label{eq:consonsig}
{\rm{for}} <\Sigma>=M_P,\quad p_3=(-4,-7)\\
{\rm{for}} <\Sigma>=M_G,\quad p_3=(-1,-2).
\end{eqnarray}
We can conclude that for zero $n_2$, from Eq.~(\ref{eq:srhnd1}), since $n_3 > 0$, so must $\sigma$ be positive. Then we can write the power $p_2-p_3$ ($m_{\nu_2}/m_{\nu_3}\sim\vep^{p_2-p_3}$) as follows:
\begin{equation}
  \label{eq:2}
  p_2-p_3 = -2(l'_2 + n_2)- (2n_2 + \sigma) +(2n_3 + \sigma) \mp 2(l'_2 + n_3).
\end{equation}
The uncertainty in the final sign comes from whether $|l'_2| > |n_3|$. If this is the case then we get:
\begin{eqnarray}
\label{eq:difn3n2}
p_3-p_2=4(n_2-n_3).
\end{eqnarray}
Otherwise we end up with
\begin{equation}
p_2-p_3= -4(l'_2+n_2)
\end{equation}
The second form is of no use to us, since we know that $-l_2'$ is big for the models we are considering, and since $n_2$ is small
we can not get an acceptable mass ratio for $m_{\nu_2}$ to $m_{\nu_3}$. For the first form, Eq.~(\ref{eq:difn3n2}), we need $n_2 \ne 0$,
because substituting \eq{eq:n3zeta} into  \eq{eq:difn3n2} we have $p_2-p_3=2p_{13}-4n_2$ and we need $p_{13}\geq 1$ so for $n_2=0$ we have $p_2-p_3\geq 2$. 

With the above requirements then we can see that the parameters $n_3$ and $n_2$ do not depend on $r'_d$. The only parameter which 
depends on this is $\sigma$, through Eq.~(\ref{eq:2}), using the fact that $l'_2 = -r'_d/2$. This also fixes the scale at 
which the $U(1)$ should be broken. So, independently of $r'_d$, we have the following solutions
\begin{eqnarray}
p_{12}=\frac{1}{4},\ p_{13}=1, \ p_2-p_3=\frac{3}{2} & \Rightarrow & \ n_2=\frac{1}{8},\ n_3=\frac{1}{2};\nn \\ 
p_{12}=\frac{1}{2},\ p_{13}=1, \ p_2-p_3=1    & \Rightarrow & \ n_2=\frac{1}{4},\ n_3=\frac{1}{2}.\label{eq:solsn3n2}
\end{eqnarray}
We can write the approximate expressions of mixings and masses in terms of the above results and the coefficients $a^\nu_{ij}$ of $O(1)$,
\begin{eqnarray}
\label{eq:mixmasresn}
t^\nu_{23}&=&\frac{a^{\nu}_{23}}{a^{\nu}_{33}},\quad
t^\nu_{13}\ \ = \ \ \frac {a^\nu_{13}\vep^{|2n_3|}}{\sqrt{a^{\nu \ 2}_{33} +a^{\nu \ 2}_{23} }},\quad
t^\nu_{12}\ \ = \ \ \frac{a^\nu_{12}\vep^{|2n_2|}}{(c_{23}a^\nu_{22}-s_{23}a^\nu_{32})},\nn\\
\frac{m_{\nu_2}}{m_{\nu_3}}&=& \frac{c^{\nu\ 2}_{23}}{c^{\nu\ 2}_{12}}\frac{(a^{\nu}_{22}-a^{\nu}_{32}t_{23})^2}{(a^{\nu \ 2}_{33} +a^{\nu \ 2}_{23})}\vep^{|4(n_3-n_2)|},\quad
m_{\nu_3}\ \ = \ \ \frac{v^2}{\langle \Sigma \rangle}(a^{\nu \ 2}_{33} +a^{\nu \ 2}_{23})\vep^{|p_3|}.
\end{eqnarray}
As we have seen above, the charges $\sigma$ are constrained by the differences $r'_d$, the requirements of 
Eq.~(\ref{eq:2}) and the solutions to \eq{eq:solsn3n2}, which have the same value for $n_3$, so for these two sets of solutions we have the same value for $\sigma$. We write down these solutions for $<\Sigma>=M_P$ in Table (\ref{tbl:solGSTMP}) and for  $<\Sigma>=M_G$ in Table (\ref{tbl:solGSTMG}).
\begin{table}[ht]
  \centering
  \begin{tabular}{|l|c|cc|ccc|}
    \hline
Sol.& $r'_d$ & $n_2$ & $n_3$ & $p_3$ & $\sigma$ & $M_3$ [GeV] \\
    \hline
    {\bf 1}&11&$\frac{1}{8}$&$\frac{1}{2}$ &  (-4,-7)&(14,16)&$O(10^{10}),O(10^8)$ \\
    {\bf 1}&11&$\frac{1}{4}$&$\frac{1}{2}$& (-4,-7)&(14,16)&$O(10^{10}),O(10^8)$ \\
    {\bf 2}&10&$\frac{1}{8}$&$\frac{1}{2}$&(-4,-7)&(13,14)&$O(10^{11}),O(10^9)$ \\
    {\bf 2}&10&$\frac{1}{4}$&$\frac{1}{2}$&(-4,-7)&(13,14)&$O(10^{11}),O(10^9)$ \\
    {\bf 3}&$\frac{41}{4}$&$\frac{1}{8}$&$\frac{1}{2}$&(-4,-7)&$(\frac{53}{4},\frac{57}{4})$&$O(10^{10})$,$O(10^8)$ \\
    {\bf 3}&$\frac{41}{4}$&$\frac{1}{4}$&$\frac{1}{2}$&(-4,-7)&$(\frac{53}{4},\frac{57}{4})$&$O(10^{10})$,$O(10^8)$ \\
    \hline
  \end{tabular}
  \caption{$\Sigma$ at $M_P$ for the solutions satisfying the GST relation.}
  \label{tbl:solGSTMP}
\end{table}
\begin{table}[ht]
  \centering
  \begin{tabular}{|l|c|cc|ccc|}
    \hline
    Sol.&$r'_d$ & $n_2$ & $n_3$ & $p_3$ & $\sigma$ & $M_3$ [GeV] \\
    \hline
    {\bf 1}&11& $\frac{1}{8}$&$\frac{1}{2}$&(-1,-2)&$(10,11)$&$O(10^8)$ \\
    {\bf 1}&11& $\frac{1}{4}$&$\frac{1}{2}$&(-1,-2)&$(10,11)$&$O(10^8)$ \\
    {\bf 2}&10& $\frac{1}{8}$&$\frac{1}{2}$&(-1,-2)&$(10,11)$&$O(10^9), O(10^{10})$ \\
    {\bf 2}&10& $\frac{1}{4}$&$\frac{1}{2}$&(-1,-2)&$(10,11)$&$O(10^9), O(10^{10})$ \\
    {\bf 3}&$\frac{41}{4}$& $\frac{1}{8}$&$\frac{1}{2}$&(-1,-2)&$(\frac{37}{4},\frac{41}{4})$&$O(10^9), O(10^8)$ \\
    {\bf 3}&$\frac{41}{4}$& $\frac{1}{4}$&$\frac{1}{2}$&(-1,-2)&$(\frac{37}{4},\frac{41}{4})$&$O(10^9), O(10^8)$ \\
    \hline
  \end{tabular}
  \caption{$\Sigma$ at $M_G$ for the solutions satisfying the GST relation.}
  \label{tbl:solGSTMG}
\end{table}

 The solutions presented here satisfy the conditions of the single neutrino right-handed dominance, Eq.~(\ref{eq:srhnd1}), 
which relate second and third families. For the first and second family we need similar conditions, which are safely satisfied whenever
$2n_1>2n_2>-\sigma$ for $(2n_i+\sigma)$ positive. Thus $n_1$ is not completely determined but we can choose it to be a negative number between $-\sigma/2$ and $0$.

Now that we have determined the conditions that the charges $l'_i$ and $n_i$ need to satisfy in order to produce 
SRHND solutions we can determine the $e_i$ and $l_i$ charges, which are in agreement with the cancellation of
 anomalies, Eqs.(\ref{eq:A3}-\ref{eq:A1p}), and that determines the matrices $Y^e$, $Y^u$ and $Y^d$.
In Section \ref{sec:su5q} we presented the conditions that the fermion mass matrices $Y^u$,  $Y^d$, 
$Y^e$ and $Y^{\nu}$ need to satisfy in order to give acceptable predictions for mass ratios and mixings but 
without specifying the charges. 
The charges are then determined from $r'_d$ and $k_d$. We start by rewriting $k_d$ using the $SU(5)$ charge relations,
and the fact that $l'_i \equiv l_i + h_u$:
\begin{equation}
  \label{eq:25}
  k_d = q_3 + d_3 + h_d = e_3 + l_3 - h_u = e_3 + (l'_3 - h_u ) - h_u 
\end{equation}
Then we use the fact that $k_u = 0 = 2e_3 + h_u$, and we can solve for $e_3$ in terms of $k_d$ and $r'_d$ ( using Eq.~(\ref{eq:soll1l2}):
\begin{equation}
  \label{eq:26}
  e_3 = \frac{2 k_d + r'_d}{10}
\end{equation}
Once we have $e_3$, and $l'_3$, we can get $l_3$ since $h_u = - 2e_3$. From there, we can calculate the other charges from
$s,s',r,r'$ using Eq.~(\ref{eq:11}) and Eq.~(\ref{eq:24}).

The charges calculated in this way are laid out in Table \ref{tbl:sol-leptch-GST}.


%
\begin{table}[ht]
  \centering
  \begin{tabular}{|l|c|c|cc|ccccc|c|}
    \hline
    Sol.& $r'_d$ & $k_d$ & $n_2$ & $n_3$ &$e_1$ & $e_2$ & $e_3$ & $l_1$ & $l_3$ & Fit \\
    \hline 
    {\bf 1}&11& $\frac{-5}{2}$ & $\frac{1}{8}$&$\frac{1}{2}$&$\frac{187}{20}$ & $\frac{-33}{20}$ & $\frac{3}{5}$ & $\frac{67}{10}$ & $\frac{-43}{10}$ & - \\
    {\bf 1}&11& $\frac{-5}{2}$ & $\frac{1}{4}$&$\frac{1}{2}$&$\frac{187}{20}$ & $\frac{-33}{20}$ & $\frac{3}{5}$ & $\frac{67}{10}$ & $\frac{-43}{10}$ & - \\
    {\bf 2}&10& $\frac{-5}{2}$ & $\frac{1}{8}$&$\frac{1}{2}$& $8$ &  $-2$ & $0$  & $\frac{15}{2}$ & $\frac{-5}{2}$ & - \\  
    {\bf 2}&10& $\frac{-5}{2}$ & $\frac{1}{4}$&$\frac{1}{2}$&$8$ &  $-2$ & $0$  & $\frac{15}{2}$ & $\frac{-5}{2}$ & 1 \\  
    {\bf 3}&$\frac{41}{4}$ & $-2$ &$\frac{1}{8}$&$\frac{1}{2}$  & $\frac{63}{8}$ & $\frac{9}{8}$ & $\frac{25}{8}$ & $\frac{51}{8}$ & $\frac{-31}{8}$ & - \\
    {\bf 3}&$\frac{41}{4}$ & $-2$ &$\frac{1}{4}$&$\frac{1}{2}$  & $\frac{63}{8}$ & $\frac{9}{8}$ & $\frac{25}{8}$ & $\frac{51}{8}$ & $\frac{-31}{8}$ & - \\
\hline
  \end{tabular}
  \caption{Charged lepton charges for the $SU(5)$ type solutions with $u=v=0$ satisfying the GST relation. The fits are discussed in section \ref{sec:fitsmasses}}
  \label{tbl:sol-leptch-GST}
\end{table}

\subsection{Solutions for the extended $SU(5)$ case with $u+v \ne 0$}
\label{sec:solutions-su5-like}

For this class of solutions, it is clear from Eq.~(\ref{eq:chargrelgensu5like}) and Eq.~(\ref{eq:gyukpar}) that the quark sector results will
be unchanged. This happens since $s,s',r,r'$ are blind to whether the family charges are related by the $SU(5)$ relation, or the extended $SU(5)$ relation.
$k_u$ must always be zero,
and the parameterization happens to leave $k_d$ unchanged. Since $k_e$ is not unchanged, as discussed in section~\ref{sec:genrzsu5like},
we need to find $k_e$ in order to know the structure of the electron Yukawa matrix.


It is helpful to rewrite $k_e$ and $k_d$, from the form in Eq.~(\ref{eq:8}) by using Eqs.~(\ref{eq:hd},~\ref{eq:1}) and $k_u = 0$:
\begin{eqnarray}
k_d&=&l_3+3e_3+u+\frac{4}{3}m,\nn\\
\label{eq:kd_ke_general}
k_e&=&l_3+3e_3+u+m,
\end{eqnarray}
where we have written $u+v=m$, as we will discuss in Section \ref{sbsec:susycppr} $m$ can be determined such a that the effects of the breaking of $U(1)$ in the $\mu$ term are of order $\leq m_{3/2}$. But on the other hand we need to keep the observed relation at low energies $m_b=O(m_{\tau})$, so either $m$ has to remain small or be negative to achieve $|k_d|=O(|k_e|)$.
In the present case the $Y^d$ matrix has exactly the same form as in \eqs{eq:charggensu5like} and $Y^e$ has the form
\begin{eqnarray}
Y^{e}=
\left(
\begin{array}{ccc}
\vep^{|s'+r'_{d}+k_e|}&\vep^{|s+r'_{d}+k_e|}&\vep^{|r'_d+k_e|}\\
\vep^{|s'+r_{d}+k_e|}&\vep^{|s+r_{d}+k_e|}&\vep^{|r_d+k_e|}\\
\vep^{|s'+k_e|}&\vep^{|s+k_e|}&\vep^{|k_e|}
\end{array}
\right).
\label{eq:chleptmatuvn0}
\end{eqnarray}
%
%

With $l_2=l_3$,
which determines the solutions of the charges $e_i$ and $l_i$
compatible with the condition $r_d=l_2-l_3=0$, the discussion 
follows exactly as  Section \ref{sec:neutrino-sector-gst} because there we have not referred to other parameters than to 
$k_d$, $r$, $r'$, $s$ and $s'$ without specifying their relations with the charges cancelling the anomalies.


In this case, the analysis that leads to Eq.~(\ref{eq:26}) must be repeated, but accounting for the
fact that instead of the $SU(5)$ relation between the charges, we must instead use the extended
$SU(5)$ relation between the charges. In this case, we find that:
\begin{equation}
  \label{eq:27}
  k_d = 3e_3 + l_3 + u+\frac{4}{3}(u + v) - 2 h_u = 5 e_3 + l'_3 + \frac{10}{3} u + \frac{4}{3} v
\end{equation}
where we have used that $l'_i=l_i-2e_3-u$. $l'_i$ is defined in such a way that $l_i+n_j+h_u=l'_i+n_j$. Using again the fact that $l'_3 = l'_2 = -\frac{r'_d }{2}$, we find that:
\begin{equation}
  \label{eq:28}
  e_3 = \frac{1}{10}(2 k_d + r'_d - \frac{20}{3} u - \frac{8}{3} v )
\end{equation}

Using these results, and the values of $s,s',r_f, r'_f$, we can find the charges 
in Table~\ref{tbl:sol-leptch-GSTuvne0}.

\begin{table}[ht]
  \centering
  \begin{tabular}{|l|c|c|cc|c|ccccccc|c|}
    \hline
    Sol.    & $r'_d$        &  $k_d$         & $u$ & $v$ & $k_e$ &$e_1$ & $e_2$ & $e_3$ & $l_1$ & $l_3$ & $n_2$ & $n_3$ & Fit \\
    \hline
    {\bf 1} & 11            & $\frac{-5}{2}$ & $-\frac{13}{2}$ & $7$  & $\frac{8}{3}$ &$\frac{709}{60}$ & $\frac{49}{60}$ 
                                                               & $\frac{46}{15}$ & $\frac{77}{15}$ & $\frac{-88}{15}$ &
                                                               $\frac{1}{8}$ & $\frac{1}{2}$ & - \\
    {\bf 1} & 11            & $\frac{-5}{2}$ & $\frac{-13}{2}$ & $7$  & $\frac{8}{3}$ &$\frac{709}{60}$ & $\frac{49}{60}$ 
                                                               & $\frac{46}{15}$ & $\frac{77}{15}$ & $\frac{-88}{15}$ & 
                                                               $\frac{1}{4}$ & $\frac{1}{2}$ & - \\
    {\bf 2} &10             & $\frac{-5}{2}$ & $-13$ & $\frac{27}{2}$             & $\frac{8}{3}$ & $\frac{407}{30}$ 
                                       &  $\frac{107}{30}$ & $\frac{167}{30}$  & $\frac{47}{15}$ & $\frac{-103}{15}$& 
                                       $\frac{1}{8}$& $\frac{1}{2}$ &2 \\  
    {\bf 2} &10             & $\frac{-5}{2}$ & $-13$ & $\frac{27}{2}$             & $\frac{8}{3}$ & $\frac{407}{30}$ 
                                       &  $\frac{107}{30}$ & $\frac{167}{30}$  & $\frac{47}{15}$ & $\frac{-103}{15}$& 
                                       $\frac{1}{4}$ & $\frac{1}{2}$ &- \\  
    {\bf 3} &$\frac{41}{4}$ & $-2$ & $\frac{-10}{3}$ & $\frac{23}{6}$           &$\frac{7}{3}$  & $\frac{363}{40}$ 
                                       & $\frac{3}{40}$ & $\frac{73}{40}$ & $\frac{653}{120}$ & $\frac{-577}{120}$& 
                                       $\frac{1}{8}$&$\frac{1}{2}$ &3 \\
    {\bf 3} &$\frac{41}{4}$ & $-2$ & $\frac{-10}{3}$ & $\frac{23}{6}$           &$\frac{7}{3}$  & $\frac{363}{40}$ 
                                       & $\frac{3}{40}$ & $\frac{73}{40}$ & $\frac{653}{120}$ & $\frac{-577}{120}$& 
                                       $\frac{1}{4}$&$\frac{1}{2}$& - \\
    \hline
  \end{tabular}
  \caption{Charged lepton charges for the extended $SU(5)$
    solutions with $m=u+v=1/2$ satisfying the GST relation. The fits are discussed in section \ref{sec:fitsmasses}}
  \label{tbl:sol-leptch-GSTuvne0}
\end{table}

\section{$SU(5)$ solutions not satisfying the GST relation}
\label{sec:su5-solutions-not-GST}
\subsection{The quark sector}
\label{sec:quark-sector-not-GST}
As we can see the GST relation puts a constraint on the opposite signs
of $s$ and $s'$ and on the difference of $r'_d=l_1-l_3$. If we do not
impose these requirements, allowing all the numbers $s,\ s^\prime,\ r,
\ r^\prime$ and $k_d$ to have the same sign, positive or negative, we
can factorize the $k_d$ factor out of the $Y^d$ matrix and so can
write the down matrix in the form
\begin{eqnarray}
Y^d=
\vep^{|k_d|}
 \left[
\begin{array}{ccc}
\vep^{|s'+l_1-l_3|}&\vep^{|s'|}&\vep^{|s'|}\\
\vep^{|s+l_1-l_3|}&\vep^{|s|}&\vep^{|s|}\\
\vep^{|l_1-l_3|}&1&1
\end{array}
\right].
\end{eqnarray}
In this case we do not have the restriction $|s+l_1-l_3|=|s'|$ so the parameter $l_1-l_2$ is not fixed
 by these conditions. In these cases $k_d$ is not constrained so it can acquire a value in the range 
$\sim (0,3)$ for different values of $\tan\beta$. In these cases all positive or all negative charges, 
the cases which reproduce quark masses and mixings are for
\begin{eqnarray}
\label{eq:solsspnongst}
|s|=2,\ |s'|=3 \quad {\rm{or}} \quad |s|=2,\ |s'|=4.
\end{eqnarray}
For $|s|=2, \ |s'|=3$ we have
\begin{eqnarray}
Y^d=\vep^{|k_d|}
\label{eq:ydinl1l3A}
 \left[
\begin{array}{ccc}
\vep^{|3+l_1-l_3|}& \vep^{|3|}&\vep^{|3|}\\
\vep^{|2+l_1-l_3|}& \vep^{|2|}& \vep^{|2|}\\ 
\vep^{|l_1-l_3|}&1&1
\end{array}
\right].
\end{eqnarray}
For $|s|=2, \ |s'|=4$ we have
\begin{eqnarray}
\label{eq:ydinl1l3B}
Y^d=\vep^{|k_d|}
 \left[
\begin{array}{ccc}
\vep^{|4+l_1-l_3|}& \vep^{|4|}& \vep^{|4|}\\
\vep^{|2+l_1-l_3|}& \vep^{|2|}& \vep^{|2|}\\ 
\vep^{|l_1-l_3|}&1&1
\end{array}
\right].
\end{eqnarray}
From  \eq{eq:ydinl1l3A} and  \eq{eq:ydinl1l3B} we can check if certain differences of leptonic charges can yield a suitable quark phenomenology. From \eq{eq:yukeigen} we can see that in the cases of having all charges $l_i$ and $e_i$ either positive or negative, then all the terms contributing to the first eigenvalue of $Y^d$, $y_1$, will have the same power, as we mentioned earlier. So the difference $r'_d$ here is constrained to reproduce an appropriate ratio
$m_d/m_s$.  Let us take here for definitiveness  the  case for positive charges (the negative charges case is completely analogous). Thus for $s=2$, $s'=3$, we have
\begin{eqnarray}
\frac{m_d}{m_s}\sim \frac{\vep^{3+r'_d}}{\vep^2}\sim (\lambda^2, \lambda^{3/2})
\end{eqnarray}
so in this case we have $r'_d=1,\ 3/2$. 
For the case $s=2$, $s'=4$, we have
\begin{eqnarray}
\frac{m_d}{m_s}\sim \frac{\vep^{4+r'_d}}{\vep^2}\sim (\lambda^2, \lambda^{3/2})
\end{eqnarray}
we do not want $r'_d=0$ as it will give somewhat large contribution from the $(3,1)$ element of the $Y^d$ matrix to the eigenvalues. So for this case $r'_d\approx 1/2$. 
In this case we have the following matrices for  \eq{eq:ydinl1l3A} 
\begin{eqnarray}
\label{eq:sol1-2nongst}
Y^d\!=\!
 \left[\begin{array}{ccc}
\epsilon^{4}&\epsilon^{3}&\epsilon^{3}\\
\epsilon^{3}&\epsilon^{2}&\epsilon^{2}\\
\epsilon^{1}&1&1
\end{array}
\right]\vep^{k_d},\
Y^d\!=\!
 \left[\begin{array}{ccc}
\epsilon^{9/2}&\epsilon^{4}&\epsilon^{4}\\
\epsilon^{7/2}&\epsilon^{2}&\epsilon^{2}\\
\epsilon^{3/2}&1&1
\end{array}
\right]\vep^{k_d},
\end{eqnarray}
respectively for $r'_d=1,\ 3/2$.
For \eq{eq:ydinl1l3B} we have 
\begin{eqnarray}
\label{eq:sol3nongst}
Y^d\!=\!
 \left[\begin{array}{ccc}
\epsilon^{9/2}&\epsilon^{4}&\epsilon^{4}\\
\epsilon^{5/2}&\epsilon^{2}&\epsilon^{2}\\
\epsilon^{1/2}&1&1
\end{array}
\right]\epsilon^{k_d},
\end{eqnarray}
for $r'_d=1/2.$ These solutions work for $k_d\in (0,3)$, depending on the value of $\tan\beta$, these matrices yield acceptable phenomenology in both charged the lepton sector and d quark sector.

\subsection{The neutrino sector}
\label{sec:neutrino-sector-non-GST}
%

As we have seen in Section \ref{sec:quark-sector-not-GST}, in these cases $r'_d$ is constrained to be 
$r'_d\in(1,3/2)$ for $(s,s')=(2,3)$ and  $r'_d\approx 1/2$ for $(s,s')=(2,4)$ but let us leave it 
unspecified for the moment. We consider here the case of all the parameters related to $l_i$ and $e_i$ 
positive. In this case we require that all the neutrino charges, $n_i$ to be negative but $\sigma$ positive.
We proceed as in Section (\ref{sec:neutgst}), in order to identify the charges $l'_i$, $n_i$ and $\sigma$.
In principle we need $\vep^{|l'_1+n_2|}=\vep^{|l'_2+n_2|}$ but now we require $l'_1,l'_2\geq 0$ so now the appropriate solution to this would be
\begin{eqnarray}
l'_1=r'_d,\ l'_2=0,\quad n_2=\frac{-r'_d}{2}.
\end{eqnarray}
However in this case, as in the case of section (\ref{sec:neutgst}), we will only be able to produce $m_{\nu_2}/m_{\nu_3}\sim \vep^2$. So we work with a solution of the form \eq{eq:needp12}. For this case we then have 
\begin{eqnarray}
\label{eq:soln2Ngst}
l'_1=r'_d,\ l'_2=0,\ n_2=\frac{p_{12}-r'_d}{2}.
\end{eqnarray}
Note that in this case the charges $l_i$ are positive because $l_2=k_d-3e_3$ and
here $e_3=k_d$. For $t_{13}$ we also make use of the parameterization of \eq{eq:needp13}. Assuming that $|r'_d | > |n_3|$,
\begin{eqnarray}
\label{eq:soln3Ngst}
n_3=\frac{p_{13}-r'_d}{2}.
\end{eqnarray}
In order to achieve an appropriate ratio for $m_{\nu_2}/m_{\nu_3}$ we need now the conditions 
$2n_3+\sigma>0$, $2n_2+\sigma>0$, $l'_2+n_2<0$, $l'_2+n_3<0$, for one of the last two inequalities the equality can be satisfied, but not for both. For this case, we have
also  $p_2-p_3=4(n_3-n_2)$ and using \eq{eq:soln2Ngst} and \eq{eq:soln3Ngst}  we have
$p_2-p_3=2(p_{13}-p_{12})$. We can also choose the parameters $p_{12}$, $p_{13}$ and $p_2-p_3$ as in
\eq{eq:solsn3n2} but now $n_3$ and $n_2$ are given by  \eq{eq:soln2Ngst} and \eq{eq:soln3Ngst}.
Thus we have
\begin{eqnarray}
\nonumber
&& p_{12}=\frac{1}{4},\ p_{13}=1, \ p_2-p_3=\frac{3}{2}\\
&&
\rightarrow \ n_2=\frac{1}{8}-\frac{r'_d}{2}<0 ,\ n_3=\frac{1}{2}-\frac{r'_d}{2}<0 \Rightarrow r'_d\geq 1, \label{eq:n3n2rdNgstA}\\ 
\nonumber
&&
p_{12}=\frac{1}{2},\ p_{13}=1, \ p_2-p_3=1\\
&&
\rightarrow \ n_2=\frac{1}{4}-\frac{r'_d}{2}<0,\ n_3=\frac{1}{2}-\frac{r'_d}{2}<0 \Rightarrow r'_d\geq 1. \label{eq:n3n2rdNgstB}
\end{eqnarray}
In Section (\ref{sec:su5q}) we determined the approximate values for $r'_d$. For $(s,s')=(2,3)$ we can have $r'_d=1,3/2$ while for $(s,s')=(2,4)$ we have 
$r'_d\approx 1/2$, which however is not in agreement with the conditions of \eq{eq:n3n2rdNgstA} and \eq{eq:n3n2rdNgstB}. The approximate expressions of mixings and masses in terms of the above results and the coefficients $a^\nu_{ij}$ of $O(1)$ are as in \eq{eq:mixmasresn}, except for $t^\nu_{13}$ and $t^\nu_{12}$ which now read
\begin{eqnarray}
\label{eq:mixangngst}
t^\nu_{13}\ \ = \ \ \frac {a^\nu_{13}\vep^{|r'_d+n_3|-|n_3|}}{\sqrt{a^{\nu \ 2}_{33} +a^{\nu \ 2}_{23} }},\quad
t^\nu_{12}\ \ = \ \ \frac{a^\nu_{12}\vep^{(|r'_d+n_2|-|n_2|)}}{(c_{23}a^\nu_{22}-s_{23}a^\nu_{32})}.\nn\\
\end{eqnarray}
We have listed the possible solutions of Table (\ref{tbl:solNGSTMP}) for \eq{eq:n3n2rdNgstA} at $\langle \Sigma\rangle=M_P$ and in Table (\ref{tbl:solNGSTMG}) for $\langle \Sigma\rangle=M_G$.
\begin{table}[htbp]
  \centering
  \begin{tabular}{|cc|cccc|}
    \hline
$r'_d$ & $n_2$ &$n_3$& $p_3$ & $\sigma$ & $M_3  [GeV]$ \\
    \hline
$1$ & $\frac{-3}{8}$ & $0$ & $(-\frac{9}{2},-\frac{5}{2})$ & $(\frac{9}{2},5)$ & $O(10^{15})$  \\
$\frac{3}{2}$ & $\frac{-5}{8}$ & $\frac{-1}{4}$ &  $(-\frac{17}{4},-\frac{19}{4})$ & $(5,\frac{11}{2})$ & $O(10^{15})$\\
$1$ & $\frac{-1}{4}$&1& (-6,-7)&(6,7)&$O(10^{15})$,  $O(10^{14})$\\
$\frac{3}{2}$ & $\frac{-1}{2}$&$\frac{-1}{4}$& (-6,-7)&($\frac{27}{4}$,$\frac{31}{4}$) &$O(10^{14})$, $O(10^{15})$\\
\hline
\end{tabular}
  \caption{$\Sigma$ at $M_P$ for the solutions not satisfying the GST relation.}
  \label{tbl:solNGSTMP}
\end{table}
\begin{table}[htbp]
  \centering
  \begin{tabular}{|cc|cccc|}
    \hline
$r'_d$ & $n_2$&$n_3$ & $p_3$ & $\sigma$ & $M_3$ [GeV] \\
    \hline
$1$ & $\frac{-3}{8}$&$0$ & $(0,-\frac{1}{2})$ & $(0,\frac{1}{2})$ & $O(10^{15})$ \\
$\frac{3}{2}$ & $\frac{-5}{8}$&$\frac{-1}{4}$&  $(-\frac{1}{4},-\frac{3}{4})$ &$(1,\frac{3}{2})$ &$O(10^{15})$,\\
$1$ & $\frac{-1}{4}$&$0$&1& (-1,-2)&$O(10^{18})$\\
$\frac{3}{2}$ & $\frac{-1}{2}$&$\frac{-1}{4}$& (-1,-2)&($\frac{7}{4}$,$\frac{11}{4}$)&$O(10^{18})$,$O(10^{17})$\\
 \hline
  \end{tabular}
  \caption{$\Sigma$ at $M_G$ for the solutions not satisfying the GST relation.}
  \label{tbl:solNGSTMG}
\end{table}
\begin{table}[ht]
  \centering
  \begin{tabular}{|c|cccccc|c|}
    \hline
$r'_d$ & $k_d$ &$e_1$ & $e_2$ & $e_3$ & $l_1$ & $l_3$ & Fit\\
\hline
1&2& $\frac{17}{5}$ & $\frac{12}{5}$ & $\frac{2}{5}$ & $\frac{9}{5}$ & $\frac{4}{5}$ & 4 \\
$\frac{3}{2}$& $2$ & $\frac{17}{5}$ & $\frac{12}{5}$ & $\frac{2}{5}$ & $\frac{7}{10}$ & $\frac{4}{5}$ & 5\\
1& $3$ & $\frac{18}{5}$ & $\frac{13}{5}$ & $\frac{3}{5}$ & $\frac{7}{10}$ & $\frac{-3}{10}$ & -\\
$\frac{3}{2}$& $3$ & $\frac{18}{5}$ & $\frac{13}{5}$ & $\frac{3}{5}$ & $\frac{6}{5}$ & $\frac{-3}{10}$ & -\\
    \hline
  \end{tabular}
  \caption{Charged lepton $U(1)_X$ charges for the solutions $u=v=0$ not satisfying the GST 
relation. The fits are discussed in section \ref{sec:fitsmasses}}
  \label{tbl:sol-lepchar-NGST}
\end{table}

\subsection{Solutions for the extended $SU(5)$ case with $u+v \ne 0$}
\label{sec:solutions-su5-like-non-GST}

We do not present any charges for this class of solutions, but here is how the charges would be calculated.
In this case, the analysis is carried out in the same way as in section \ref{sec:solutions-su5-like}. The only subtlety
is that the relation linking $l'_2$ to $r'_d$ is different. Instead, we have, from Eq.~(\ref{eq:soln2Ngst}) that $l'_2 = l'_3 = 0$.
Putting this result into Eq.~(\ref{eq:27}), we achieve:
\begin{equation}
  \label{eq:29}
  e_3 = \frac{1}{10}(2 k_d + \frac{4}{3} u - \frac{8}{3} v ).
\end{equation}

From $e_3$ and $l'_3$, the other charges may be calculated using the known values of $s,s',r_d,r'_d,u \, \mathrm{and}\, v$,
by using the extended $SU(5)$ charge relations, Eq.~(\ref{eq:1}) and the simplified parametrization, Eq.~(\ref{eq:16}).

\section{The non-$SU(5)$ Cases}
\label{sec:non-su5-cases}

\subsection{Solutions for $u=v=0$,  in the Pati-Salam case\label{sec:patisalamq}}
With $l_2=l_3$, in this case we have $s=q_2-q_3=l_2-l_3=0$ then the charges of the matrices are
\begin{eqnarray}
C(Y^{u, \nu})&=&
 \left[
\begin{array}{ccc}
  l_1-l_3+e_1-e_3 & l_1-l_3+e_2-e_3  & l_1-l_3 \\
  e_1-e_3         & e_2-e_3          & 0\\
  e_1-e_3         & e_2-e_3          & 0
\end{array}
\right]\nn\\
C(Y^{d, l})&=& 
 \left[
  \begin{array}{ccc}
    l_1 + l_3 + e_1 + e_3 & l_1 + l_3 + e_2 + e_3 & l_1 + l_3 + 2e_3 \\
    2l_3 + e_1 + e_3      & 2l_3 + e_2 + e_3      & 2l_3 + 2e_3 \\
    2l_3 + e_1 + e_3      & 2l_3 + e_2 + e_3      & 2l_3 + 2e_3
  \end{array}
\right]
\end{eqnarray}
In this case the $U(1)_X$ symmetry does not give an appropriate description of fermion masses and mixings, however it can be combined with non-renormalizable operators of the Pati-Salam group, \cite{King:2000ge}, in order to give a good description of the fermion phenomenology.
\subsection{Solutions for $u+v=0$}
One trivial example of non- $SU(5)$ cases was given in section (\ref{sec:yukawa-textures-uv-nonzero}) for the solution $u+v=0$. 
We proceed as in the section (~\ref{sec:su5q})- in order to analyze the appropriate phenomenology. We are interested 
in the cases $l_2=l_3$, this together with the condition of $O(1)$ top Yukawa coupling give us the following matrices of charges, 
which are derived with the appropriate substitutions in \eq{eq:yu-umvn0}-\eq{eq:ye-umvn0},
\begin{eqnarray}
C(Y^d)&=&
 \left[
\begin{array}{ccc}
  l_1 + e_1 &
  \frac{4(l_3-l_1)}{3} + e_1 - e_3 &
  e_2 - e_3 \\
  \frac{l_3-l_1}{3} + e_2 - e_3 &
  e_2 - e_3 &
  e_2 - e_3 \\
  \frac{l_3-l_1}{3} &
  0 &
  0
\end{array}
\right],\nn\\
C(Y^u)&=&
 \left[
\begin{array}{ccc}
  l_1 + e_1  &
  \frac{2(l_3-l_1)}{3} + e_3 - e_1 &
  \frac{2(l_3-l_1)}{3} + e_3 - e_1 \\
  \frac{l_3 - l_1}{3} + e_3 - e_3 &
  e_3 - e_2 &
  e_3 - e_2 \\
  \frac{l_3 - l_1}{3} &
  0 & 0
\end{array}
\right],\nn\\
C(Y^e)&=&
 \left[
\begin{array}{ccc}
  l_1+e_1 & l_1 + e_2 & l_1 + e_3 \\
  e_2-e_3 & e_2 - e_3 & 0 \\
  e_2-e_3 & e_2 - e_3 & 0
\end{array}
\right].\nn\\
\end{eqnarray}
Due to the form of the charges in the up and down quark matrices, first at all  we would need two expansion parameters: $\epsilon^u$ and $\epsilon^d$. But with this structure alone it is not possible to account simultaneously for appropriate mass ratios of the second to third family of quarks and for an appropriate $V_{cb}$ mixing. So in this case just with a $U(1)$ it is  not possible to explain fermion masses and mixings in the context of the single neutrino right-handed dominance, SNRHD.

%
\section{Numerical fits of masses and mixings}
\label{sec:fitsmasses}
\subsection{Fitted examples}
In this section we present numerical fits to some of the examples detailed in Sections 
(\ref{sec:su5-solut-satisfy-GST},\ref{sec:su5-solutions-not-GST}) and we compare the results  
with a fit of a generic $SU(3)$-like case \cite{King:2001uz}. 
The simplest way to construct the lepton Yukawa matrices from the charges is to first calculate $h_{u,d}$. We extract $h_d$ from $k_e$, $l_3$ and $e_3$ from  $k_e = l_3 + e_3 + h_d$. In general, we can use Eq.~(\ref{eq:kd_ke_general}) to obtain:
\begin{equation}
  \label{eq:22}
  h_u + h_d = m = 3(k_d - k_e)
\end{equation}
This is then enough to construct the lepton Yukawas from the appropriate line of the table (\ref{tbl:sol-leptch-GST} or \ref{tbl:sol-leptch-GSTuvne0}) of the lepton and Yukawa family charges. Below we specify the  examples that we have chosen to fit.
\subsubsection*{Fit 1: $SU(5)$ type solution ($u=v=0$): example satisfying the GST relation}
\label{sec:fit-1}

This takes GST solution 2, (Eq.~(\ref{eq:tex2gst})) in the $SU(5)$ type case, with $u=v=0$. The charges $l_i$, $e_i$, and $n_{2,3}$ are read off from the
fourth line of Table \ref{tbl:sol-leptch-GST}. In principle, the value of $\sigma$ would be read off from either Table \ref{tbl:solGSTMG} ( for neutrino
masses generated at the GUT scale ) or Table \ref{tbl:solGSTMP} ( for neutrino masses geneated at the Planck scale). However, these tables allow for
a range of $\sigma$; for this fit, we take $\sigma = 21/2$ for GUT scale neturino mass generation, and $\sigma = 29/2$ for Planck scale neutrino mass
generation.

Then, up to $\sigma$ and $n_1: -\sigma/2 \le n_1 \le 0$, the Yukawa and Majorana matrices are:

\begin{eqnarray}
  \label{eq:5}
  Y^u \! &=&\! 
   \left[
    \begin{array}{ccc}
      a^u_{11} \epsilon^{16} & a^u_{12} \epsilon^{6} & a^u_{13} \epsilon^{8} \\
      a^u_{21} \epsilon^{6} & a^u_{22} \epsilon^{4} & a^u_{23} \epsilon^{2} \\
      a^u_{31} \epsilon^{8} & a^u_{32} \epsilon^{2} & a^u_{33}
    \end{array}
  \right]\!,
\quad \quad \quad
  Y^d \ = \ 
   \left[
    \begin{array}{ccc}
      a^d_{11} \epsilon^{31/2} & a^d_{12} \epsilon^{11/2} & a^d_{13} \epsilon^{11/2} \\
      a^d_{21} \epsilon^{11/2} & a^d_{22} \epsilon^{9/2} & a^d_{23} \epsilon^{9/2} \\
      a^d_{31} \epsilon^{15/2} & a^d_{32} \epsilon^{5/2} & a^d_{33} \epsilon^{5/2}
    \end{array}
  \right]
  \nn \\
\label{eq:7}
  Y^e \! &=&\! 
   \left[
    \begin{array}{ccc}
      a^e_{11} \epsilon^{31/2} & a^e_{12} \epsilon^{11/2} & a^e_{13} \epsilon^{15/2} \\
      a^e_{21} \epsilon^{11/2} & a^e_{22} \epsilon^{9/2} & a^e_{23} \epsilon^{5/2} \\
      a^e_{31} \epsilon^{11/2} & a^e_{32} \epsilon^{9/2} & a^e_{33} \epsilon^{5/2}
    \end{array}
  \right]\!,
  \
  Y^\nu = 
   \left[
    \begin{array}{ccc}
      a^\nu_{11} \epsilon^{|n_1 + 5/2|} & a^\nu_{12} \epsilon^{11/4} & a^\nu_{13} \epsilon^{12/4} \\
      a^\nu_{21} \epsilon^{|n_1 - 15/2|} & a^\nu_{22} \epsilon^{29/4} & a^\nu_{23} \epsilon^{28/4} \\
      a^\nu_{31} \epsilon^{|n_1 - 15/2|} & a^\nu_{32} \epsilon^{29/4} & a^\nu_{33} \epsilon^{28/4}
    \end{array}
  \right]
 \nn \\
  \label{eq:9}
  M_{RR}\! &=&\!
   \left[
    \begin{array}{ccc}
      \epsilon^{|2n_1+\sigma|} & \epsilon^{|1/4 + n_1 + \sigma|} & \epsilon^{|1/2+ n_1 + \sigma|} \\
      . & a^N_{22} \epsilon^{|1/2 + \sigma| } & \epsilon^{|3/4 + \sigma|} \\
      . & . & \epsilon^{|1 + \sigma|}
    \end{array}
  \right] \left<\Sigma\right>
\end{eqnarray}

\subsubsection*{Fit 2: Extended $SU(5)$ solution ($u+v \ne 0$) satisfying the GST relation}
\label{sec:fit-2}

This takes GST solution 2, (Eq. (\ref{eq:tex2gst})), in the extended $SU(5)$ case with $u+v\ne0$. The charges $l_i$, $e_i$ and $n_{2,3}$ are read
off from the third line of Table \ref{tbl:sol-leptch-GSTuvne0}. The values of $\sigma$ taken are $\sigma = 19/2$, $\sigma = 29/2$ for GUT scale
and Planck scale neutrino mass generation respectively. Again, $n_1$ is taken to lie in the region $-\sigma/2 \le n_1 \le 0$.
{\footnote{The difference between Fit 1 and Fit 2 is that the charges (Tables (\ref{tbl:sol-leptch-GST}) and 
(\ref{tbl:sol-leptch-GSTuvne0}) respectively) are determined in a different way and hence the value of the 
effective parameter expansion $\vep$ is different}. 
\begin{eqnarray}
  \label{eq:10}
  Y^u \! &=&\! 
   \left[
    \begin{array}{ccc}
      a^u_{11} \epsilon^{16} & a^u_{12} \epsilon^{6} & a^u_{13} \epsilon^{8} \\
      a^u_{21} \epsilon^{6} & a^u_{22} \epsilon^{4} & a^u_{23} \epsilon^{2} \\
      a^u_{31} \epsilon^{8} & a^u_{32} \epsilon^{2} & a^u_{33}
    \end{array}
  \right]\!,
  \quad \quad \quad
  Y^d \ = \  
   \left[
    \begin{array}{ccc}
      a^d_{11} \epsilon^{31/2} & a^d_{12} \epsilon^{11/2} & a^d_{13} \epsilon^{11/2} \\
      a^d_{21} \epsilon^{11/2} & a^d_{22} \epsilon^{9/2} & a^d_{23} \epsilon^{9/2} \\
      a^d_{31} \epsilon^{15/2} & a^d_{32} \epsilon^{5/2} & a^d_{33} \epsilon^{5/2}
    \end{array}
  \right]
  \nn \\
\label{eq:3}
  Y^e &=& 
   \left[
    \begin{array}{ccc}
      a^e_{11} \epsilon^{46/3} & a^e_{12} \epsilon^{16/3} & a^e_{13} \epsilon^{22/3} \\
      a^e_{21} \epsilon^{16/3} & a^e_{22} \epsilon^{14/3} & a^e_{23} \epsilon^{8/3} \\
      a^e_{31} \epsilon^{16/3} & a^e_{32} \epsilon^{14/3} & a^e_{33} \epsilon^{8/3}
    \end{array}
  \right]\!,
  \
  Y^\nu =
   \left[
    \begin{array}{ccc}
      a^\nu_{11} \epsilon^{|n_1 +5|} & a^\nu_{12} \epsilon^{\frac{41}{8}} & a^\nu_{13} \epsilon^{\frac{11}{2}} \\
      a^\nu_{21} \epsilon^{|n_1 - 5|} & a^\nu_{22} \epsilon^{\frac{39}{8}} & a^\nu_{23} \epsilon^{\frac{9}{2}} \\
      a^\nu_{31} \epsilon^{|n_1 -5|} & a^\nu_{32} \epsilon^{\frac{39}{8}} & a^\nu_{33} \epsilon^{\frac{9}{2}}
    \end{array}
  \right]
  \nn\\
  M_{RR} \!&=&\!
   \left[
    \begin{array}{ccc}
      \epsilon^{|2n_1+\sigma|} & \epsilon^{|1/8 + n_1 + \sigma|} & \epsilon^{|1/2 + n_1+\sigma|} \\
      . & a^N_{22} \epsilon^{|1/4+\sigma|} & \epsilon^{|5/8+\sigma|} \\
      . & . & \epsilon^{|1+\sigma|}
    \end{array}
  \right] \left<\Sigma\right>
\end{eqnarray}

\subsubsection*{Fit 3: Extended $SU(5)$ solution ($u+v\neq 0$), satisfying the GST relation}
\label{sec:fit-3}

This takes GST solution 3, ( Eq. (\ref{eq:sol3gst})), in the extended $SU(5)$ case with $u+v\ne0$. The charges $l_i$, $e_i$ and $n_{2,3}$ are read off from
the fifth line of table \ref{tbl:sol-leptch-GSTuvne0}. The values of $\sigma$ taken are $\sigma = 39/4$, $\sigma = 55/4$ for GUT and Planck scale neutrino
mass generation respectively. $n_1$ lies in the region $-\sigma/2 \le n_1 \le 0$.
\begin{eqnarray}
\label{eq:18}
Y^u \! &=&\! 
   \left[
    \begin{array}{ccc}
      a^u_{11} \epsilon^{38/4} & a^u_{12} \epsilon^{22/4} & a^u_{13} \epsilon^{29/4}\\
      a^u_{21} \epsilon^{22/4} & a^u_{22} \epsilon^{14/4} & a^u_{23} \epsilon^{7/4}\\
      a^u_{31} \epsilon^{29/4} & a^u_{32} \epsilon^{7/4} & a^u_{33}
    \end{array}
  \right]\!,
 \quad \quad \quad
   Y^d = 
   \left[
    \begin{array}{ccc}
      a^d_{11} \epsilon^{62/4} & a^d_{12} \epsilon^{21/4} & a^d_{13} \epsilon^{21/4}\\
      a^d_{21} \epsilon^{21/4} & a^d_{22} \epsilon^{15/4} & a^d_{23} \epsilon^{15/4}\\
      a^d_{31} \epsilon^{33/4} & a^d_{32} \epsilon^{8/4} & a^d_{33} \epsilon^{8/4}
    \end{array}
  \right]
  \nn\\
  Y^e &=& 
   \left[
    \begin{array}{ccc}
      a^e_{11} \epsilon^{46/3} & a^e_{12} \epsilon^{19/3} & a^e_{13} \epsilon^{97/12}\\
      a^e_{21} \epsilon^{61/12} & a^e_{22} \epsilon^{47/12} & a^e_{23} \epsilon^{13/6}\\
      a^e_{31} \epsilon^{61/12} & a^e_{32} \epsilon^{47/12} & a^e_{33} \epsilon^{13/6}
    \end{array}
  \right]\!,
  \
  Y^\nu =
   \left[
    \begin{array}{ccc}
      a^\nu_{11} \epsilon^{|n_1+\frac{41}{8}|} & a^\nu_{12} \epsilon^{\frac{21}{4}} & a^\nu_{13} \epsilon^{\frac{45}{8}}\\
      a^\nu_{21} \epsilon^{|n_1-\frac{41}{8}} & a^\nu_{22} \epsilon^{5} & a^\nu_{23} \epsilon^{\frac{37}{8}}\\
      a^\nu_{31} \epsilon^{|n_1-\frac{41}{8}} & a^\nu_{32} \epsilon^{5} & a^\nu_{33} \epsilon^{\frac{37}{8}}
    \end{array}
  \right]
  \nn \\
  \label{eq:17}
  M_{RR} \!&=&\!
   \left[
    \begin{array}{ccc}
      \epsilon^{|2n_1+\sigma|} & \epsilon^{|1/8 + n_1 + \sigma|} & \epsilon^{|1/2 + n_1+\sigma|} \\
      . & a^N_{22}\epsilon^{|1/4+\sigma|} & \epsilon^{|5/8+\sigma|} \\
      . & . & \epsilon^{|1+\sigma|}
    \end{array}
  \right] \left<\Sigma\right>
\end{eqnarray}
\subsubsection*{Fit 4: $SU(5)$ ($u=v=0$) solution not satisfying the GST relation}
Here we present a solution non satisfying the GST relation of the form of \eq{eq:ydinl1l3A} for $l_1-l_3=1$, which corresponds to the set of charges of the first line of Table (\ref{tbl:sol-lepchar-NGST}). We also fix here the expansion parameter $\vep=0.19$, using the FI term. The high energy Yukawa and Majorana matrices are:
\begin{eqnarray}
Y^u \! &=&\! 
   \left[
    \begin{array}{ccc}
      a^u_{11} \epsilon^{6} & a^u_{12} \epsilon^{5} & a^u_{13} \epsilon^{3} \\
      a^u_{21} \epsilon^{5} & a^u_{22} \epsilon^{4} & a^u_{23} \epsilon^{2} \\
      a^u_{31} \epsilon^{3} & a^u_{32} \epsilon^{2} & a^u_{33}
    \end{array}
  \right]\!,
  \quad \quad \quad
  Y^d \! = \!  
   \left[
    \begin{array}{ccc}
      a^d_{11} \epsilon^{4} & a^d_{12} \epsilon^{3} & a^d_{13} \epsilon^{3} \\
      a^d_{21} \epsilon^{3} & a^d_{22} \epsilon^{2} & a^d_{23} \epsilon^{2} \\
      a^d_{31} \epsilon & a^d_{32}  & a^d_{33}
    \end{array}
  \right]\epsilon^{|k_d|}\nn\\
 Y^e\!&=&\!
 \left[
    \begin{array}{ccc}
      a^e_{11} \epsilon^{4} & a^e_{12} \epsilon^{3} & a^e_{13} \epsilon \\
      a^e_{21} \epsilon^{3} & a^e_{22} \epsilon^{2} & a^e_{23}  \\
      a^e_{31} \epsilon^{3} & a^e_{32}\epsilon^{2}  & a^e_{33}
    \end{array}
  \right]\epsilon^{|k_d|},
\quad 
 Y^\nu \!=\! 
   \left[
    \begin{array}{ccc}
      a^\nu_{11} \epsilon^{|n_1 + 1|} & a^\nu_{12} \epsilon^{5/8} & a^\nu_{13} \epsilon \\
      a^\nu_{21} \epsilon^{|n_1-3/8|} & a^\nu_{22} \epsilon^{3/8} & a^\nu_{23}  \\
      a^\nu_{31} \epsilon^{|n_1|} & a^\nu_{32} \epsilon^{3/8} & a^\nu_{33} 
    \end{array}
  \right],
 \nn \\
    M_{RR}\! &=&\!
 \left[
    \begin{array}{ccc}
       \epsilon^{|2n_1 + \sigma|} &  \epsilon^{|-3/8 + n_1 + \sigma|} &  \epsilon^{|n_1+\sigma|} \\
       . &  a^N_{22}\epsilon^{|-3/4 + \sigma|} & \epsilon^{|-3/8 + \sigma|} \\
       . & . &\epsilon^{|\sigma|}
    \end{array}
  \right] \left<\Sigma\right>.
\label{eq:31}
\end{eqnarray}
\subsubsection*{Fit 5: $SU(5)$ ($u=v=0$) solution not satisfying the GST relation}
Here we present another solution non satisfying the GST relation of the form of \eq{eq:ydinl1l3A} for $l_1-l_3=3/2$, which corresponds to the set of charges of the second line of Table (\ref{tbl:sol-lepchar-NGST}). We also fix here the expansion parameter $\vep=0.185$, using the FI term. The high energy Yukawa and Majorana matrices are:
\begin{eqnarray}
Y^u \! &=&\! 
   \left[
    \begin{array}{ccc}
      a^u_{11} \epsilon^{6} & a^u_{12} \epsilon^{5} & a^u_{13} \epsilon^{3} \\
      a^u_{21} \epsilon^{5} & a^u_{22} \epsilon^{4} & a^u_{23} \epsilon^{2} \\
      a^u_{31} \epsilon^{3} & a^u_{32} \epsilon^{2} & a^u_{33}
    \end{array}
  \right]\!,
  \quad \quad \quad
  Y^d \! = \!  
   \left[
    \begin{array}{ccc}
      a^d_{11} \epsilon^{9/2} & a^d_{12} \epsilon^{3} & a^d_{13} \epsilon^{3} \\
      a^d_{21} \epsilon^{7/2} & a^d_{22} \epsilon^{2} & a^d_{23} \epsilon^{2} \\
      a^d_{31} \epsilon^{3/2} & a^d_{32}  & a^d_{33}
    \end{array}
  \right]\epsilon^{|k_d|}\nn\\
 Y^e\!&=&\!
 \left[
    \begin{array}{ccc}
      a^e_{11} \epsilon^{9/2} & a^e_{12} \epsilon^{7/2} & a^e_{13} \epsilon^{3/2} \\
      a^e_{21} \epsilon^{3} & a^e_{22} \epsilon^{2} & a^e_{23}  \\
      a^e_{31} \epsilon^{3} & a^e_{32}\epsilon^{2}  & a^e_{33}
    \end{array}
  \right]\epsilon^{|k_d|},
\quad 
 Y^\nu \!=\! 
   \left[
    \begin{array}{ccc}
      a^\nu_{11} \epsilon^{|n_1 + 1|} & a^\nu_{12} \epsilon^{5/8} & a^\nu_{13} \epsilon \\
      a^\nu_{21} \epsilon^{|n_1-3/8|} & a^\nu_{22} \epsilon^{3/8} & a^\nu_{23}  \\
      a^\nu_{31} \epsilon^{|n_1|} & a^\nu_{32} \epsilon^{3/8} & a^\nu_{33} 
    \end{array}
  \right],
 \nn \\
    M_{RR}\! &=&\!
 \left[
    \begin{array}{ccc}
       \epsilon^{|2n_1 + \sigma|} &  \epsilon^{|-5/8 + n_1+\sigma|} &  \epsilon^{|-1/4+n_1+\sigma|} \\
       . &   a^N_{22}\epsilon^{|-5/4+\sigma|} & \epsilon^{|-7/8+\sigma|} \\
       . &  . &\epsilon^{|-1/2+\sigma|}
    \end{array}
  \right] \left<\Sigma\right>.
\label{eq:32}
\end{eqnarray}
\subsection{Details of the fitting method}
\label{sec:deta-fitt-meth}
One of the purposes of these fits is to compare which solution fits the data best while constraining the  abritary coefficients to remain at $O(1)$.
We therefore choose a minimization routine to find these $O(1)$ coefficients and compare the numerical values for the different solutions.
 In the quark sector we use eight experimental inputs in order to determine the parameters (coefficients or phases):
\begin{eqnarray}
\label{eq:fitparquarks}
V_{ub}/V_{cb}, \quad V_{td}/V_{ts}, \quad V_{us},\quad {\rm{Im}}\{J\},\quad m_u/m_c,\quad  m_c/m_t,\quad
m_d/m_s,\quad m_s/m_b.
\end{eqnarray}
We explain in the Appendix (\ref{ap:compinf}) how this fit is performed, the important point is that we can only fit eight parameters and the rest need to be fixed. The minimization algorithm has been optimized to fit the solutions satisfying the GST relation because the number of parameters is close to eight. We also fit examples of the non GST solutions but since there are more free parameters in this cases (mainly phases) it is un-practical to make a fit by fixing so many free parameters. So we present particular examples in these cases which do not necessarily correspond to the best $\chi^2$.

In the lepton sector we perform two fits, one for the coefficients of the charged lepton mass matrix  and the other for the coefficients of the neutrino mass matrix. We do not perform a combined fit for the coefficients of $Y^\nu$ and $Y^e$ because the uncertainties in these sectors are quite different. While the uncertainties in the masses of the charged leptons is very small, the uncertainties in lepton mixings and quantities related to neutrino masses are still large, such that we cannot determine the parameters involved to a very good accuracy.

The quantities used for the fit of the coefficients of the charged lepton  mass matrix are
\begin{eqnarray}
\frac{m_e}{m_\mu},\quad \frac{m_\mu}{m_\tau},
\end{eqnarray}
such that we can just determine two parameters, $a^e_{12}$ and $a^e_{22}$, but for the cases presented here this is enough.
In order to do the fit for the coefficients of the neutrino mass matrix we use  the observables
\begin{eqnarray}
\label{eq:fitparneuts}
t^l_{23},\quad t^l_{13},\quad t^l_{12},\quad \frac{|m_{\rm{sol}}|}{|m_{\rm{atm}}|},\quad  m_{\nu_3}
\end{eqnarray}
where we relate $t^l_{23}$ to the atmospheric mixing, $t^l_{12}$ to the solar mixing and  $t^l_{13}$ to the reactor mixing. In this case we are going to be able to fit just five parameters.  For this reason and because the uncertainties in the above observables are significantly bigger than the uncertainties in the quark sector, the fits of the coefficients of the neutrino mass matrix have large errors and they may leave a room for other solutions once the experimental uncertainties improve. Since we only have an upper bound for the reactor angle, $t^l_{13}$, we fit the solutions in the neighborhood of this upper bound.
\subsection{Results of the fits}
\subsubsection{Fit 1: $SU(5)$ ($u=v=0$) example satisfying the GST relation}

This is a $SU(5)$ type solution, and hence $u=v=0$, which satisfies the GST relation. The textures are as laid out in Eq.~(\ref{eq:9}). 

\subsubsection*{Quark sector}
We can use the expressions \eq{eq:yukeigen} and \eq{eq:mixsgeral} adapted to the solution of \eq{eq:sol2gst} in order to fit the Yukawa coefficients, 
along with the appropriate phases entering into the expressions of mixings. The expansion parameter $\vep$ is determined with the Fayet-Iliopoulos term 
and the appropriate charges cancelling the anomalies, for this case its value is $\vep=0.183$. The parameters that we 
fit are the real parameters
\begin{eqnarray}
\label{eq:fitpargst}
a^u_{12},\quad a^u_{23},\quad a^d_{22},\quad a^d_{12},\quad a^d_{13},\quad a^d_{23},\quad  a^d_{32},\quad \cos(\Phi_2),
\end{eqnarray}
which enter in the expressions of mixings and masses, \eq{eq:yukeigen}-\eq{eq:Vsasyu1}. Note that in these expressions the coefficients $a^f_{ij}$ can be complex but for the fit we choose them real and write down explicitly the phases.  We are free to choose the parameters to fit. However we need to check which are the most relevant parameters to test the symmetry. Thus we follow this as a guideline to choose the parameters to fit and leave other parameters fixed.  Due to the form of \eq{eq:sol2gst} the mixing angles in the $(2,3)$ sector of both matrices contribute at the same order in the $V_{\rm{CKM}}$ matrix mixing, $s^Q_{23}=|a^d_{23}-a^u_{23}e^{i\Phi_{X_{23}}}|\vep^2$, so we have decided to put a phase here. In the $s^u_{12}$ diagonalization angle and the second eigenvalue of $Y^u$ the combination $a^u_{22}e^{i\Phi_3}-a^{u\ 2}_{23}$ appears, so we have chosen as well to include a phase difference there. The fixed parameters are then
\begin{eqnarray}
\label{eq:fixpargst}
a^u_{22},\quad \Phi_1, \quad \Phi_3,\quad \Phi_{X_{23}},
\end{eqnarray}
where $\Phi_1$ has the form of \eq{eq:phi1} and the phases $\Phi_3$ and $\Phi_{X_{23}}$ can be written as {\footnote{In terms of the $\beta_i$ phases appearing in the diagonalization matrices, \eq{eq:pardimatL}, we have $\Phi_1=-\beta^{u \ L}_3$, $\Phi_2=-\beta^{d\ L}_3$ and $\Phi_{X_{23}}=(\beta^{d\ L}_2-\beta^{d\ L}_1)-(\beta^{u\ L}_2-\beta^{u\ L}_1)$.}}
\begin{eqnarray}
\label{eq:phi3}
\Phi_3=\phi^u_{22}-2\phi^u_{23},\quad \Phi_{X_{23}}=(\phi^d_{33}-\phi^d_{23})-(\phi^u_{33}-\phi^u_{23}) .
\end{eqnarray}
The results of the fit in the quark sector appear in the second column of  Table (\ref{tabl:sol2gst}). 
\begin{table}[ht]
\begin{center}
\begin{tabular}{|r|l||l|l|}
\hline
\multicolumn{4}{|c|}{{Quark Fitted Parameters}}\\ \hline
\multicolumn{2}{|c|}{{GST sol. 2}}&
\multicolumn{1}{|c|}{{GST sol. 2, $u,v\neq 0$}}&
\multicolumn{1}{|c|}{{GST sol. 3, $u,v\neq 0$}}
\\ \hline
Parameter & BFP Value & BFP Value & BFP Value \\
$a^{u}_{12}$& $2.74\pm 0.61$ & $1.04\pm 0.19$ &  $2.74\pm 0.71$\\
$a^u_{23}$& $1.68\pm 0.17 $ &   $1.34\pm 0.13$ &  $1.41\pm 0.18$ \\
$a^d_{22}$& $1.08\pm 0.18$ & $1.05\pm 0.11$ & $0.70\pm 0.23$ \\
$a^d_{12}$& $0.93\pm 0.15$ & $0.55\pm 0.20$ & $0.74\pm 0.13$ \\
$a^d_{13}$& $0.29\pm 0.21$ & $0.30\pm 0.14$ & $0.74\pm 0.17$\\
$a^d_{23}$& $0.79\pm 0.10$ &  $0.70\pm 0.13$& $0.66\pm 0.35$\\
$a^d_{32}$& $0.48\pm 0.17$ &  $1.28\pm 0.32$ & $1.28\pm 0.58$\\
$\cos(\Phi_2)$& $0.454\pm 0.041$ &  $0.456\pm 0.041$ & $0.547\pm 0.424$\\ \hline
\multicolumn{4}{|c|}{{Quark Fixed Parameters}} \\ \hline
$\vep$ & $0.183$ & $0.217$ & $0.154$ \\
$a^{u}_{22}$& $1$& $1$ &  $1.4$ \\
$\cos(\Phi_3)$ & $0.8$ & $0.83$ & $0.8$\\
$\cos(\Phi_{X_{23}})$& $1$ & $1$ & $1$\\
$\Phi_1$ & $\pi/2$ & $\pi/2$ & $\pi/2$ \\ \hline
\multicolumn{3}{|c|}{{$\chi^2$}} & \\ \hline
$\chi^2$ & $1.47$ & $2.41$ & $4.32$\\
\hline
\end{tabular}
\end{center}
\caption{\footnotesize{Quark fitted parameters for the examples of Section \ref{sec:su5-solut-satisfy-GST}). The second column corresponds to 
the Solution 2 in the $SU(5)$ ($u=v=0$) case, the third column to the Solution 2 in the $u \ne -v \neq 0$ case. 
The fourth column presents the fit to the Solution 3 in the $u\neq -v \neq 0$ case.}}
\label{tabl:sol2gst}
\end{table}
Given these results we can think that the structure of Yukawa matrices has the following form
\begin{eqnarray}
Y^u= \left[
\begin{array}{ccc}
* &y_{12}e^{i\Phi_1}&y_{13}\\
y_{12}e^{i\Phi_1}&y_{22}e^{i\Phi_3}&y_{23}\\
y_{13} &y_{23}&1 
\end{array}
\right],\quad
Y^d= \left[
\begin{array}{ccc}
*& y_{12}e^{i\Phi_2} & y_{13}e^{i\Phi_2}\\
\left[y_{21}e^{i\Phi^R_2}\right] & y_{22} & y_{23}\\
*&y_{32}&1
\end{array}
\right],
\end{eqnarray}
where $y_{ij}$ denote real elements and we have associated the phases $\Phi_i$ to particular elements of the matrices. Note that we need three phases to determine the amount of CP violation experimentally required because in all the fits we found $\Phi_{X_{23}}=0$. If this phase was not zero then it could have been associated to the $Y^d_{23}$ element. The entries marked with $*$ cannot be determined because they are not restricted by masses and mixings, due to the structure of the Yukawa matrices. The value of $y_{21}e^{i\Phi^R_2}$ is determined indirectly  because we need to satisfy the GST relation so $t^R_{12}=t^L_{12}$ for both up and quark sectors.
\subsubsection*{Lepton sector}
We have fixed the coefficients of $Y^d$ in the quark sector and now we can use the results for the charged lepton matrix $Y^e$. The masses of the charged lepton are obtained through the $SU(5)$  relations, ensuring the correct value of charged lepton masses, once the masses of the d-quarks are in agreement with experimental information. Thus in this case we perform a fit just for coefficients of the neutrino mass matrix, $Y^\nu$, using the ratio of neutrino mass differences (solar to atmospheric), the mass of the heaviest neutrino and  the lepton mixings, which have a contributions from both the charged leptons and the neutrinos.  Here the relevant parameter that we need from the quark sector is  $a^d_{32}$ because the tangent of the angle diagonalizing $Y^e$ on the left is related to  this parameter: $t^e_{23}=a^e_{23}\propto a^d_{32}$.   Since this is an $O(1)$ mixing we have to take it into account for the results of the $U_{MNS}$ mixings, thus we have
\begin{eqnarray}
\label{eq:t23lept}
t^l_{23}=\frac{|c^e_{23}s^\nu_{23}e^{-i\phi_{X_{23}}}-s^e_{23}c^{\nu}_{23} |}{|s^\nu_{23}s^e_{23}+c^{\nu}_{23}c^e_{23}e^{i\phi_{X_{23}}}| },
\end{eqnarray}
where we use the expression \eq{eq:tan23} to determine $s^\nu_{23}$ and $c^{\nu}_{23}$, and the approximation $t^e_{23}=a^d_{32}$; $\phi_{X_{23}}$ is a phase relating $e$ and $\nu$  mixings in the $(2,3)$ sector \cite{NuMngsPhases}. We denote the  $U_{MNS}$ angles by the superscript $l$ and by $e$ and $\nu$ the charged lepton and neutrino mixings respectively. The mixings $t^l_{13}$ and  $t^\nu_{12}$  are essentially given by the neutrino mixings, \eqs{eq:mixmasresn}, so we fit these mixings according to \eq{eq:tan13} and \eq{eq:tan12} respectively. We note from Table (\ref{tabl:sol2gstneut}) that in the lepton sector we need two phases, $\phi_{X_{23}}$ and $\phi^{\nu}$. The phase $\phi_{X_{23}}$ can be associated to the charged lepton sector and we can put it in the $Y^e_{23}$ entry. The second phase, $\phi^{\nu}$ can be assigned to $Y^{\nu}_{22}$.
We fit  the mass ratio and the heaviest neutrino state  using their expressions appearing  in \eqs{eq:mixmasresn}. The results for this fit appear in the second column of Table (\ref{tabl:sol2gstneut}).
\begin{table}[ht]
\begin{center}
\begin{tabular}{|r|c|c|c|c|c|c|}
\hline
\multicolumn{7}{|c|}{{Neutrino Fitted Parameters}}\\ \hline
\multicolumn{3}{|l|}{{$\!\!\!$Parameter$\!\!\!$ ~~ GST sol. 2}}&
\multicolumn{2}{|c|}{{GST sol. 2, $u,v\neq 0$}}&
\multicolumn{2}{|c|}{{GST sol. 3, $u,v\neq 0$}}
\\ \hline
 & $M_P$ & $M_G$ & $M_P$ & $M_G$ & $M_P$ & $M_G$ \\
 & $\!\!$ BFP value & $\!\!$ BFP value & $\!\!$ BFP value  & $\!\!$ BFP value   & $\!\!$ BFP value & $\!\!\!$ BFP value    \\
$a^\nu_{23}$& $\!\!\!\!\!0.75\pm 0.79\!\!\!\!$ & $\!\!\!\!0.67\pm 0.61\!\!\!$ & $\!\!0.21\pm 0.25\!\!\!$ & $\!\!\!\!0.85\pm 0.27\!\!\!\!$ & $\!\!\!0.30\pm 0.18\!\!\!$ & $\!\!\!\!0.40\pm 0.15\!\!\!\!$ \\
$a^\nu_{13}$& $\!\!\!\!\!1.41\pm 1.32\!\!\!\!$ & $\!\!\!\!\!1.36\pm 1.10\!\!\!$ & $\!\!0.97\pm 0.47\!\!\!$ & $\!\!\!\!1.25\pm 0.63\!\!\!\!$ & $\!\!\!1.02\pm 0.50\!\!\!$ & $\!\!\!\!1.45\pm 0.70\!\!\!\!$  \\
$a^\nu_{12}$& $\!\!\!\!\!2.23\pm 0.92\!\!\!\!$ & $\!\!\!\!\!2.10\pm 0.81\!\!\!$ & $\!\!1.25\pm 0.29\!\!\!$ & $\!\!\!\!2.08\pm 0.69\!\!\!\!$ &  $\!\!\!1.35\pm 0.34\!\!\!$ &  $\!\!\!\!1.97\pm 0.45\!\!\!\!$ \\
$a^\nu_{22}$& $\!\!\!\!\!1.84\pm 1.37\!\!\!\!$ & $\!\!\!\!\!1.96\pm 1.92\!\!\!$ & $\!\!1.23\pm 1.41\!\!\!$ & $\!\!\!\!1.98\pm 0.79\!\!\!\!$ & $\!\!\!1.48\pm 1.31\!\!\!$ & $\!\!\!\!2.26\pm 1.6\!\!\!\!$ \\
$a^\nu_{32}$& $\!\!\!\!1.47\pm 1.93\!\!\!\!$ & $\!\!\!\!\!0.98\pm 0.91\!\!\!$ & $\!\!0.65\pm 0.70\!\!\!$ & $\!\!\!\!1.53 \pm 0.75\!\!\!\!$ & $\!\!\!0.53 \pm 0.78\!\!\!$ & $\!\!\!\!0.56 \pm 0.98\!\!\!\!$ \\
\hline
\multicolumn{7}{|c|}{{Neutrino Fixed Parameters}}\\ \hline
$\vep$ & $0.183$ & & $0.217$ &  & $0.154$ & \\
$a^e_{23}$ & $a^d_{32}=0.48$ & & $-1.6$ &   & $1.2$ &\\
$a^\nu_{33}$ & $1$ &  & $1$ &   & $0.7$ & $1$\\
$\sigma$        & $29/2$ & $21/2$ & $29/2$ & $19/2$  & $55/4$ & $39/4$ \\ 
$\!\!\!c(\phi_{\!X_{23}})\!\!\!$ & $0.29$ &  & $0.29$ &  & $1$ & $0.5$\\
$\!\!\!c(\!\phi^{\nu}\!)\!\!$ & $-1$ & & $-0.5$ &  & $0.86$ & $1$\\ \hline
$\!\!\!\!(n_2,n_3)\!\!\!$ & \multicolumn{2}{|c|}{{$(1/4,1/2)$}}&
\multicolumn{4}{|c|}{{$(1/8,1/2)$}}\\
\hline
\multicolumn{7}{|c|}{{$\chi^2$}} \\ \hline
$\chi^2$ & $0.44$ & $0.12$ & $1.67$ & $0.49$ & $2.16$& $0.72$\\ 
\hline
\end{tabular}
\end{center}
\caption{\footnotesize{Neutrino fitted parameters for the examples of Section \ref{sec:su5-solut-satisfy-GST}. The second column corresponds to the 
Solution 2 in the $SU(5)$ ($u=v=0$) case, the third column to the Solution 2 in the $u + v \neq 0$ case. The fourth column 
presents the fit to the Solution 3 in the $u + v \ne 0$ case. Here $c(y)$ is the cosine of the respective parameter.}}
\label{tabl:sol2gstneut}
\end{table}

\subsubsection{Fit 2 and Fit 3: Extended $SU(5)$ solutions with $u + v \neq 0$ satisfying the GST relation}
These are both extended $SU(5)$ solutions, with $u+v \ne 0$, satisfying the GST relation.
Fit 2 corresponds to the textures laid out in Eq.~(\ref{eq:3}), and Fit 3 corresponds to the textures laid out
in Eq.~(\ref{eq:17}).

\subsubsection*{Quark sector}

This section is completely analogous to the previous one, the only difference is in the value of $\vep$. We present here two examples. The first example corresponds to the first solution of \eq{eq:sol2gst}, which we called Solution 2, and corresponds to $\vep=0.217$ according to the charges of the third row of \eq{tbl:sol-leptch-GSTuvne0}. The second example corresponds to the first solution of \eq{eq:sol3gst}, which has been called Solution 3 and corresponds to $\vep=0.154$, according to the charges of the fourth row of \eq{tbl:sol-leptch-GSTuvne0}.  The fitted and fixed parameters are also those of the previous example, \eq{eq:fitpargst} and \eq{eq:fixpargst} respectively. The results for the quark fitting are presented in the third and fourth column of Table (\ref{tabl:sol2gst}), respectively, so we can compare directly with the previous case.
\subsubsection*{Lepton sector}
This case is different from the Section \ref{sec:fit-1} because now we do not have the $SU(5)$ relations. Instead the parameter 
$k_e$ is different from $k_d$, as explained in Section (\ref{sec:su5q}), and hence $Y^e\neq (Y^d)^T$. In this case we perform two fits, one for the coefficients of the charged lepton mass matrix, $Y^e$ and another for the coefficients of the  neutrino mass matrix, $Y^\nu$.

For the Solution 2, taking into account the value of the charges, the second row of Table (\ref{tbl:sol-leptch-GSTuvne0}), and that $m=u+v=1/2$ we have $k_e=-8/3$. We note in this case that since we need $m_b\approx m_{\tau}$, which are given by 
\begin{eqnarray}
\label{eq:mbmtau}
m_b&=&m_t\vep^{|k_d|},\quad k_d=l_3+3e_3+u+4(u+v)/3\nn\\
m_\tau&=&m_t\vep^{|k_e|},\quad k_e=l_3+3e_3+u+(u+v),
\end{eqnarray}
we expect the sum $(u+v)$ to remain small.

Now the coefficients  $a^e_{23}$ and $a^d_{32}$ are not related but we can fix $a^e_{23}$ in the neutrino sector such that it is in agreement with the results from neutrino oscillation. We have performed a fit using the experimental information of the parameters of \eq{eq:fitparneuts}. Here we have also used the expression \eq{eq:t23lept} in order to fit the atmospheric angle, the expressions \eq{eq:tan13} and \eq{eq:tan12} to fit $t^l_{13}$ and $t^l_{12}$ (reactor and solar angle respectively) and the mass ratio and the heaviest neutrino state  using their expressions appearing  in \eqs{eq:mixmasresn}. The results for this fit appear in the third column of Table (\ref{tabl:sol2gstneut}). 

Once the parameter $a^e_{23}$ has been fixed we fit the parameters of the charged lepton mass matrix, of the form \eq{eq:chleptmatuvn0} and the other parameters as in the first solution of \eq{eq:sol2gst}. In this case the relevant parameters are $a^e_{12}$ and $a^e_{22}$. However if we just fit the expressions
\begin{eqnarray}
\frac{m_e}{m_{\mu}}&=&\frac{|a^e_{12}|^2}{|(a^e_{22}-a^e_{23}a^e_{32})|^2}\vep^{4/3}=s^{e\ 2}_{12},\nn \\
\frac{m_\mu}{m_\tau}&=&(a^e_{22}-a^e_{23}a^e_{32})\vep^2,
\end{eqnarray}
the coefficients $a^e_{12}$ and $a^e_{22}$ are not quite $O(1)$ so we have to make use of a coefficient, $c$ such that $(a^e_{22}-a^e_{23}a^e_{32})\rightarrow \ (a^e_{22}-a^e_{23}a^e_{32})/c $, e.g.  $c=3$, in order to have acceptable values for charged lepton masses. This fit is presented in the second column of Table (\ref{tabl:sol2gscluvn0}). In this case the extra-coefficient needed for the fit is not really justified in the context of just a single $U(1)$ symmetry.
 
For the Solution 3, we have $m=1/2$, $k_e=-13/6$, according to the charges of the  third row of Table (\ref{tbl:sol-leptch-GSTuvne0}). The fit of the coefficients of the neutrino mass matrix are completely analogous for Solution 2 and they appear in  the third column of Table (\ref{tabl:sol2gstneut}). The relevant parameters for the charged lepton sector are
\begin{eqnarray}
\frac{m_e}{m_{\mu}}&=&\frac{|a^e_{12}|^2}{|(a^e_{22}-a^e_{23}a^e_{32})|^2}\vep^{29/6}=s^{e\ 2}_{12},\nn \\
\frac{m_\mu}{m_\tau}&=&(a^e_{22}-a^e_{23}a^e_{32})\vep^{7/4}.
\end{eqnarray}
For this case {\it there is no need} to invoke another coefficient as for the Solution 2. $O(1)$ coefficients in this case can account for the masses and mixings in the leptonic sector. Once the coefficient $a^e_{23}$ is fitted in the charged lepton sector   then we need to use this parameter as a fixed parameter in the fit for the neutrino sector but in this case the fit is not as good as for the previous solution.  The results are presented in the third column of Table (\ref{tabl:sol2gscluvn0}).
\begin{table}[ht]
\begin{center}
\begin{tabular}{|r|l|l|}
\hline
\multicolumn{3}{|c|}{{Charged lepton Fitted Parameters}}\\ \hline
\multicolumn{2}{|c|}{{GST sol. 2, $u,v\neq 0$}}&
\multicolumn{1}{|c|}{{GST sol. 3, $u,v\neq 0$}}
\\ \hline
Parameter & BFP Value &  BFP Value \\
$a^e_{12}$& $0.56\pm 0.006$ & $2.88 \pm 0.032$ \\
$a^e_{22}$& $0.92\pm 0.013$ & $1.87 \pm 0.013$\\
\hline
\multicolumn{3}{|c|}{{Charged lepton Fixed Parameters}}\\ \hline
$\vep$ & $0.217$ & $0.154$\\
$a^e_{23}$ & $-1.6$ & $1.2$ \\ 
$a^e_{32}$ & $1.8$ & $1.2$ \\ \hline
\multicolumn{3}{|c|}{{$\chi^2$}} \\ \hline
$\chi^2$ & $0.05$ & $1.2\times 10^{-5}$ \\
\hline
\end{tabular}
\end{center}
\caption{\footnotesize{Charged lepton fitted parameters for the examples of Section \ref{sec:su5-solut-satisfy-GST}
The second column corresponds to the Solution 2 in the $u + v \neq 0$ case. The fourth column presents the fit to the Solution 3 in the $u + v \ne  0$ case.}}
\label{tabl:sol2gscluvn0}
\end{table}
\subsubsection{Fit 4: $SU(5)$ type ($u=v=0$) solution not satisfying the GST relation}

This is a $SU(5)$ type solution, hence $u=v=0$, which doesn't satisfy the GST relation. The charges are as laid out
in Eq.~(\ref{eq:31}). 

\subsubsection*{Quark sector}
Here we also use the expressions \eq{eq:yukeigen} and \eq{eq:mixsgeral} adapted to the solution \eq{eq:ydinl1l3A} for $r'_d=l_1-l_3=1$ and check the fit with an exact numerical solution, which agrees with the fit to \eq{eq:yukeigen} and \eq{eq:mixsgeral} within a $5\%$ error.
Since the fit can just fit eight parameters, in this case it is not possible to select out ``the best fit'', according to the criteria that we have used for the previous fits, so we present the following solution for the coefficients of the up and down Yukawa matrices:
\begin{eqnarray}
\label{eq:ngstsol1}
a^u&=&
 \left[
\begin{array}{ccc}
0.42 & 0.58 e^{-i\pi/2} & 0.51\\
0.58 e^{-i\pi/2} & 0.9  e^{-i\pi} &0.43e^{-i\pi/2} \\
0.51 &0.43e^{-i\pi/2} & 1 \\
\end{array}
\right],\nn\\
a^d&=&
 \left[
\begin{array}{ccc}
e^{-i0.5} & 0.8 & 0.29 e^{i 0.48}\\
1.63 e^{-i1.49} & 0.86 e^{-i1.2} &0.55e^{-i0.7} \\
e^{-i0.79} &0.4e^{-i0.5} & e^{-i3.05} \\
\end{array}
\right].
\end{eqnarray}
For this fit we have $\chi^2=2.31$.
\subsubsection*{Lepton sector}
In the lepton sector, once we have done the fit to the quark masses, the $SU(5)$ relations produce acceptable values for the charged lepton masses, what we need to care about are the mixings for the neutrino sector. According to the expressions for the mixings in the $(1,2)$ and $(1,3)$ neutrino sector, \eq{eq:mixangngst}, now $t^\nu_{13}=a^\nu_{13}\vep/\sqrt{a^{\nu \ 2}_{33}+a^{\nu\ 2}_{23}}$ and 
 $t^\nu_{12}=a^\nu_{12}\vep^{1/4}/(c^{\nu}_{23}a^{\nu}_{22}-s^{\nu}_{23}a^{\nu}_{32}))$, for $(n_2,n_3)=(-3/8,0)$. On the other hand the mixings in the charged lepton sector go as $t^e_{12}=|a^d_{21}+3a^d_{23}a^d_{31}/a^d_{33}|\vep/3|a^d_{22}+3a^d_{32}a^d_{23}|$ and $t^e_{13}=a^d_{31}\vep/|a^d_{33}+|a^{d}_{32}|^2|$, so here these contributions are important to the $U_{MNS}$ $s^l_{12}$ and $s^l_{13}$ mixings, identified respectively to the solar and reactor mixings, for example for $s^l_{13}$ we have
\begin{eqnarray}
s^l_{13}&=&|c^{e}_{12}c^{e}_{13}s^\nu_{13}-c^\nu_{13}(e^{i(\beta^{e}_1-\beta^\nu_1)}c^\nu_{23}(c^{e}_{12}c^{e}_{23}s^{e}_{13}+e^{i\beta^{e}_3}s^{e}_{12}s^{e}_{23})\nn \\
& &-e^{i(\beta^{e}_2-\beta^{\nu}_2)}s^{\nu}_{23}(e^{i\beta^{e}_3}s^{e}_{12}c^{e}_{13}-c^{e}_{12}s^{e}_{13})s^{e}_{23})|.
\end{eqnarray}
The mixing $s^l_{23}$ is driven by the neutrino mixing $s^\nu_{23}$
\begin{eqnarray}
s^l_{23}c^l_{13}\approx |e^{i(\beta^{e}_2-\beta^\nu_2)}s^\nu_{23}c^{e}_{12}c^{e}_{23}-e^{i(\beta^{e}_1-\beta^\nu_1)}s^{e}_{23}c^\nu_{13}c^\nu_{23}|.
\end{eqnarray}
Despite all the contributions to the mixings $s^l_{13}$ and $s^l_{12}$ we can reproduce the observed masses and mixings in the neutrino sector with $O(1)$ coefficients and with out any phase in this sector, we just use the phases of the right handed quark matrix, which are given by
\begin{eqnarray}
\beta^{e}_1&=&{\rm{ArcTan}}\left[\frac{\sin(\phi^d_{33})}{\cos(\phi^d_{33})+|a^d_{32}|^2}\right]-\phi^d_{31}\nn\\
\beta^{e}_2&=&(\phi^d_{32}-\phi^d_{33})+\beta^{dR}_1\nn\\
\beta^{e}_3&=&(\phi^d_{22}-\phi^d_{21})-\beta^{dR}_2,
\end{eqnarray}
and are specified in \eq{eq:ngstsol1}.
The results of this fit are given in the second row of in Table (\ref{tabl:sol2nongstneut}).
\begin{table}[ht]
\begin{center}
\begin{tabular}{|r|l|l|}
\hline
\multicolumn{3}{|c|}{{Neutrino Fitted Parameters}}\\ \hline
\multicolumn{2}{|c|}{{Non GST sol. 1}}&
\multicolumn{1}{|c|}{{Non GST sol. 2}}\\
\hline
Parameter & BFP Value &  BFP Value\\
$a^\nu_{23}$& $1.6\pm 0.8$ & $2\pm 0.9$\\
$a^\nu_{13}$& $1.4\pm 0.7$ & $0.9\pm 0.3$ \\
$a^\nu_{12}$& $1\pm 0.6$ & $1.6\pm 0.3$ \\
$a^\nu_{22}$& $0.67\pm 0.27$ & $0.5 \pm 0.4$ \\
\hline
\multicolumn{3}{|c|}{{Neutrino Fixed Parameters}}\\ \hline
$\vep$ & $0.19$ & $0.185$\\
$a^e_{23}$ & $-3a^d_{32}=-1.2$ & $-3a^d_{32}=-1.25$\\
$a^\nu_{33}$ & $1$  & $1$  \\
$a^N_{22}$ & $2$ & $2$\\
$\sigma$        & $(4.5,0.5)$ & $(5,1)$ \\ 
$(n_2,n_3)$ &$(-3/8,0)$ & $(-5/8,-1/4)$\\
\hline
\multicolumn{3}{|c|}{{$\chi^2$}} \\ \hline
$\chi^2$ & $(5.09,4.77)$ & $(4.78,3.79)$\\ 
\hline
\end{tabular}
\end{center}
\caption{\footnotesize{Neutrino fitted parameters for two of the non GST examples of Section \ref{sec:su5-solutions-not-GST}
The second and third columns correspond respectively to solution 1 and 2 in the non GST $SU(5)$ ($u=v=0$) cases, 
for the first one we have used $r'_d=1$ and for the second $r'_d=3/2$. While we have fitted in the first case 
$t^\nu_{13}$ to saturate its current upper limit, we have allowed for the second case to be smaller than it. 
The first entry for $\sigma$ corresponds to the fit using $M_P$ and the second entry using $M_G$; analogously for $\chi^2$.}}
\label{tabl:sol2nongstneut}
\end{table}
\subsubsection{Fit 5: $SU(5)$ type ($u=v=0$) solution not satisfying the GST relation}

This is a $SU(5)$ type solution, and hence $u=v=0$ which doesn't satisfy the GST relation.
The textures are as laid out in Eq.~(\ref{eq:32}).

\subsubsection*{Quark sector}
Here we present the following solution for the case $r_d=l_1-l_3=\frac{3}{2}$, in this case the coefficients of the up and down Yukawa matrices:
\begin{eqnarray}
\label{eq:ngstsol1_5}
a^u&=&
 \left[
\begin{array}{ccc}
0.5& 0.6 e^{-i\pi/2} & 0.5\\
0.6 e^{-i\pi/2} & e^{-i\pi} &0.43e^{-i\pi/2} \\
0.5 &0.43e^{-i\pi/2} & 1 \\
\end{array}
\right],\nn\\
a^d&=&
 \left[
\begin{array}{ccc}
1& 0.72 & 0.29 e^{i 0.49}\\
1.82 e^{-i2.28} & 0.76 e^{-i1.12} &0.55e^{-i0.71} \\
e^{-i1.57} &0.4e^{-i0.41} & e^{-i2.951} \\
\end{array}
\right].
\end{eqnarray}
For this fit we have $\chi^2=2.10$.
\subsubsection*{Lepton sector}
The analysis of this fit is completely analogous to the Fit 4, the results of the fitting procedure is presented in the second column of Table \ref{tabl:sol2nongstneut}.
\subsubsection{Top and bottom masses and $\mathbf{\tan\beta}$}
For these cases $\tan\beta$ and $a^d_{33}$ are a prediction, once the coefficient $a^u_{33}$ is fixed through the value of $m_t$, $m_t=Y^u_{33}v/\sqrt{2}$. The values of $a^u_{33}$, $a^d_{33}$ and $\tan\beta$ for the cases presented in this section are given in Table (\ref{tabl:tanbetares}). We can see that for a natural value of $a^u_{33}=1$ we have acceptable values for $\tan\beta$ (which should be $>2$) and $a^d_{33}$ in any of the cases presented.
\begin{table}[ht]
\begin{center}
\begin{tabular}{|r|l|l|l|}
\hline
\multicolumn{4}{|c|}{{$\tan\beta$, $a^u_{33}$ and $a^d_{33}$}}\\ \hline
\multicolumn{1}{|r|}{Parameter} &
\multicolumn{1}{|l|}{{GST sol. 2, $u\!=v\! =\! 0$}}&
\multicolumn{1}{|l|}{{GST sol. 2, $u,v\neq 0$}}&
\multicolumn{1}{|l|}{{GST sol. 3, $u,v\neq 0$}}
\\ \hline
$a^u_{33}$ & $(1,1.34)$ & $(1.1.3)$ & $(1,1.3)$\\
$a^d_{33}$  & $(5.33^{-1.13}_{+2.81},2.40^{-0.12}_{+0.13})$ & $(3.49^{-0.73}_{+1.84},1.62^{-0.09}_{+0.08})$ & $(3.23^{-0.68}_{+1.70}, 1.62^{-0.09}_{+0.10})$ \\
$\tan\beta$ & $(3.00^{-0.66}_{+4.82}, 1.00^{-0.06}_{+0.06})$ & $(3.00^{-0.66}_{+1.61}, 1.07^{-0.07}_{+0.07})$ & $(3.00^{-0.45}_{+1.61}, 1.07^{-0.07}_{+0.07})$ \\
\hline
($\epsilon$, $|k_d|$) & ($0.183$, $5/2$) &  ($0.217$, $5/2$) & ($0.154$, $2$)\\ \hline
\multicolumn{1}{|r|}{Parameter} &
\multicolumn{1}{|l|}{{$\!\!$Non GST sol. 1, $u\!=v\! =\!0\!\!\!$}}&
\multicolumn{2}{|l|}{$\!\!$Non GST sol. 2, $u\! = v\! =\!0\! $}\\ \hline
$a^u_{33}$ & $(1,1.2)$ & $(1,1.2)$ & \\
$a^d_{33}$ & $(2.12^{-0.44}_{+0.98},1.1^{-0.03}_{+0.08})$ & $(2.11^{-0.45}_{+0.87},1.3^{-0.06}_{+0.09})$  & \\
$\tan\beta$ & $(3^{-0.66}_{+1.61},1.3^{-0.1}_{+0.1})$ & $(3^{-0.78}_{+1.32},1.2^{-0.2}_{+0.2})$  & \\
\hline
($\epsilon$, $|k_d|$) & ($0.19$, $2$) & ($0.185$, $2$) & \\ \hline
\end{tabular}
\end{center}
\caption{\footnotesize{Value of $a^d_{33}$ and $\tan\beta$ for the different models presented, once $a^u_{33}$ is fixed using $m_t$.}}
\label{tabl:tanbetares}
\end{table}
\subsection{Comparison to the $SU(3)$ case}
In this section we present the comparison to a generic $SU(3)$ case
\cite{King:2001uz}. 
What we fit are the $O(1)$ coefficients of a Yukawa matrices of the form
\begin{eqnarray}
Y^f= \left[
\begin{array}{ccc}
\vep^8_f &\vep^3_f& \vep^3_f\\
\vep^3_f &\vep^2_f& \vep^2_f\\
\vep^3_f &\vep^2_f& 1
\end{array}
\right],
\end{eqnarray}
where we allow two different expansion paramaters $\vep_u$ and $\vep_d$ and complex phases to reproduce the CP violation phase. It is enough to consider one different phase in each of the $Y^u$ and $Y^d$ matrices. Here we put the phases on $Y^d_{13}$ and $Y^u_{12}$ \cite{Ross:2004qn}, but we have the freedom to use other choices.  We have used here as well the method of minimization that we have used for the $U(1)$ cases. The results of these fits are consistent with previous determination of these parameters, \cite{Roberts:2001zy, Ross:2004qn}, taking into account the change induced by the change of the value used here for the  parameter $m_c/m_s=15.5\pm 3.7$ and the different methods used for the determination of coefficients{\footnote{In \cite{Roberts:2001zy, Ross:2004qn} $m_c/m_s=9.5\pm 1.7$.}}. The fits presented here are the fits with the lowest possible $\chi^2$ because of the minimization procedure.
\begin{table}[t]
\begin{center}
\begin{tabular}{|r|l||l|}
\hline
\multicolumn{3}{|c|}{{ Quark fitted Parameters, SU(3)-like case}}\\ \hline
Parameter & BFP Value $\pm \sigma$  & BFP Value $\pm \sigma$   \\
$a{'u}_{22}$& $1.11\pm 0.55$  & $1.11\pm 0.07$ \\
$a^d_{12}$& $0.66\pm 0.32$  & $2.45\pm 0.20 $ \\
$a^d_{13}$& $0.10\pm 0.12$  & $0.91\pm 0.15$ \\
$a^d_{22}$& $0.74\pm 0.10 $  & $1.77\pm 0.09$ \\
$a^d_{23}$& $0.45\pm 0.29 $  & $1.18\pm 0.12$ \\
$\epsilon^u$& $0.05\pm 0.007$  & $0.05\pm 0.007$ \\
$\epsilon^d$& $0.25\pm 0.03$ &  $0.16 \pm 0.02$ \\
$\cos(\Phi_2)$& $0.516\pm 0.1$  & $0.450\pm 0.045$ \\ \hline
\multicolumn{3}{|c|}{{ Quark Fixed Parameters, SU(3)-like case}}\\ \hline
$\Phi_1^*$ & $-1.25 \approx=-0.8\pi/2$   &  $1.120 \approx 0.7\pi/2$ \\ \hline
\multicolumn{3}{|c|}{{$\chi^2$}} \\ \hline
$\chi^2$ & $0.972$ & $0.974$\\
\hline
\end{tabular}
\end{center}
\caption{\small{Fitted and fixed parameters for the $SU(3)$-like case.}}
\label{tabl:paru1su3}
\end{table}
We have not included here for the $SU(3)$ case a fit in the neutrino sector because in the $SU(3)$-like cases the neutrino sector requires more assumptions than in the analogous $U(1)$ cases.

Another important difference between the $SU(3)$ and the $U(1)$ cases presented here is that in the first one there are 
two parameter expansions $\vep^u$ and $\vep^d$ which have been fitted while in the $U(1)$ cases there is only one expansion 
parameter which can be fixed by relating the $U(1)$ symmetry to the cancellation of anomalies and the Fayet-Iliopoulos term. 
This has allowed that more $O(1)$ coefficients have been able to be fitted.

By comparing Tables (\ref{tabl:sol2gst}) and (\ref{tabl:paru1su3}) we can see that according to the minimization 
procedure and the criteria of $O(1)$ coefficients, the second case of the $SU(3)$ solution fits better the data.  
However the $U(1)$ solutions also have a good fit and taking into account the fact that for the neutrino sector we 
just have added the SRHND conditions, the fits in both of the $U(1)$ cases presented are good. We can therefore consider 
that $U(1)$ symmetries are still an appealing description of the fermion masses and mixings observed. Note that although 
the Solution 3 in the $u\neq v\neq 0$ does not fit the data as well as the Solution 2 (in either case, $u=v=0$ or not) in 
the quark sector, it does reproduce masses and mixings in the charged lepton sector. We have for this case $Y^e\neq (Y^d)^ T$ 
but we have $m_b\approx m_{\tau}$ without introducing ad-hoc $O(1)$ coefficients in order to reproduce the appropriate mixings.
\begin{table}[ht]
\begin{center}
\begin{tabular}{|l|c|c|c|}
\hline
\multicolumn{4}{|c|}{{ Comparison}}\\ \hline
  &   $U(1)$ (GST)& $U(1)$ (Non-GST)&  $SU(3)$-like \\
\# of expansion pars.       &1& 1&2\\
\# of free pars.(quark sector) &12& $>$18 &10\\
GST relation &yes & no & yes\\
prediction for $\tan\beta$ & small& small &no\\
lepton sector &o.k.&o.k&o.k\\
simple flavour charges &no &yes & yes\\
 \hline
\end{tabular}
\end{center}
\caption{\small{Some criteria of comparison. Here the number of free parameters corresponds to the number of coefficients, phases and parameter expansions that need to be adjusted or determined in the fits.}}
\label{tabl:compar}
\end{table}

Given the results of these fits we need further criteria in order to compare models based in anomalous $U(1)$ models and non-Abelian models, such as $SU(3)$. These other criteria may be found in the predictions that the models presented here can give in the
supersymmetric sector.

\section{Flavour issues in SUSY flavour symmetry models}
\label{sec:susyconst}

Since the flavour symmetry is expected to be broken at a high energy scale,
non supersymmetric models will have a hierarchy problem, since the cutoff
of the theory must at least be of the order of the flavour symmetry breaking scale.
Supersymmetric models with soft
breaking parameters around the TeV scale do not have this problem. For
this reason flavour symmetries are almost exclusively considered in the context
of one of the minimal supersymmetric models, or one of the popular SUSY unified
theories. The soft Lagrangian parameters are strongly constrained by the
supersymmetric flavour problem and the supersymmetric CP problem. 

The supersymmetric flavour problem needs the soft scalar mass squared matrices
to be diagonal to good approximation at high energy scales, since the off-diagonal
elements contribute to one-loop flavour violating decays such as the highly 
constrained $\mu\rightarrow e \gamma$ in the lepton sector and $b\rightarrow s\gamma$ in
the quark sector. It also requires that the trilinear couplings are aligned well to
the corresponding Yukawa matrix, since off diagonal elements in the trilinears in
the mass eigenstate basis also contribute to highly constrained decays. 
The supersymmetric CP problem is related to the phases of the parameters in
the soft Lagrangian. The general requirement is that these phases need to be
small for the majority of soft breaking parameters.

The reason that these problems are relevant in the context of family symmetries
is that in general, the existence of the family symmetry and the fields that break
it can give dangerous contributions to the soft Lagrangian parameters. It would
be remiss to look at these models but not check whether CP violation or flavour
violation is likely to rule them out. The starting point for investigating these
problems is to consider the hidden sector part of the theory, which leads to the
size and phases of the vevs of the fields which break the $U(1)$ symmetry, $\theta$
and $\overline\theta$.

\subsection{The flavon sector}
\label{sec:flavons}

We start by considering the values of the expansion parameters $\epsilon$ and $\overline{\epsilon}$.
They are defined by:
\begin{equation}
  \label{eq:epsilon}
  \epsilon \equiv \frac{\left<\theta\right>}{M} \;\; \; \overline{\epsilon} = \frac{\left<\overline{\theta}\right>}{M},
\end{equation}
where $\theta$ and $\overline{\theta}$ are scalars which break the $U(1)_F$ symmetry, and have charges of $1,-1$ respectively
under the symmetry. We wish to arrange that $\epsilon = \overline{\epsilon}$, which entails arranging that the
potential is minimized by $<\theta> = <\overline{\theta}>$. This would be simple if the $U(1)$ were non-anomalous, and thus
missing a Fayet-Iliopoulis term. If we set the $\theta$ sector of the superpotential to be:
\begin{equation}
  \label{eq:33}
  W_\theta = S(\theta \overline{\theta} - M_\theta^2)
\end{equation}
We introduce a new field, $X$, which has charge $q_X$ under $U(1)$. $q_X$ will be unspecified, but some number such that
when $<X> \ne 0$, it doesn't contribute to the fermion mass operators ( or, at the very least, it doesn't contribute at
leading order). Then, if we give $\theta$ and $\overline{\theta}$ the same soft mass
\footnote{This requirement may seem somewhat strong, but we also wish to minimize flavour violation coming from the D-term
associated with $U(1)$, which is proportional to $m_\theta^2 - m_{\overline\theta}^2$, and will provide a non-universal
contribution to the scalar masses. This contribution will lead to off diagonal elements in the SCKM basis which can easily
be dangerously large with regard to flavour violation.
}
, and require that $X$ doesn't get
a soft mass, we end up with a hidden sector potential:
\begin{equation}
  \label{eq:34}
  V = | \theta \overline{\theta} - M_\theta^2 |^2 + \frac{g^2}{2} \left( |\theta|^2 + |\overline{\theta}|^2 - q_X |X|^2 + \xi^2\right)^2 
  + m^2(\theta^2 + {\overline{\theta}^2}).
\end{equation}
If we minimize this potential with respect to $\theta, \overline{\theta}$ and $X$, we end up with the following constraints:
\begin{eqnarray}
  \label{eq:35}
  \frac{\partial V}{\partial \theta} &= 0 =& 2 \overline\theta ( \theta \overline\theta - M_\theta^2) + {g^2}\theta
  ( |\theta|^2 - |\overline\theta|^2 + q_X |X|^2 + \xi^2 ) + 2 m^2 \theta\\
  \label{eq:36}
  \frac{\partial V}{\partial \overline\theta} &= 0 = &
  2 \theta ( \theta \overline\theta - M_\theta^2 ) + {g^2} \overline \theta  ( |\theta|^2 - |\overline\theta|^2 + q_X |X|^2 + \xi^2)
  + 2m^2 \overline\theta \\
  \label{eq:37}
  \frac{\partial V}{\partial X} & = 0 = & \frac{g^2}{2} 2 X  ( |\theta|^2 - |\overline\theta|^2 + q_X |X|^2 + \xi^2 )  
\end{eqnarray}

Since $X$ doesn't have a mass term, it would be massless unless $<X> \ne 0$. Therefore, of the two solutions of Eq.~(\ref{eq:37}),
we have to take $X \ne 0$. From this, we see that:
\begin{equation}
  \label{eq:38}
  |\theta|^2 - |\overline\theta|^2 + q_X |X|^2 + \xi^2 = 0
\end{equation}
Substituting Eq.~(\ref{eq:38}) into Eq.~(\ref{eq:35}) and Eq.~(\ref{eq:36}), and multiplying by $\overline\theta$ and $\theta$ respectively,
we find:
\begin{eqnarray}
  \label{eq:39}
  0 &=&  \theta\overline\theta(\theta\overline\theta - M^2_\theta) + m^2 \theta^2 \\
  0 &=&  \theta\overline\theta(\theta\overline\theta - M^2_\theta) + m^2 {\overline\theta}^2
\end{eqnarray}
From this, we can deduce that either $\theta = \overline\theta = 0$ or $|\theta| = |\overline\theta| = M_\theta$. The potential is minimized
by the second solution if $m^2 < 2 M_\theta^2$. As we expect $M_\theta$ to be a GUT scale mass, and $m$ to be a TeV scale soft mass term, we
find, that as desired, that we will have:
\begin{eqnarray}
  \label{eq:40}
  <\theta> = <\overline\theta> \Rightarrow \epsilon = \overline\epsilon
\end{eqnarray}

This allows us to consider Yukawa textures without having to keep track of whether the overall charge for each
term is positive or negative.

\subsubsection{Getting $\epsilon$ from the Fayet-Iliopoulos term}
\label{sec:epsilon-from-FI}

The GST requirement leads to needing flavon fields with opposite charges. under $U(1)_F$. Were this not the case, we would have an
elegant way of generating $<\theta>$. Consider a simple case where $\theta$ doesn't have a superpotential mass term, but does
have a soft mass:
\begin{equation}
  \label{eq:41}
  V = \frac{g^2}{2} ( -|\theta|^2 + \xi^2 )^2 + m^2_\theta \theta^2
\end{equation}
Then, without the need for an explicit mass term in the superpotential, we would find that minimizing the potential with respect
to $\theta$ would lead to:
\begin{equation}
  \label{eq:42}
  <\theta> = \xi \sqrt{1 + \frac{m_\theta^2}{\xi^2}} \approx \xi
\end{equation}
Where the final approximation is due to the fact that we expect $\xi^2$ to be much larger than $m^2_\theta$. 
So we have managed to set $<\theta>$ from $\xi$, which can be predicted from string theory. So this allows one to
predict the flavon vev, rather than having to put it in by hand. 

This provides a motivation for trying to set up the case where $<\theta>$ and $<\overline\theta>$ could both be
set by the FI term. However, it doesn't seem possible to make this work without adding in either an extra symmetry,
or extra matter. Even then, trying to arrange things so that $<\theta> = <\overline\theta> = z \xi$, with $z$ some
real number is difficult. 

\subsection{Yukawa Operators}

Since the net $U(1)$ charge can be either positive or negative and we have $\epsilon = \overline\epsilon$, an effective potential has the following form:
\begin{eqnarray}  
  \nonumber
  W = \sum_{f=u,d;\;ij} & Q^i f^{c\;j} H_f &  a^f_{ij} \epsilon^{|q_i + f_j + h_f|} \\
  \label{eq:effectsup}
  + \sum_{f=e,n;\;ij} & L^i f^{c\;j} H_f & a^f_{ij} \epsilon^{|l_i + f_j + h_f|}.  
\end{eqnarray}
We cannot say anything in particular about the K\"ahler potential. We can assume that the phases responsible for CP violation only appear in the flavour sector.
Then observable CP violating phases will be put into the Yukawa couplings indirectly from the effective superpotential of Eq.~(\ref{eq:effectsup}). In general 
we can consider an effective K\"ahler potential of the form:
\begin{eqnarray}
K=K_o(t_\alpha)-\ln(S+\bar{S}+\delta_{GS})+ \sum_i f_i (t_\alpha) \theta_i \bar{\theta_i}+...+\sum_{ij} K^{\Phi}_{ij}\Phi^i\bar{\Phi}^j
\end{eqnarray}
where $K_o$ is the K\"ahler potential of the moduli fields,  $t_\alpha=T_\alpha+\bar{T}_\alpha$, $S$ is the dilaton, 
$f _i(t_\alpha)$ are possible functions of these moduli fields e.g. $f(t)=\Pi^p_{\alpha=1}t_{\alpha}^{n(\alpha)_{ij^*}}$. But we cannot specify 
the form of the K\"ahler metric. 
It may be that the K\"ahler metric is canonical, in which case $K^{\Phi}_{ij^*}=\delta_{ij^*}$. Such a form has a good change of leading to
acceptable phenomenology, since the scalar mass matrices will be proportional to the identity at the appropriate high energy scale. When
rotating the scalar mass matrices to the super-CKM (SCKM) basis at the high energy scale, the transformation will leave the mass matrices invariant.
Flavour violation tends to be proportional to off-diagonal elements in the scalar mass matrices in the SCKM basis, so any flavour violation will
be due to RG effects, and will therefore be suppressed. On the other hand, the K\"ahler metric could have off-diagonal structure, in which case
the risk of flavour violating effects would be high, and the case where the K\"ahler metric is diagonal but non-universal is potentially very interesting since flavour changing effects are induced in general by the SCKM rotation.

\subsection{The SUSY CP problem}
\label{sbsec:susycppr}

\subsubsection{The $\mu$ problem}

In order to avoid the $\mu$ problem, a symmetry or other mechanism to protect $\mu$ from unwanted contributions needs to be introduced.
The $\mu$ parameter can have contributions from the superpotential, (expected to be at the Planck scale) and from the K\"ahler potential,
via the Giudice-Masiero mechanism \cite{Giudice:1988yz} or other mechanisms \cite{Casas:1992mk,Kim:1994eu}, $\mu = \mu_W + \mu_K$. The charges of the fields $H_u$ and $H_d$ under the flavour symmetry 
can be chosen in such a way that $\mu_W(M_P)$ is forbidden in the superpotential. Then another field, $S$ can be introduced, so that the term 
$\lambda S H_uH_d$ is allowed in the K\"ahler potential, which generates an effective $\mu = O(m_{3/2})$. 
Note that in the cases that we have found for $u+v\neq 0$ there is no $\mu_W$ at $M_P$. In general for a theory containing two flavon fields with opposite charges, once the flavour symmetry 
is broken below the Planck scale, the contributions to the $\mu$ term are:
\begin{eqnarray}
\label{eq:mubreaku3s}
\epsilon^{|u+v|} H_u H_d \mu_W + \epsilon^{|u+v|} H_u H_d \mu_K
\end{eqnarray}
Thus, even if the $\mu$ term is missing from the superpotential at renormalizable level, it will be generated by non-renormalizable
operators once the family symmetry is broken. However, it will appear suppressed by a factor of $\epsilon^{|u+v|}$. To get an sufficient
suppression, either $|u+v|$ must be large or $\epsilon$ must be small.
Obviously, since the same factor $\epsilon^{|u+v}|$ appears suppressing both superpotential and K\"ahler potential $\mu$ contributions,
there is no extra constraint from considering the second term in Eq.~(\ref{eq:mubreaku3s}).

However, $|u+v|$ is related to the anomaly cancellation conditions considered in Section \ref{sec:anomconst}. There are two possibilities
for having small $|u+v|$. The first is to have small expansion parameters, $\epsilon$; however if $\epsilon$ becomes too small, it makes
predicting the fermion mass hierarchy very difficult. The second is to accept a contribution to $\mu$ that is larger than order O($m_{3/2}$);
however phenomenologically, the total $\mu$ should not be much bigger than the $O(m_{3/2})$. It is, however, possible to apply a new discrete
symmetry to disallow the superpotential $\mu$ term, which never allows any flavon corrections to generate it. 

\subsubsection{Electric dipole moment constraints}
The electric dipole moments (EDMs) constrain the form of the trilinear couplings, $(Y^A_{f})_{ij}$. The trilinear couplings are
defined through $(Y^A_{f})_{ij}H_{f}Q_i f^c_j$. Here we need to ensure that there is not a large contribution from the phases found 
in the trilinear terms to the CP violating phases. In the context of flavour symmetries it is usually postulated that the only phases 
appearing in the theory are in the Yukawa couplings and any other phase will enter as a consequence of a dependence in the Yukawa couplings. 
Then to check if the model gives contribution below the bounds one needs to compare the diagonal elements of the Yukawa couplings 
with the diagonal elements of the trilinear couplings, in the SCKM basis. The trilinear terms in general can be written as:
 \begin{eqnarray}
 \label{eq:trilinears}
 \mathbf{(Y^A_{f})}_{ij}=Y^{f}_{ij}F^a\partial_a\left(\tilde{K}+\ln(K^f_f K^i_i K^j_j) \right)
+F^a\partial_a  {Y}^{f}_{ij}
 \end{eqnarray}
We can always write the first term in a ``factorisable'' form \cite{Kobayashi:2000br}, such that if the Yukawa couplings, 
\eq{eq:effectsup}, are the only source of CP violation then the first term does not give any contribution at the leading order.
For the second term, which involves the derivative in terms of the flavon fields, if the flavon field is the only field with $F^\theta\neq 0$ then  the 
diagonal trilinear couplings in  the SCKM basis are real at leading order in the flavon fields \cite{Ross:2002mr}. 
Thus there is not an $O(1)$ contribution to the CP phases from this sector. 

One can check this simply by writing the last term of \eq{eq:trilinears} in the SCKM basis:  
\begin{eqnarray}
  \nonumber
  (F^a\partial_a({\hat Y}^f))^{\rm{SCKM}}_{ij} &=& F^a(V^\dagger_L)_{ik}(\partial_a V_L)_{kj}(Y_{\rm{Diag}})_{jj}+\\
  \label{eq:45}
  &&F^a(\partial_aY_{\rm{Diag}})_{ij}+F^a(Y_{\rm{Diag}})_{ii}
  (\partial_a V_R)_{ir}(V^\dagger_R)_{rj}
\end{eqnarray}
  Where $V^\dagger_L$ and $V^\dagger_R$ diagonalize the Yukawa matrix: $Y_{\rm{Diag}}=V^\dagger_L Y V^\dagger_R$. The leading term of the Eq.~(\ref{eq:45})
is the second term and it is at most of order $\theta$. 
If another field has non-zero F-term, $F^X\neq 0$ then all the quantities appearing in \eq{eq:trilinears} can be written as a expansion 
in $X$ and $\theta/M=\vep$:
\begin{equation}
  (Y_{\rm{Diag}})_{ii}=(a_{ii}+b_{ii}X)\vep^{p_{ii}}\label{eq:46}.
\end{equation}
We are assuming that only  the matter sector in 
\eq{eq:effectsup} has phases leading to CP violation, so the term $b_{ii}X\vep^{p_{ii}}$ is real and hence so is:
\begin{equation}
  F^X(\partial_a Y_{\rm{Diag}})_{ii}=F^X b_{ii}\theta^{p_{ii}}\label{eq:47}
\end{equation}
\subsection{SUSY flavour problem}
In addition to the F term contribution to the soft masses we have to add the D term contributions
 \begin{eqnarray}
(M^2)_{ij}=(M^2)_{F\ ij}+(M^2)_{D\ ij}.
 \end{eqnarray}
If the K\"ahler metric is diagonal in the basis where the symmetry is broken both contributions are diagonal and proportional to the K\"ahler metric. 
For example, consider universal SUGRA: $(M^2)_{F\ ij}=K_{ij}m^2_o$. However, even if we assume that the first term is indeed proportional to 
the  K\"ahler metric, the D-term will not in general be proportional to the K\"ahler metric:

\begin{equation}
\label{eq:48}
(M^2)_{D\ ij}= \sum_N g_N X_{N\ \theta_a} K_{ij^*}(\theta_a)m^2_{D},\;\;\; m^2_{D}=O(m^2_{3/2})
\end{equation}
The main problems for FC processes for these kind of theories are the contributions to the trilinear couplings from the anomalous D-term 
contribution to the soft masses \cite{Chung:2003fi}. For the last issue there is no real solution so far but one can ameliorate the problem by making all the
scalars heavier, which is a simply mass suppression.

In order to study all the possible consequences of models with the superpotential structure of \eq{eq:effectsup}, 
we can parameterize the K\"ahler metric according to the different contributions it may have, 
assuming a broken underlying symmetry with at least two flavon fields with opposite charges. 
Once this is done we can then study their consequences. As mentioned earlier, this analysis is beyond the scope of this paper, so
we just mention how extreme and dangerous situations may arise and we leave the analysis for a future reference \cite{inpreparation}. Some authors have studied possible consequences of flavour models for FC effects but very specific assumptions need to be assumed due to the many unknown supersymmetric parameters \cite{Babuetal, Ciuchini:2003rg,Masina:2003wt}.

The most strict bound for flavour changing processes is coming from the decay $\mu\rightarrow\ e \ \gamma$ \cite{Hisano:1995cp}-\cite{Masina:2002mv} and given the fact that we 
have a large mixing angle in the left handed sector of the charged lepton matrices it is crucial to determine under which conditions we can 
produce a suppressed effect.  Also the constraints given  by the process $B\rightarrow\ \Phi \ K_S$ may select out some of the possibilities presented.
\subsubsection{Non minimal sugra and diagonal K\"ahler metric}
Consider, for example, the case for which at the scale at which the flavour symmetry is broken, the K\"ahler metric is diagonal. For this case, we also
want the soft scalar mass matrices diagonal but not proportional to the unit matrix, due to possible different D term contributions. Since the general case it is difficult to handle we consider the case where $M^2_{\tilde f \ 1}-M^2_{\tilde f \ 2}$ is small and $M^2_{\tilde f \ 1}- M^2_{\tilde f \ 3}>0$.
In order to estimate the flavour changing processes we need to take into account the effects from renormalization group equations (RGE's) and then at the electroweak scale make the transformation to the basis where the fermions are diagonal. Here we consider the case of leptons, since we are interested in determining
$\delta^{l}_{ij}$ and in particular $\delta^{l}_{12}$ which is the most constrained parameter due to $B(\mu\rightarrow e\ \gamma)$.

We make an estimation of the contributions from the renormalization $\beta$ functions in this case, such that at the scale where the dominant right handed neutrino it is decoupled we can write the soft masses as
\bea
\label{eq:massren}
M^2_{\tilde L\ ij}(M_{Y})\approx M^2_{\tilde L \ ij}(M_X)-\frac{1}{16\pi^2} \ln\left(\frac{M_X}{M_Y} \right)(\beta^{(1)}_{M^2_{\tilde L\ ij}})
\eea
for $M_X=M_{\rm{G}}$ or $M_{\rm{P}}$, GUT or Planck scales respectively, and for 
$M_Y=M_{RR\ 3}$ in this case and considering just one loop corrections. The $\beta$ functions of $M^2_{\tilde L\ ij}$, from $M_X$ to $M_{RR\ 3}$ receive the contributions from the MSSM particles plus the contribution from right-handed neutrinos.
At $M_3$ we then run from that scale to the electroweak symmetry breaking scale with the appropriate $\beta$ function and matter content. In the case of SNRHD scenario and the form of the Yukawa matrices that we have considered in Section (\ref{sec:fitsmasses}) we can make the following approximations for the $\beta$ functions{\footnote{For the MSSM see for example \cite{Martin:1993zk}, when including right handed neutrinos, see for example \cite{Hisano:1995cp}.}}:
\bea
\label{eq:betasMSSM}
\left(\beta^{(1)}_{ M^2_{\tilde L\ ii} }\right)^{MSSM}\!\!\!\!\!\!\!&\!\approx\!&\!\!
2\left[(m^2_{M^2_{\tilde L\ ij}} +m^2_{\tilde H_d})\left(|Y_{2i}|^2+|Y_{3i}|^2\right) + m^2_{\tilde e_2}(1+a^2)(\left|Y_{2i}|^2+r^2_{\tilde e_{23}}|Y_{3i}|^2\right)\right]\nn\\
&& -6g^2_2|m_2|^2-\frac{6}{5}g^2_1|m_1|^2-\frac{3}{5}g^2_1 S\nn\\
\left(\beta^{(1)}_{ M^2_{\tilde L\ ij} }\right)^{MSSM}\!\!\!\!\!\!\!&\!\approx\!&\!\! (2m^2_{\tilde H_d}+m^2_{\tilde L\ i} +m^2_{\tilde L\ j})
\left( Y^{e *}_{2i}Y^{e*}_{2j} + Y^{e *}_{3i}Y^{e*}_{3j}\right)+\nn\\
&&+ 2m^2_{\tilde e_2}(1+a^2)\left(Y^{e *}_{2i}Y^{e*}_{2j} + r^2_{\tilde e_{23}} Y^{e *}_{3i}Y^{e*}_{3j} \right)
\eea
where we have assumed that the trilinear terms can be written as 
$A^f_{ij}=aY^f_{ij}M^2_{\tilde e}$, and $M^2_{\tilde e}$ is not necessarily diagonal. The parameter $S$, defined as $S=m^2_{\tilde H_u}-m^2_{\tilde H_d}+\rm{Tr}\left[M^2_{\tilde Q} -M^2_{\tilde L}-2 M^2_{\tilde u}+ M^2_{\tilde d}+ M^2_{\tilde e} \right]$, does not generate big contributions as long the masses involved remain somewhat degenerate. The $\beta$ functions generated by the dominant right-handed neutrino can be approximated by
\bea
\label{eq:betasMR3}
\left(\beta^{(1)}_{ M^2_{\tilde L\ ij} }\right)^{\nu_{M_3}}\!\!\!\!\!&\!\approx\!&\!\!2Y^{\nu *}_{3i}Y^{\nu}_{3j}\left[m^2_{\tilde L 3} +  m^2_{\tilde \nu 3}(1+b^2)+m^2_{\tilde H_u} \right]
\eea
From $M_X=M_3$ to $M_Y=M_{\rm{S}}$ -the supersymmetry breaking scale-, we consider   $\left(\beta  ^{(1)}_{ M^2_{\tilde L\ ij }}\right)^{MSSM}$. For this estimation we ignore the effect from $M_{\rm{S}}$ down to  the electroweak scale. At this scale we then transform the renormalized $M^2_{\tilde L}$ in the basis where the charged leptons are diagonal. Since there is a large mixing angle $(s^{e_L}_{23})$ in the left sector of $Y^e$ we are interested here only in estimating $(M^2_{\tilde L})_{LL}$. We can use the parameterization of Appendix A in order to make this transformation, i.e.
\bea
  \label{eq:49}
  Y^f_{\rm{diag}}=V^{f\dagger}_{L} Y^f V^f_{R},\quad
  (M^2_{\tilde L})'_{LL}=V^{f\dagger}_{L}  M^2_{\tilde L}  V^f_{L},
\eea
for $V^f_{L,R}$ as parameterized in \eq{eq:pardimatL}, with the $\beta$ phases as follow
\bea
\{\beta^{e_L}_1,\beta^{e_L}_2,\beta^{e_L}_3\}=\{\phi^{e}_{X_{23}},0,0\},\quad \phi^{e}_{X_{23}}=\beta^{e_L}_1-\beta^{e_L}_2.
\eea
Using these approximations, we obtain the following results
\begin{eqnarray}
(M^2_{\tilde L})^{\prime}_{12}&=&
s^{e_L}_{12}(c^{e_L}_{23} {m^2_{\tilde L\ 22 }} - {m^2_{\tilde L\ 11 }}) + \nn\\
&+&(c^{e_L}_{12})^2e^{-i\beta_{3L}}\left(c^{e_L}_{23} e^{-i\beta_{2L}}   {m^2_{\tilde L\ 12 }}-2t_{12}c^{e_L}_{23}s^{e_L}_{23}  e^{i\beta_{3L}} \rm{Re}\{ {m^2_{\tilde L\ 23}} e^{-i\chi}\}\right.\nn\\
&&-\left. s^{e_L}_{23}e^{-i\beta_{1L}}   {m^2_{\tilde L\ 13 }} \right),\nn\\
(m^2_{\tilde L})^{\prime}_{13}&=&
c^{e_L}_{23} s^{e_L}_{23}s^{e_L}_{12}e^{i\beta_3L}( {m^2_{\tilde L\ 22 }} - {m^2_{\tilde L\ 33 }}   ) +\nn\\
&+&  c^{e_L}_{12}c^{e_L}_{23} \left(\left( 
e^{-i\chi} c^{e_L}_{23}t_{12}  {m^2_{\tilde L\ 23 }}
-e^{i\chi} t_{12}t_{23}s^{e_L}_{23} \beta^*_{m^2_{\tilde L\ 23 }}
\right)\right.+\nn\\
&+&\left.t_{23}e^{i\chi} {m^2_{\tilde L\ 12 }}+   {m^2_{\tilde L\ 13 }}  \right)
\nn,\\
(m^2_{\tilde L})^{\prime}_{23}&=&
c^{e_L}_{23}s^{e_L}_{23}e^{i\beta_{3L}}\left({m^2_{\tilde L\ 22 }}- {m^2_{\tilde L\ 33 }}     \right)
  \nn\\
&&+e^{i\beta_{3L}}c^{e_L}_{12}\left((c^{e_L}_{23})^2e^{-i\chi}  {m^2_{\tilde L\ 23 }}   -  (s^{e_L}_{23})^2e^{i\chi} \beta^*_{m^2_{\tilde L\ 23 }}   \right)  ,\nn\\
(m^2_{\tilde L})^{\prime}_{11}&=&
(s^{e_L}_{12})^2\left((c^{e_L}_{23})^2 m^2_{\tilde L\ 22 }+ (s^{e_L}_{23})^2 m^2_{\tilde L\ 33 }   \right) 
\nn\\
(m^2_{\tilde L})^{\prime}_{22}&=&
(c^{e_L}_{12})^2\left((c^{e_L}_{23})^2  m^2_{\tilde L\ 22 }+(s^{e_L}_{23})^2  m^2_{\tilde L\ 33 }\right)-(c^{e_L}_{12})^2c^{e_L}_{23}s^{e_L}_{23} 2\rm{Re}\{ {m^2_{\tilde L\ 23}} e^{-i\chi}\}\nn\\ 
(m^2_{\tilde L})^{\prime}_{33}&=&
(c^{e_L}_{13})^2\left((s^{e_L}_{23})^2 m^2_{\tilde L\ 22 }+ (c^{e_L}_{23})^2 m^2_{\tilde L\ 33 } \right)+
c^{e_L}_{23}s^{e_L}_{23}\left( 2\rm{Re}\{ {m^2_{\tilde L\ 23}} e^{-i\chi}\}   \right)\nn\\
 \end{eqnarray}
here the soft masses $m^2_{\tilde L {ij}}$ are the soft masses at $M_{\rm{S}}$, renormalized from $M_X=M_{\rm{G}}, M_{\rm{P}}$ down to $M_3$ with the appropriate contributions from the dominant right handed neutrino, \eq{eq:massren}, and \eq {eq:betasMSSM}-\eq{eq:betasMR3} and then from $M_3$ to $M_S$ with the appropriate $\beta^(MSSM)$ functions. Thus we began with a diagonal matrix $M^2_{\tilde L}$ at $M_X$, then the RGE effects up to the scale where $M_3$ is decoupled generate a non diagonal matrix which receives more RGE contributions from $M_3$ to $M_S$. At electroweak scale we transformed to the basis where charged leptons are diagonal.
The mixing angles in this sector can be approximated as
\bea
s^{e_L}_{12}=|(a^e_{12}-t_{32}a^e_{13})|/|(a^e_{22}-a^e_{32}a^e_{23})|\epsilon^{p^e_{12}},\quad
s^{e_L}_{13}=a^e_{13}/a^e_{33}\epsilon^{p^e_{13}},\quad 
s^{e_L}_{23}=a^e_{23}/a^e_{33}
\eea
The powers $p^e_{ij}$ for the different solutions presented now correspond to 
$p^e_{12}=2/3,14/3$, $p^e_{13}=29/12,71/12$  for Fits 2 and 3 respectively.
So in this case we see that we need a big suppression of the element $(m^2_{\tilde l \ L})^{\prime}_{12}$ in order to be in agreement with the 
observed bound on $\mu\rightarrow\ e \gamma$. In the present example the suppression it is related to a bound on $( m^2_{\tilde L1}\! -\!m^2_{\tilde L2})$ and a relative big set of soft masses. The results of these estimations are presented in Table \ref{tbl:nonsugex}.

\begin{table}[ht]
  \centering
  \begin{tabular}{|c|c|c|c|}
\hline
\multicolumn{4}{|c|}{Estimation of $\delta_{ij}$ for the Fit.3 of Section \ref{sec:fitsmasses}.}\\
\hline
 Paramter & Ex. I & Ex. II & Ex. III\\
\hline
$m_{\tilde L 1}[\rm{GeV}]$ & 520 & 520 &520\\
$m_{\tilde L 2}[\rm{GeV}]$ & 530 & 530 & 570\\
$m_{\tilde L 3}[\rm{GeV}]$ & 500 & 500 & 230\\
$m_{\tilde e 1}[\rm{GeV}]$ & 520 & 520 & 520\\
$m_{\tilde e 2}[\rm{GeV}]$ & 530 & 530 & 550\\
$m_{\tilde e 3}[\rm{GeV}]$ & 500 & 300& 300\\
$M_1[\rm{GeV}]$ & 500 & 500 & 500\\
$M_2[\rm{GeV}]$ & 2$M_1$ & 700 &700\\
$M_{H_d}[\rm{GeV}]$ & 510 & 510 &510\\
$M_{H_u}[\rm{GeV}]$ & 510 & 510&510\\
$M_{\rm{S}}$ & 1000 & 1000&1000\\
\hline
$\overline{m}_{\tilde l}$ &514 &486&456\\
\hline
$x=m^2_{\tilde \gamma}/m^2_{\tilde l}$&\multicolumn{3}{|c|}{$0.3$}\\ 
$|(\delta^l_{LL})^E_{12}|$ &$4.3\times 10^{-3}$ & 
$5.6\times 10^{-3}$ &$1.4\times 10^{-3}$\\
$|(\delta^l_{LL})^B_{12}|$ &\multicolumn{2}{|c|}{$O(10^{-1})$}& $O(10^{-2})$\\ 
$|(\delta^l_{LL})^E_{13}|$ &$1.7\times 10^{-3}$ & $1.8\times 10^{-3}$&$1.9\times 10^{-2}$\\
$|(\delta^l_{LL})^E_{23}|$ &$5.7\times 10^{-2}$ & $6.4\times 10^{-2}$&$6.3\times 10^{-1}$\\
$|(\delta^l_{LL})^B_{23}|$ &\multicolumn{2}{|c|}{$O(10^{-1})$}& $O(10^{-1})$\\ 
\hline
  \end{tabular}
  \caption{Estimation of $|\delta^l_{ij}|^E$ in the fit 3 presented for the non minimal sugra example and its comparison to the observed bounds $|\delta^l_{ij}|^E$ \cite{Hisano:1995cp}-\cite{Masina:2002mv}.}
  \label{tbl:nonsugex}
\end{table}
As we can see from the results of Table (\ref{tbl:nonsugex}) the estimation of $|(\delta^l_{LL})^E_{ij}|$ is less dependent on the relation among the original soft mass terms $m^2_{\tilde L i}$ than on the value taken for the average s-lepton mass, which indeed needs to be large.
Here we note that this is just an estimation on the conditions that $B(\mu\rightarrow e\ \gamma)$ imposes on the soft masses, but with out fully checking whether or not appropriate masses for all the MSSM parameters can be obtained. In the following we consider a numerical investigation in the minimal sugra case.
\subsubsection{Numerical Investigation of $B(\mu\rightarrow e\ \gamma)$ in minimal sugra}
\label{sec:numer-invest-fits}
The presence of a right-handed neutrino fields leads to RG lepton flavour violation. Since the masses of the right handed neutrinos are so light for the GST solutions, fits 1-3, we attempted a numerical analysis for all of the fits of Section (\ref{sec:fitsmasses}) using the same modified version of SOFTSUSY \cite{Allanach:2001kg} as used in \cite{King:2003kf}. 

In order to get a good handle, we have embedded the flavour model fits into a string-inspired mSUGRA type scenario, with no D-term contribution to the scalar masses. This scenario was chosen because it is expected to be the embedding with the lowest flavour violation. In the scenario,
$A_0, m^2_0, M_{1/2}$ are all related to a gravitino mass $m_{3/2}$.

 As $n_1$ was only constrained to be between $-\sigma/2$ and $0$, we allow it to vary within this
range. We define the model at the GUT scale as:
\begin{equation}
  \label{eq:23}
  m^2_0 = \frac{1}{4} m_{3/2}^2 
  \;\;,\;\;
  A^0 = \sqrt{\frac{3}{4}} m_{3/2}\;\;\;
  M_{1/2} = \sqrt{\frac{3}{4}} m_{3/2}.
\end{equation}
This setup of the soft parameters corresponds to benchmark point A in \cite{King:2003kf}.
The results are as follows, for Fit 1 the code being used can not generate any low energy data for this fit so we do not find any safe $B(\mu\rightarrow e \gamma)$ region using the conditions presented above.
The Fit 2  has $\mathrm{BR}({\mu\rightarrow e\gamma}) <= 10^{-30}$ which is unattainably low, thus this fit is plausible within the context of the minimal sugra conditions that have  been specified. 
The smallness of the branching ratio for fit 2 comes about because with no RG running, in mSUGRA this rate would be exactly zero. The RG flavour violation will come from terms proportional to ${Y^\nu}^\dag Y^\nu$, whose elements are tiny ( the largest is $O(10^{-14})$ ).

The Fit 3 generates a tachyonic s-electron for the full $(m_{3/2}, n_1)$ range.  This is not to say that this fit will always have a tachyonic s-electron in other, less trivial embeddings. 
Fits 4 and 5 produce regions below and above the experimental limits on $B(\mu\rightarrow e\gamma)$, the graphs for these fits appear in Tables (\ref{fig:br_meg_fit_4a}-\ref{fig:br_meg_fit_5b}).
\begin{figure}[htbp]
  \centering
  \input{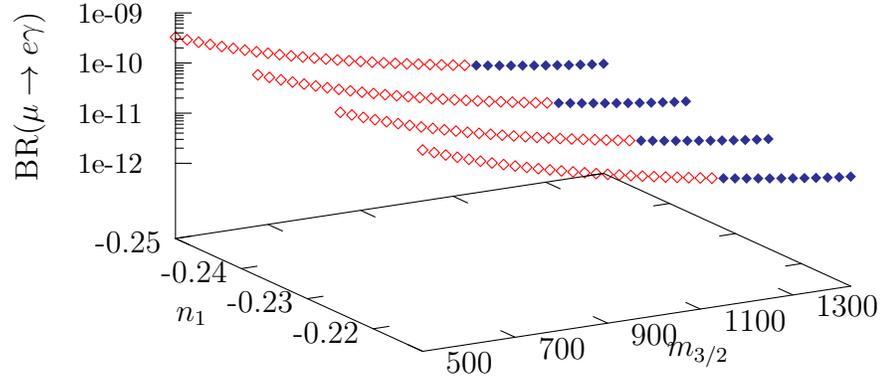}
  \caption{$\mathrm{BR}(\mu\rightarrow e\gamma)$ for fit 4, with $\left<\Sigma\right> = O(M_G)$. The solid points are below the experimental
limit of $1.1 \cdot 10^{-11}$, and the hollow points are above.}
  \label{fig:br_meg_fit_4a}
\end{figure}

\begin{figure}[htbp]
  \centering
  \input{fit4b.tex}
  \caption{$\mathrm{BR}(\mu\rightarrow e\gamma)$ for fit 4, with $\left<\Sigma\right> = O(M_{Pl})$.  The solid points are below the experimental
limit of $1.1 \cdot 10^{-11}$, and the hollow points are above}
  \label{fig:br_meg_fit_4b}
\end{figure}

\begin{figure}[htbp]
  \centering
  \input{fit5a.tex}
  \caption{$\mathrm{BR}(\mu\rightarrow e\gamma)$ for fit 5, with $\left<\Sigma\right> = O(M_G)$.  The solid points are below the experimental
limit of $1.1 \cdot 10^{-11}$, and the hollow points are above}
  \label{fig:br_meg_fit_5a}
\end{figure}

\begin{figure}[htbp]
  \centering
  \input{fit5b.tex}
  \caption{$\mathrm{BR}(\mu\rightarrow e\gamma)$ for fit 5, with $\left<\Sigma\right> = O(M_{Pl})$.  The solid points are below the experimental
limit of $1.1 \cdot 10^{-11}$, and the hollow points are above}
  \label{fig:br_meg_fit_5b}
\end{figure}

\section{Conclusions}
\label{sec:conclusions}
In summary, we began our analysis
by reviewing the Green-Schwartz (GS) conditions for anomaly
cancellation for theories based on a 
$U(1)$ family symmetry. We then used these conditions
to fix the charges of all the quark, lepton
and Higgs fields and studied possibilities where the Higgs mass 
$\mu$ term is either present or absent in the original
superpotential. The solutions which we constructed do not necessarily require
an underlying Grand Unified Theory (GUT) but may be
consistent with unification because of the GS conditions. 
Regardless of the presence of an explicit unified gauge group,
the explicit solutions can produce matrices of the form that are
identical to those that would be expected in 
an $SU(5)$ case or Pati-Salam unified theory, for example.

The flavour structure of the resulting Yukawa matrices is 
controlled by the charges of the quarks and leptons under the 
$U(1)$ family symmetry gauge group.
We have determined these charges which are consistent with
anomaly cancellation, and studied cases
which can reproduce quark Yukawa matrices satisfying
the Gatto-Sartori-Tonin (GST) relation, as well as other cases
which do not satisfy the GST relation. 
We find the GST relation to be an
appealing description of the value of the element $V_{us}$,
and the GST relation provides a useful criterion 
for classifying flavour models. 
In our view, having the Cabibbo angle emerging automatically
from a flavour model should have a similar status to gauge
coupling unification in a high scale model. 
Having classified the solutions in terms of the GST condition,
we then further classify
the solutions according to which of them can produce the
observed mixings in the lepton sector, and those that are consistent
with a sub-class of solutions based on the SRHND or sequential dominance
scenario with the further condition that the charges of the lepton
doublets for the second and third family are equal, $l_2=l_3$.  
We find that the GST solutions combined with SRHND results in 
highly fractional charges. 
On the other hand non-GST solutions with SRHND results in simpler
charges, and we have therefore studied both sorts of examples.

We have presented three numerical examples of solutions satisfying the
GST relation and two examples of non-GST solutions in order to compare
how well these solutions fit the experimental information while
maintaining $O(1)$ coefficients. For the GST solutions, one of these
examples corresponds to a model that can be thought of as coming from an
underlying $SU(5)$ and for which a $\mu$ term is allowed in the
superpotential. It is well known that in this case, given the relation
$Y^e=Y^{d \ T}$, there should be a Clebsch-Gordan coefficient
different in the charged lepton $(2,3)$ sector and in the $(2,3)$
d-quark sector in order to produce appropriate mixings in the
context of the $U(1)$ flavour symmetry and the GUT theory. Two other
GST examples are presented for which the $\mu$ term is not allowed and
which are not consistent with an underlying $SU(5)$, or other GUT theory. In
these cases $Y^e\neq (Y^d)^T$ but it is possible to maintain the
relation $m_\tau\approx m_b$ and in one of them just the $O(1)$
coefficients of the underlying $U(1)$ theory can account for the
appropriate mixings in the charged lepton and d-quark sector.
The non-GST cases also give a good description of masses and mixings,
although in this case we need to rely on further coefficients,
possible Clebsch-Gordan coefficients from an underlying GUT, in order
to achieve a good phenomenological description.
 
For the above examples we have provided detailed numerical fits 
of the $O(1)$ coefficients required to reproduce the observed
masses and mixings in both quark and lepton sectors.
The purpose of performing such fits
is to compare how well the different models can fit the data, 
and to try to determine quantitatively the best possible model
corresponding to the best possible fit. 
Although in the cases just mentioned the solutions which fit the
data best are the solutions consistent with an underlying $SU(5)$ theory, the
other two fits are quite plausible and represent interesting
possibilities which cannot be excluded. 
Since all the models constructed
have good agreement with the fermion masses and mixings, we clearly 
need further criteria in order to discriminate between the different
classes of $U(1)$ family symmetry models. 

One may ask the more general question whether family symmetries based on
abelian or non-abelian gauge groups are generically preferred?
In order to address this question, we have extended the fit to include
a generic symmetric form of quark and lepton mass matrices that can be
understood in the context of a theory based on $SU(3)$ family
symmetry. We have found that overall the generic $SU(3)$ family
symmetry produces Yukawa matrices which tend to fit the data better, 
although the effect is not decisive, and 
one cannot draw a strong conclusion based solely on fits to
fermion masses and mixings (or the way they can be reproduced). 
We have therefore enumerated 
some other possible criteria that are important in order to
further discriminate among different flavour theories. 
Including the effects from the supersymmetric sector provides 
an additional way to discriminate among different
theories based on their different predictions for 
soft masses and the resulting flavour changing processes and CP violation.  
We have presented two frameworks in which
these processes can be studied in the context of flavour theories. The
first is a non-minimal sugra scenario where family symmetries
may render the K\"ahler metric diagonal at the flavour symmetry breaking
scale, with off-diagonal elements arising only due to RG contributions and
the non-degeneracy of soft masses. The second framework is a minimal sugra
scenario for which a numerical exploration of $\mu\rightarrow e\ \gamma$
was performed. The results of this analysis
shows marked differences between the different models presented. Of the
GST cases only one survives the test of $B(\mu\rightarrow e\ \gamma)$
while for all of the non-GST cases presented there exist regions
compatible with the $B(\mu\rightarrow e\ \gamma)$ experimental limit.

In conclusion, 
at the present time, phenomenological analyses provide some guidance
about what family symmetry approaches may be valid, but do not yet allow
one to draw any firm conclusion. More specific assumptions or data in
the supersymmetric sector are needed in order to further discriminate 
between classes of models based on different family symmetry, 
unification or GST criteria.

\section*{Acknowledgments}
L. V-S. would like to thank the School of Physics and Astronomy at the U of Southampton for its hospitality during a visit last year. S.K. would like to thank the MCTP for its hospitality during August 2004 when this work was under development. The work of G. K. and L. V-S. is supported by the U. S. A. Department of Energy.

%
\newpage
\appendix
\section{Conventions for the Yukawa diagonalization matrices\label{ap:diagmat}}
We diagonalize the Yukawa matrices, $Y^f$, with the unitary matrices $V^f_L$ and $V^f_R$ such that $Y^f_{\diag}=V^{f\dagger}_LY^f V^{f}_R$. $V^f$ can be parameterized as
\begin{equation}
\label{eq:pardimatL}
V^{f\dagger}=
 \left[
\begin{array}{ccc}
e^{i\alpha^f_1}&0&0\\
0&e^{i\alpha^f_2}&0\\
0&0&e^{i\alpha^f_3}
\end{array}
\right]R^{f T}_{12}R^{f T}_{13}
 \left[
\begin{array}{ccc}
1&0&0\\
0&e^{i\beta^f_3}&0\\
0&0&1
\end{array}
\right]
R^{f T}_{23}
 \left[
\begin{array}{ccc}
1&0&0\\
0&e^{i\beta^f_2}&0\\
0&0&e^{i\beta^f_1}
\end{array}
\right],
\end{equation}
where a plane $R^f$ rotation has the form: 
\begin{equation}
R^f_{23}=
 \left[
\begin{array}{ccc}
1&0&0\\
0&c^f_{23}&-s^f_{23}\\
0&s^f_{23}&c^f_{23}
\end{array}
\right].
\end{equation}
In this notation, the CKM matrix is $V=V^{u\dagger}_L V^{d}_L$.

\section{Comparison to experimental information\label{ap:compinf}}
The experimental information determining $V_{\rm{CKM}}$, usually put in terms of the Wolfenstein parameters $A$, $\lambda$, $\rho$ and $\eta$, is extracted mainly from 
semileptonic decays of B mesons, CP violation in the K system, 
$B^0_{d,s}-\bar{B^0_{d,s}}$ oscillations, and CP asymmetries in various $B$ decays. 
We use a fit, based in a Bayesian approach (see for example \cite{Roberts:2001zy} and \cite{Caravaglios:2000an}),  of the the parameters $A$, $\lambda$, $\rho$ and $\eta$ 
including all the available information. Once we have done this, we compare the predictions of the mass textures with {\it the fitted parameters} because these include in a statistical way the experimental information from all the experiments considered. 
In the limit where we neglect all supersymmetric contributions to these
observables, the fitted values for $\rhobar$ and $\etabar$ are 
\begin{equation}
\label{rhoeta2003}
\rhobar=0.199^{+0.053}_{-0.049},\quad \etabar=0.328^{+0.037}_{-0.036},
\end{equation}
In order to compare to the $V_{\rm{CKM}}$ prediction, as given by  the $U(1)$ symmetries, we need to choose four elements, or combinations of them, we choose
\begin{equation}
\label{eq:vckmcndsF}
\frac{|V_{ub}|}{|V_{cb}|},\quad \frac{|V_{td}|}{|V_{ts}|},\quad |V_{us}|,\quad {\rm{Im}} \{ J \},
\end{equation}
where $J$ is the Jarlskog invariant. We choose these parameters because they can be put neatly in terms of the Wolfenstein parameters
\begin{eqnarray}
\frac{|V_{ub}|}{|V_{cb}|}=\frac{\lambda}{c_{\lambda}}\sqrt{\rhobar^2+\etabar^2},\quad \frac{|V_{td}|}{|V_{ts}|}&=&\frac{\lambda}{c_{\lambda}}\sqrt{(c_{\lambda}-\rhobar)^2+\etabar^2},\quad |V_{us}|=\lambda,\nn\\
{\rm{Im}}\{ J\}&=&A^2\lambda^6\etabar.
\end{eqnarray}
To include the information of the quark masses we use
\begin{eqnarray}
\label{eq:qksmasscndsF}
\frac{m_{u}}{m_{c}},\quad \frac{m_{c}}{m_{t}},\quad
\frac{m_{d}}{m_{s}},\quad \frac{m_{s}}{m_{b}}.
\end{eqnarray}
The ratios $\frac{m_{u}}{m_{c}}$ and $\frac{m_{d}}{m_{s}}$ can be determined from the best measured ratios of  the following mass ratios and the $Q$ parameter, which is determined accurately from chiral perturbation theory;
\begin{equation}
\frac{m_{u}}{m_{d}},\quad \frac{m_{c}}{m_{s}},\quad
Q=\frac{m_s/m_d}{\sqrt{1-(\frac{m_u}{m_d})^2}}
\end{equation}
We note here that a change in $m_s$ with respect to previous similar fits \cite{Roberts:2001zy} has an impact in the coefficients determined for the $SU(3)$ symmetry, although it is consistent with previous determinations if we consider the errors involved. We have used here $m_c/m_s=15.5\pm 3.7$ in contrast to $m_c/m_s=9.5\pm 1.7$ as used in \cite{Roberts:2001zy}.
\begin{table}[ht]
  \centering
  \begin{tabular}{|c|c|c|}
    \hline
Parameter & Exp. value & Value at $M_X$\\
\hline
$\frac{V_{ub}}{V_{cb}}$ & $(9.16\pm 0.67)\times 10^{-2}$&\\
$\frac{V_{td}}{V_{ts}}$ & $0.1989\pm 0.0093$&\\
$V_{us}$ &  $0.224\pm 0.0036$&\\
${\rm{Im}}\{ J\}$ & $(2.88\pm 0.4)\times 10^{-5}$& $(1.4\pm 0.5)\times 10^{-5}$\\
$\frac{m_u}{m_c}$ & $(1.9\pm 0.19)\times 10^{-3}$& \\
$\frac{m_c}{m_t}$ & $(7.5\pm 1.7)\times 10^{-3}$ & $(2.6\pm 1.8)\times 10^{-3}$ \\ 
$\frac{m_d}{m_s}$ & $(5.2\pm 0.35)\times 10^{-2}$ &\\
$\frac{m_s}{m_b}$ & $(6.4\pm 2.3)\times 10^{-2}$& $(4.5\pm 2.4)\times 10^{-2}$ \\
\hline
\end{tabular}
\caption{Experimental values used for the fit of numerical coefficients in the quark Yukawa matrices}
\label{tbl:expinputs}
\end{table}
We put in Table (\ref{tbl:expinputs}) the experimental values of the parameters that we use to determine the coefficients of the Yukawa texture. From equations \eq{eq:vckmcndsF} and  \eq{eq:qksmasscndsF} we see that we can fit only {\it eight} parameters of both of the Yukawa matrices as given by the $U(1)$ symmetries, but we will see that in most cases that is sufficient in order to account for the viability of a given ansatz or symmetry for the matrices. We also need to take into account the RGE effects in going from the electroweak scale to the scale at which the $U(1)$ symmetry breaks. We assume first that this is the GUT scale, so in order to determine the values of the parameters defining the $U(1)$ symmetry (analogously for the $SU(3)$ case) we take the values of the parameters appearing in Eqs. (\ref{eq:vckmcndsF}-\ref{eq:qksmasscndsF}) at the GUT scale. One of the reasons for using mass ratios instead of just masses is because the RGE effects on the mass ratios  has less impact than for the masses.
 This fit of the parameters defining the $U(1)$ symmetry is performed with the aid of the MINUIT package adapted for ${\rm{root}}$ \cite{rootref}. In this way we are able to compare how well a symmetry is fitted to the experimental information, and compare among the fits for different symmetries.

We do a completely analogous analysis in the neutrino sector, using the following observables
\begin{eqnarray}
\tan\theta_{23},\quad \tan\theta_{13},\quad \tan\theta_{12},\quad  m_{\nu_3}, \quad \frac{m_{\nu_2}}{m_{\nu_3}}
\end{eqnarray}
and their experimental values as appear in Table (\ref{tbl:expinputsneut})
\begin{table}[ht]
  \centering
  \begin{tabular}{|c|c|}
    \hline
Parameter & Exp. value \\
\hline
$\tan\theta_{23}$& $1.07\pm 0.37$  \\
$\tan\theta_{13}$& $0.21\pm 0.1$ (u.b) \\
$\tan\theta_{12}$& $0.65\pm 0.12$  \\
$m_{\nu_3}$& $0.05\pm 0.01$   \\
$\frac{m_{\nu_2}}{m_{\nu_3}}$& $0.19\pm 0.05$   \\
\hline
\end{tabular}
\caption{Experimental values \cite{Gonzalez-Garcia:2004jd} used for the fit of numerical coefficients in the neutrino Yukawa matrix. For $\tan\theta_{13}$ we have fitted using the upper bound.}
\label{tbl:expinputsneut}
\end{table}
\subsection{Evaluation of observables with fitted parameters}
In this section we put the evaluation, at the scale $M_X$,  of the experimental inputs  using the fitted parameters in order to compare with Table \ref{tbl:expinputs}.
\begin{table}[ht]
  \centering
  \begin{tabular}{|c|l|l|l|}
    \hline
\multicolumn{1}{|c|}{{ }}&
\multicolumn{1}{|c|}{{$\!$GST sol. 2 $\!u=\!v\!=\!0\!$}}&
\multicolumn{1}{|c|}{{$\!$GST sol. 2 $\!u\!\neq\! v\! \neq 0\!$}}&
\multicolumn{1}{|c|}{{$\!$GST sol. 3 $u\!\neq\! v\! \neq\! 0\!$}}
\\ \hline
\multicolumn{1}{|c|}{{\small{Parameter}}}&
\multicolumn{3}{|c|}{{Evaluation at $M_X$}}\\ \hline
$\frac{V_{ub}}{V_{cb}}$ & $(9.15\pm 0.52)\times 10^{-2}$& $(9.12\pm 0.35)\times 10^{-2}$ & $(9.31\pm 0.27)\times 10^{-2}$ \\
$\frac{V_{td}}{V_{ts}}$ & $0.199\pm 0.081$ & $0.199\pm 0.031$ & $0.20\pm 0.078$  \\
$V_{us}$ &  $0.225\pm 0.0041$ & $0.225\pm 0.0043$ & $0.225\pm 0.0040$ \\ 
${\rm{Im}}\{ J\}$ & $(1.4\pm 0.3)\times 10^{-5}$ & $(1.4\pm 0.3)\times 10^{-5}$ & $(0.8\pm 0.6)\times 10^{-5}$ \\
$\frac{m_u}{m_c}$ & $(1.9\pm 0.21)\times 10^{-3}$ & $(1.9\pm 0.16)\times 10^{-3}$  & $(1.8\pm 0.21)\times 10^{-3}$ \\
$\frac{m_c}{m_t}$ & $(2.1\pm 0.73)\times 10^{-3}$ & $(1.9\pm 0.47)\times 10^{-3}$  & $(4.\pm 1.54)\times 10^{-3}$   \\ 
$\frac{m_d}{m_s}$ & $(4.9\pm 0.5)\times 10^{-2}$ & $(4.9\pm 0.5)\times 10^{-2}$ & $(4.9\pm 0.6)\times 10^{-2}$ \\
$\frac{m_s}{m_b}$ & $(3.0\pm 1.7)\times 10^{-2}$ & $(1.9\pm 2.1)\times 10^{-2}$ & $(4.4\pm1.3)\times 10^{-2}$\\
\hline
$\tan\theta_{23}$& $1.27\pm 0.39$ & $1.02\pm 0.41$ & $1.00\pm 0.42$ \\
$\tan\theta_{13}$& $0.21\pm 0.32$ & $0.21\pm 0.39$ & $0.27\pm 0.19$ \\
$\tan\theta_{12}$& $0.65\pm 0.41$ & $0.65\pm 0.42$ & $0.65\pm 0.32$ \\
$m_{\nu_3}$& $0.046\pm 0.05$  &  $0.023\pm0.05$  &  $0.02 \pm 0.05$\\
$\frac{m_{\nu_2}}{m_{\nu_3}}$& $0.19\pm 0.05$ & $0.16\pm 0.08$ & $0.19\pm 0.05$ \\
\hline
\multicolumn{1}{|c|}{{ }}&
\multicolumn{1}{|c|}{{$\!$Non GST sol. 1 $\!u=\!v=\!0\!$}}&
\multicolumn{1}{|c|}{{$\!$Non GST sol. 2 $\!u=\!v=\!0\!$   }}&
\multicolumn{1}{|c|}{{ }}
\\ \hline
\multicolumn{1}{|c|}{{\small{Parameter}}}&
\multicolumn{3}{|c|}{{Evaluation at $M_X$}}\\ \hline
$\frac{V_{ub}}{V_{cb}}$ & $(9.59\pm 0.49)\times 10^{-2}$& $(9.14\pm 0.65)\times 10^{-2}$& \\
$\frac{V_{td}}{V_{ts}}$ & $0.201\pm 0.073$ & $0.199\pm 0.033$ &  \\
$V_{us}$ &  $0.225\pm 0.0037$  &  $0.225\pm 0.0027$ &  \\ 
${\rm{Im}}\{ J\}$ & $(1.7\pm 0.4)\times 10^{-5}$  & $(1.4\pm 0.3)\times 10^{-5}$ &   \\
$\frac{m_u}{m_c}$ & $(1.9\pm 0.33)\times 10^{-3}$ & $(1.8\pm 0.15)\times 10^{-3}$ &\\
$\frac{m_c}{m_t}$ & $(1.1\pm 0.82)\times 10^{-3}$ & $(2.5\pm 0.56)\times 10^{-3}$ &    \\ 
$\frac{m_d}{m_s}$ & $(5.2\pm 0.3)\times 10^{-2}$ & $(4.9\pm 0.5)\times 10^{-2}$ & \\
$\frac{m_s}{m_b}$ & $(3.2\pm 1.2)\times 10^{-2}$ & $(4.5\pm 2.2)\times 10^{-2}$ &  \\
\hline
$\tan\theta_{23}$& $0.67\pm 0.41$ & &\\
$\tan\theta_{13}$& $0.19\pm 0.16$ & &\\
$\tan\theta_{12}$& $0.55\pm 0.31$ & &\\
$m_{\nu_3}$& $0.038\pm 0.04$ & & \\
$\frac{m_{\nu_2}}{m_{\nu_3}}$ & $0.13\pm 0.06$ & &\\
\hline
\end{tabular}
\caption{Evaluated values of experimental observables using the fitted parameters for the $SU(5)$ GST case, for the $u\neq -v\neq 0$ cases, the $SU(5)$ non GST case and the $SU(3)$ like cases considered in Section (\ref{sec:fitsmasses}).}
\label{tbl:gstevaluations}
\end{table}
\begin{table}[ht]
  \centering
  \begin{tabular}{|c|l|l|}
    \hline
\multicolumn{1}{|c|}{{ }}&
\multicolumn{1}{|c|}{{$SU(3)$, $\Phi_1<0$}}&
\multicolumn{1}{|c|}{{$SU(3)$, $\Phi_1>0$ }}
\\ \hline
\multicolumn{1}{|c|}{{Parameter}}&
\multicolumn{2}{|c|}{{Evaluation at $M_X$}}\\ \hline
$\frac{V_{ub}}{V_{cb}}$ & $(9.14\pm 0.65)\times 10^{-2}$& $9.13\pm 0.21)\times 10^{-2}$ \\
$\frac{V_{td}}{V_{ts}}$ & $0.199\pm 0.023$ & $0.198\pm 0.024$  \\
$V_{us}$ &  $0.225\pm 0.0027$ & $0.225\pm 0.0033$ \\
${\rm{Im}}\{ J\}$ & $(1.4\pm 0.3)\times 10^{-5}$ & $(1.4\pm 0.3)\times 10^{-5}$  \\
$\frac{m_u}{m_c}$ & $(1.8\pm 0.15)\times 10^{-3}$ & $(1.9\pm 0.15)\times 10^{-3}$  \\
$\frac{m_c}{m_t}$ & $(2.5\pm 0.56)\times 10^{-3}$ & $(2.5\pm 0.56)\times 10^{-3}$  \\ 
$\frac{m_d}{m_s}$ & $(4.9\pm 0.5)\times 10^{-2}$ & $(4.9\pm 0.5)\times 10^{-2}$ \\
$\frac{m_s}{m_b}$ & $(4.5\pm 2.2)\times 10^{-2}$ & $(4.5\pm 2.2)\times 10^{-2}$\\
\hline
\end{tabular}
\caption{Evaluated values of experimental observables using the fitted parameters for the $SU(3)$ like cases considered.}
\label{tbl:su3evaluations}
\end{table}
%
%

\end{document}